\documentclass[aps,prc,10pt,showpacs,showkeys,twocolumn,superscriptaddress,groupedaddress]{revtex4-1}
\usepackage[latin1]{inputenc}
\usepackage[english]{babel}
\usepackage{graphicx,psfrag}
\usepackage{amsmath}
\usepackage{amssymb}
\usepackage{amscd}
\usepackage{eucal}
\usepackage{color}
\usepackage{bm}
\usepackage{braket}
\usepackage{dsfont}
\newcommand{\nmax}{N_{\rm max}}
\newcommand{\hb}{\hbar\Omega}

\begin{document}
\title{Structure of the exotic $^9$He nucleus from the no-core shell model with continuum}

\author{Matteo Vorabbi}
\email{mvorabbi@triumf.ca}
\affiliation{TRIUMF, 4004 Wesbrook Mall, Vancouver, British Columbia V6T 2A3, Canada}

\author{Angelo Calci}
\affiliation{TRIUMF, 4004 Wesbrook Mall, Vancouver, British Columbia V6T 2A3, Canada}

\author{Petr Navr\'atil}
\email{navratil@triumf.ca}
\affiliation{TRIUMF, 4004 Wesbrook Mall, Vancouver, British Columbia V6T 2A3, Canada}

\author{Michael K.G Kruse}
\affiliation{Lawrence Livermore National Laboratory, P.O. Box 808, L-414, Livermore, California 94551, USA}

\author{Sofia Quaglioni}
\affiliation{Lawrence Livermore National Laboratory, P.O. Box 808, L-414, Livermore, California 94551, USA}

\author{Guillaume Hupin}
\affiliation{Institut de Physique Nucl\'eaire, CNRS/IN2P3, Universit\'e Paris-Sud, Universit\'e Paris-Saclay, F-91406, Orsay, France}

\date{\today}

\begin{abstract}

\noindent
{\bf Background:} The exotic $^9$He nucleus, which presents one of the most extreme neutron-to-proton ratios, belongs to the $N=7$ isotonic chain famous for the phenomenon of ground-state parity inversion with decreasing number of protons. Consequently, it would be expected to have an unnatural (positive) parity ground state similar to $^{11}$Be and $^{10}$Li. Despite many experimental and theoretical investigations, its structure remains uncertain. Apart from the fact that it is unbound, other properties including the spin and parity of its ground state and the very existence of additional low-lying resonances are still a matter of debate. 

\noindent
{\bf Purpose:} In this work we study the properties of $^9$He by analyzing the $n+^8$He continuum in the context of the {\it ab initio} no-core shell model with continuum (NCSMC) formalism with chiral interactions as the only input. 

\noindent
{\bf Methods:} The NCSMC is a state-of-the-art approach for the {\it ab initio} description of light nuclei. With its capability to predict properties of bound states, resonances, and scattering states in a unified framework, the method is particularly well suited for the study of unbound nuclei such as $^9$He.

\noindent
{\bf Results:} Our analysis produces an unbound $^9$He nucleus. Two resonant states are found at the energies of ${\sim}1$ and ${~\sim}3.5$~MeV, respectively, above the $n+^8$He breakup threshold. The first state has a spin-parity assignment of $J^{\pi} = {1/2}^-$ and can be associated with the ground state of $^9$He, while the second, broader state has a spin-parity of ${3/2}^-$. No resonance is found in the ${1/2}^+$ channel, only a very weak attraction.

\noindent
{\bf Conclusions:} We find that the $^9$He ground-state resonance has a negative parity and thus breaks the parity-inversion mechanism found in the $^{11}$Be and $^{10}$Li nuclei of the same $N=7$ isotonic chain.

\end{abstract}


\maketitle

\section{Introduction}
\label{sec_intro}

The study of neutron-rich nuclei, located far from the line of stability, is one of modern nuclear physics frontiers. From a theoretical perspective, these nuclei
open new questions into the importance of many-body forces at extreme neutron excesses, and challenge our current computational techniques. From an experimental perspective, these nuclei are difficult to produce in sufficient quantities and are also challenging to analyze. Nevertheless, much interest has been generated by past experiments and theoretical calculations; an interest that will be further renewed once the next generation of rare-isotope facilities such as FRIB (USA) \cite{frib} become available.

The Helium isotopes chain, $^{3-9}$He, is one of the few accessible to both detailed theoretical and experimental studies. In the case of $^9$He, the neutron to proton ratio is $N/Z=3.5$, making it one of the most neutron extreme systems studied so far. The $^9$He system is particularly interesting theoretically since it is part of a series of $N=7$ isotones in which it is believed that intruder states from the $1s0d$ shell are pushed down in energy into the $0p$ shell, promoting the possibility of a positive parity ground state. $^{11}$Be is the most famous example having an un-natural parity assignment for the ground state, which has been calculated theoretically~\cite{PhysRevLett.4.469,PhysRevLett.70.1385,SAGAWA19931,VINHMAU199533,PhysRevC.69.041302,NUNES1996171,PhysRevC.93.011305,HamamotoBe11,
PhysRevC.66.024305,PhysRevLett.101.092501,PhysRevLett.117.242501,PhysRevLett.108.142501,PhysRevC.59.1545,PhysRevC.69.064604,PhysRevC.79.054603,PhysRevC.88.011601,PhysRevC.89.014333,PhysRevC.89.064609,PhysRevLett.119.082501} as well as observed experimentally~\cite{AUTON1970305,ZWIEGLINSKI1979124,FORTIER199922,Be11Exp,IWASAKI20007,WINFIELD200148,PhysRevLett.102.062503,PhysRevC.88.064612,KWAN2014210}.
The same phenomenon is found in $^{10}$Li \cite{Thoen99,SIMON2004323,JEPPESEN2006449,AKSYUTINA430,Falou2011}, making it quite natural to hypothesize that the same trend continues for $^9$He. Both experimental~\cite{Seth87,Bohlen88,VONOERTZEN1995c129,Bohlen99,PhysRevC.67.041603,fortier2007,alfalou2007,alfalou2011,JOHANSSON201015,PhysRevC.76.021605,chen01,PhysRevC.88.034301,UBERSEDER2016323} and theoretical~\cite{POPPELIER1985120,PhysRevC.46.923,HISASHI199316,Poppelier1993,Ogloblin1995,PhysRevLett.87.082502,PhysRevC.66.044310,BARKER200442,PhysRevLett.94.052501,PhysRevC.78.044302,PhysRevC.86.044330,Volya2014,PhysRevC.91.034306,2017arXiv171004727J} efforts have been dedicated in the past to probe this hypothesis.
Experimentally the situation is still under debate and a detailed history of the experimental studies on ${}^{9}$He can be found in Ref.~\cite{PhysRevC.88.034301}.
Here we provide a brief summary of the experimental and theoretical results concerning the still open questions of the spin-parity of the ground state, and the existence of excited states.

The first experiment on ${}^9$He was performed by Seth {\it et al.} \cite{Seth87} in 1987, who found an unbound ground state at $1.13 \pm 0.10$ MeV above the neutron decay threshold with spin-parity assignment of $J^{\pi} = {1/2}^-$. Successively, other experiments were
performed~\cite{Bohlen88,VONOERTZEN1995c129,Bohlen99,PhysRevC.67.041603}
confirming the same ${1/2}^-$ unbound ground state while revising its energy to $1.27 \pm 0.10$ MeV; in particular, in Ref.~\cite{Bohlen99} this state was identified as a
narrow resonance with a width of $\Gamma = 100 \pm 60$ keV. The conclusion was that $^9$He breaks the trend of parity inversion observed in $^{11}$Be and $^{10}$Li. This was also supported by some theoretical results~\cite{POPPELIER1985120,Ogloblin1995} while contradicting other
calculations~\cite{SAGAWA19931,HISASHI199316,Poppelier1993,PhysRevLett.87.082502}.

Then in 2001 Chen {\it et al.}~\cite{chen01} observed a ${1/2}^+$ ground state corresponding to a virtual state of energy less
than $0.2$ MeV above the neutron decay threshold characterized by an S-wave scattering length of $a_0\lesssim -10$~fm, indicating for the first time parity inversion in the $^{9}$He nucleus. These results were also consistent with shell model
calculations~\cite{PhysRevC.46.923}. The presence of a ${1/2}^+$ state was also reported
in subsequent works~\cite{PhysRevC.88.034301,alfalou2007,alfalou2011,JOHANSSON201015,PhysRevC.76.021605,fortier2007}, although a smaller absolute value of the S-wave scattering length $a_0\sim -3$~fm was reported in particular in Refs.~\cite{alfalou2007,alfalou2011,JOHANSSON201015}.  The measurements of Refs.~\cite{PhysRevC.88.034301,alfalou2007,alfalou2011,JOHANSSON201015,PhysRevC.76.021605,fortier2007}  also supported the existence of a narrow ${1/2}^-$ state at the energy around $1.3$~MeV, in agreement with Ref.~\cite{Bohlen99}. In particular, in Ref.~\cite{PhysRevC.88.034301} the authors reported
the observation of a ${1/2}^+$ state at $0.18 \pm 0.085$ MeV and a ${1/2}^-$ state at $1.2 \pm 0.1$ MeV.
The only exception to these observations is given by the work of Golovkov {\it et al.}~\cite{PhysRevC.76.021605}, where the ${1/2}^-$ resonant state was found at an energy of $2.0 \pm 0.2$ MeV with a width $\Gamma \sim 2$ MeV.
This large value for the resonance width was also confirmed theoretically using {\it ab initio} variational Monte
Carlo~\cite{PhysRevC.86.044330} and continuum shell model~\cite{Volya2014} calculations.

Different from these experiments, which aimed at directly accessing $^9$He states,
Uberseder {\it et al.}~\cite{UBERSEDER2016323} recently obtained spectroscopic information on $^9$He by studying the isospin $T = 5/2$ isobaric analog states in $^9$Li through $p+^8$He elastic scattering. The authors did not observe any narrow structures within
the energy range of interest and ruled out the existence of a narrow ${1/2}^-$ state in $^9$He. They also reported the evidence of a very broad $T=5/2$ state with spin ${1/2}^+$ at an excitation energy of $17.1$ MeV in $^9$Li, which corresponds to a broad state in $^9$He at an energy of approximately $3$ MeV above the neutron decay threshold with a width of $\Gamma \sim 3$ MeV.

From all these experimental results we clearly understand that two long-standing problems affect the physics of the $^9$He system and are still unsolved.
The main problem concerns the existence of the ${1/2}^+$ state, and - if it exists - what is its energy, while the second one concerns the discrepancy between the
theoretical predictions and the experimental observations for the width of the ${1/2}^-$ state. From the theoretical point of view, the recent
calculations~\cite{PhysRevLett.94.052501,Volya2014} do not predict parity inversion and suggest a ${1/2}^-$ ground state.
On the other hand, calculations presented in Refs.~\cite{PhysRevLett.87.082502,2017arXiv171004727J} predict a ${1/2}^+$ ground state.

In this paper we study the $^9$He nucleus by analyzing the $n{+}^8$He continuum in the framework of the {\it ab initio} no-core shell model with continuum (NCSMC)~\cite{PhysRevLett.110.022505,PhysRevC.87.034326,physcripnavratil} that treats bound and unbound states in a unified way. This approach is based on a basis expansion with two key components: one describing all nucleons close together, forming the $^{9}$He nucleus, and a second one describing the neutron and $^{8}$He apart. The former part is built from an expansion over square-integrable many-body states treating all nine nucleons on the same footing. The latter part factorizes the wave function into products of $^{8}$He and neutron components and their relative motion with proper bound-state or scattering boundary conditions. As the nuclear interaction input to our calculations we adopt nucleon-nucleon (plus three-nucleon) forces from chiral EFT~\cite{Weinberg1990,Weinberg1991}. 

The paper is organized as follows: in Section~\ref{sec_formalism} we outline the formalism of our calculation, giving a brief description of the NCSMC. We also detail our selection of input chiral interactions.
In Section~\ref{sec_results} we first present our results for the binding energies of $^{4,6,8}$He and then those obtained for $n+^8$He scattering in the NCSMC formalism. Finally, in Section~\ref{sec_conclusions} we summarize our findings and draw our conclusions.

\section{Theoretical Framework}
\label{sec_formalism}

\subsection{NCSM}
The no-core shell model (NCSM) ~\cite{PhysRevLett.84.5728,PhysRevC.62.054311,BARRETT2013131} treats nuclei as systems of $A$ non-relativistic point-like nucleons interacting through realistic inter-nucleon interactions. All nucleons are active degrees of freedom. The many-body wave function is cast into an expansion over a complete set of antisymmetric $A$-nucleon harmonic-oscillator (HO) basis states containing up to  $N_{\rm max}$ HO excitations above the lowest Pauli-principle-allowed configuration: 
\begin{equation}\label{NCSM_wav}
 \ket{\Psi^{J^\pi T}_A} = \sum_{N=0}^{N_{\rm max}}\sum_i c_{Ni}^{J^\pi T}\ket{ANiJ^\pi T}\; .
\end{equation}
Here, $N$ denotes the total number of HO excitations of all nucleons above the minimum configuration,  $J^\pi T$ are the total angular momentum, parity and isospin, and $i$ additional quantum numbers. The sum over $N$ is restricted by parity to either an even or odd sequence. The basis is further characterized by the frequency $\Omega$ of the HO well. Square-integrable energy eigenstates expanded over the $N_{\rm max}\hbar\Omega$ basis, $\ket{ANiJ^\pi T}$, are obtained by diagonalizing the intrinsic Hamiltonian.

\subsection{NCSMC}
\label{sec_ncsmc}

In most experiments, the properties of $^9$He are inferred from coincidence measurements involving a neutron and a $^8$He fragment being simultaneously detected. Thus, we can model the $^9$He continuum as a state of a neutron plus a $^8$He in relative motion. In this regard, the binary cluster formulation of the NCSMC is well suited in particular at energies below the $^8$He breakup threshold of $\sim 2.14$~MeV~\cite{TILLEY2004155}.

The $^9$He wave function is represented as the generalized cluster expansion
\begin{align}
\ket{\Psi^{J^\pi T}_{A\texttt{=}9}} = &  \sum_\lambda c^{J^\pi T}_\lambda \ket{^{9} {\rm He} \, \lambda J^\pi T} \nonumber \\
& +\sum_{\nu}\!\! \int \!\! dr \, r^2 
                 \frac{\gamma^{J^\pi T}_{\nu}(r)}{r}
                 {\mathcal{A}}_\nu \ket{\Phi^{J^\pi T}_{\nu r}} \,.\label{ncsmc_wavefunc}
\end{align}
The first term consists of an expansion over NCSM eigenstates of the aggregate system ($^{9}$He) indexed by $\lambda$. These states are well suited to explain the localized correlations of the 9-body system, but are inadequate to describe clustering and scattering properties. The latter properties are addressed by the second term corresponding to an expansion over the antisymmetrized channel states
\begin{align}
\ket{\Phi^{J^\pi T}_{\nu r}} = & 
		\Big[ \!\! \left(
        		\ket{^{8} {\rm He} \, \lambda_1 J_1^{\pi_1}T_1}\ket{n \, \tfrac12^{\texttt{+}}\tfrac12}
         	\right)^{(sT)} Y_\ell(\hat{r}_{8,1}) \Big]^{(J^{\pi}T)} \nonumber\\ 
	& \times\,\frac{\delta(r{-}r_{8,1})}{rr_{8,1}} \; ,
\label{eq:rgm-state}
\end{align}
in the spirit of the resonating group method ~\cite{wildermuth1977unified,TANG1978167,FLIESSBACH198284,langanke1986,PhysRevC.77.044002}, which describe the $^8\mathrm{He} + n$ in relative motion. Here, $\vec{r}_{8,1}$ is the separation between the center-of-mass of $^{8}$He and the neutron and $\nu$ is a collective index for the relevant quantum numbers. The $^{8}$He wave function is also obtained within the NCSM with the same Hamiltonian adopted for the whole system.

 The discrete expansion coefficients  $c_{\lambda}^{J^{\pi}T}$ and the continuous relative-motion amplitudes  $\gamma_{\nu}^{J^{\pi}T} (r)$ are the solution of the generalized eigenvalue problem derived by representing the Schr\"{o}dinger equation in the model space of expansion (\ref{ncsmc_wavefunc})~\cite{physcripnavratil}. The resulting NCSMC equations are solved by the coupled-channel R-matrix method on a Lagrange mesh~\cite{Descouvemont2010,Hesse1998,Hesse2002}.

\subsection{Interaction input}
\label{sec_interaction}

The microscopic Hamiltonian can be written as
\begin{equation}
\hat{H}=\frac{1}{A}\sum_{i<j=1}^A\frac{(\hat{\vec{p}}_i-\hat{\vec{p}}_j)^2}{2m}+\sum_{i<j=1}^A \hat{V}^{NN}_{ij}+\sum_{i<j<k=1}^A \hat{V}^{3N}_{ijk}+\dots,
\label{H}
\end{equation} 
with the interaction consisting of realistic nucleon-nucleon ($NN$) and typically also three-nucleon ($3N$) and even higher-body contributions that accurately reproduce few-nucleon properties. In the NCSM and NCSMC calculations we typically employ interactions derived in the framework of chiral effective field theory (EFT)~\cite{Weinberg1990,Weinberg1991}. Chiral EFT uses a low-energy expansion in terms of $(Q/\Lambda_{\chi})^n$ that allows for a systematic improvement of the potential by an increase of the chiral order $n$. Here $Q$ relates to the nucleon momentum/pion mass and $\Lambda_{\chi}$ corresponds to the break down scale of the chiral expansion that is typically on the order of $1\,\text{GeV}$. The chiral expansion  provides a hierarchy of $NN$, $3N$, and many-nucleon interactions in a consistent scheme~\cite{OrRa94,VanKolck94,EpNo02,Epelbaum06}.

To accelerate convergence of the NCSM and NCSMC calculations, one can employ the similarity renormalization group (SRG) technique~\cite{Wegner1994,Bogner2007,PhysRevC.77.064003,Bogner201094,Jurgenson2009} to soften the chiral interaction and, in the standard scheme, keep two- and three-body SRG induced terms in all calculations, even in the case when the initial chiral $3N$ force is not included. 

We performed exploratory calculations with several chiral interactions including the next-to-next-to-next-to-leading order (N$^3$LO) $NN$ of Ref.~\cite{Entem2003} combined with the $3N$ interaction at next-to next-to leading order (N$^2$LO)~\cite{Navratil2007} as well as the N$^2$LO$_{\rm sat}$ $NN{+}3N$ interaction~\cite{EkJa15}. However, due to the technically complex task of including the $3N$ interaction in the NCSMC we were able to perform $^9$He calculations only up to $N_{\rm max}=7$ or 9. Such basis spaces turned out to be insufficient to obtain conclusive results about the behavior of the $S$-wave scattering in particular. Consequently, we decided to limit ourselves to the two-body component of the SRG-evolved $NN$ interaction. 

In particular, we employed the new $NN$ chiral potential at N$^4$LO developed by Entem, Machleidt, and Nosyk~\cite{PhysRevC.91.014002,PhysRevC.96.024004} with a cutoff $\Lambda = 500$ MeV in the regulator function introduced to deal with the infinities in the Lippmann-Schwinger equation. We did not use the bare interaction as the convergence would require a basis size well beyond our computational capabilities, but we softened the $NN$ potential via the SRG and discarded the induced three-nucleon forces. As the chiral $3N$ interaction is typically attractive in light nuclei while the SRG induced $3N$ interaction is repulsive, it is possible to find an SRG resolution scale for which the net effect of the 3N forces tends to be suppressed, and disregarding them leads to binding energies close to experiment. In any case, with this interaction, we were able to reach $N_{\rm max}=11$ (with the m-scheme dimension of $\sim$350 million for $^9$He) and understand the phase shift behavior in all partial waves as demonstrated in the next section. We note that in the basis spaces we could reach with the $NN{+}3N$ interactions, our results were qualitatively consistent at a given $N_{\rm max}$ with those obtained with the $NN$-only interaction presented in the next section.

\section{Results}
\label{sec_results}

\begin{figure}[t]
\begin{center}
\includegraphics[scale=0.35]{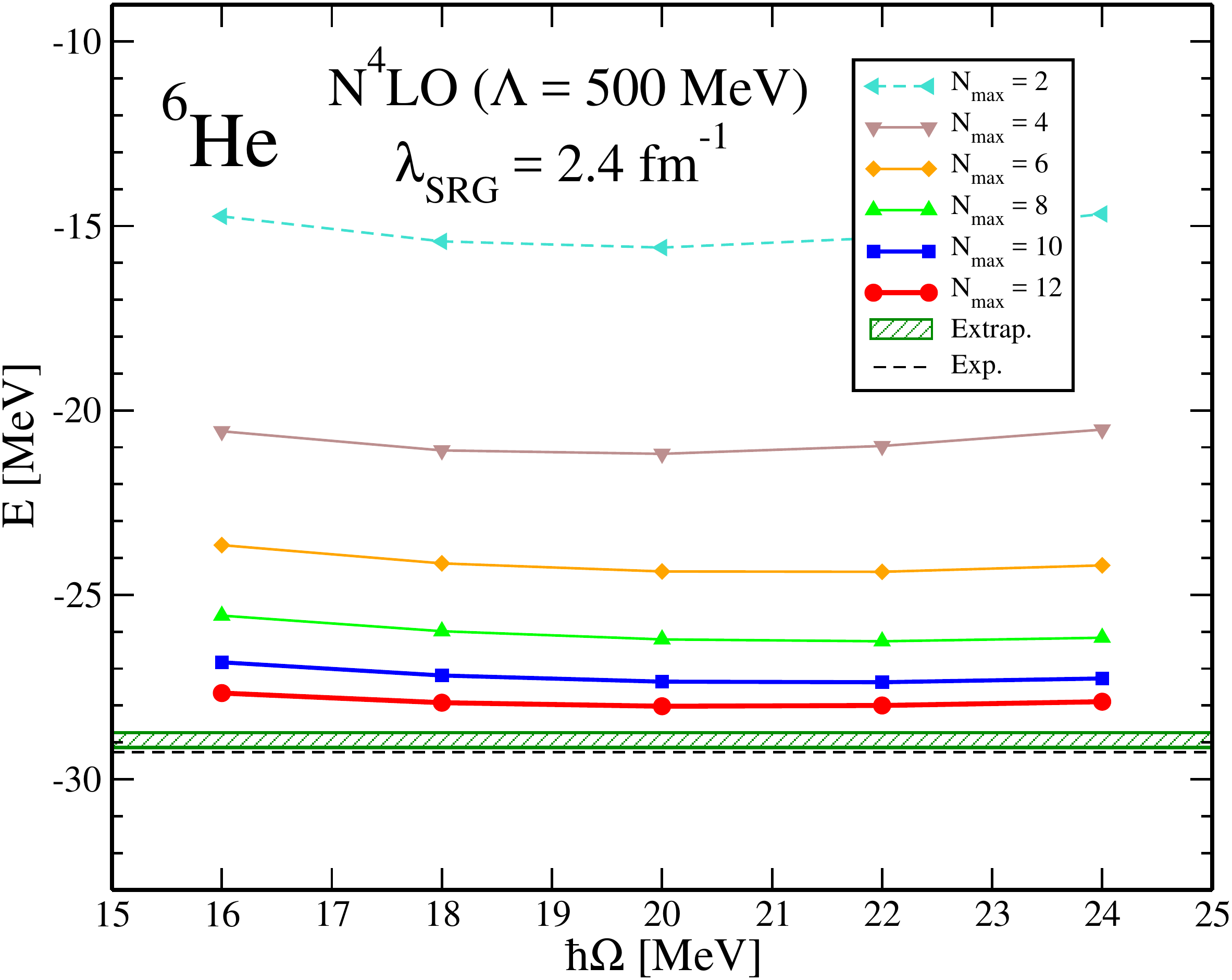}
\caption{\label{He6_hw}  (Color online) Ground-state energy of $^6$He as function of $\hb$ calculated for different values of $\nmax$ within the NCSM and using the SRG-evolved
N$^4$LO $NN$ potential~\cite{PhysRevC.91.014002,PhysRevC.96.024004} with $\lambda_{\rm SRG} = 2.4$ ${\rm fm}^{-1}$. The shaded band represents the result of the exponential extrapolation performed at $\hbar\Omega{=}20$~MeV with the estimated theoretical error.}
\end{center}
\end{figure}

\begin{figure}[t]
\begin{center}
\includegraphics[scale=0.35]{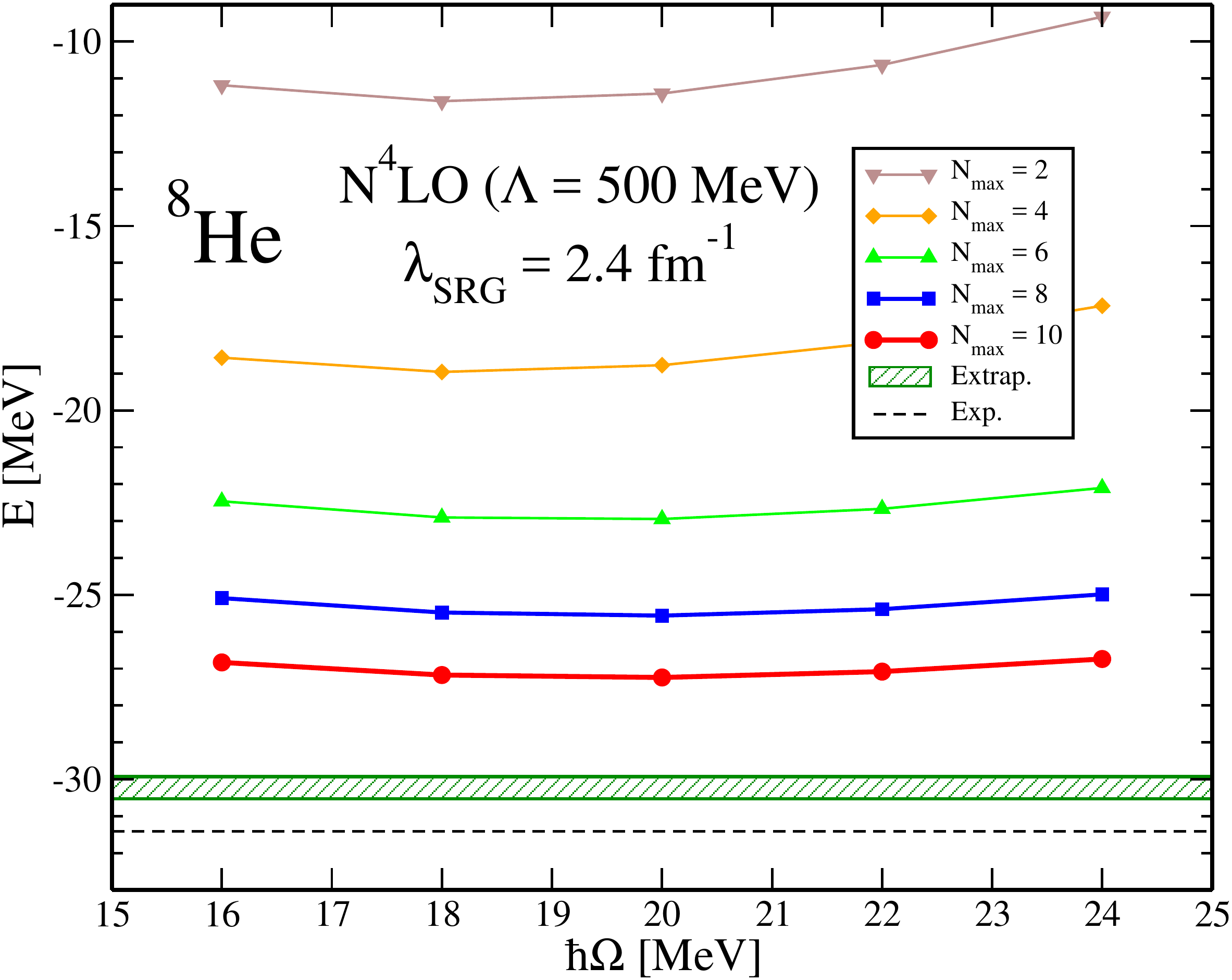}
\caption{\label{He8_hw}  (Color online) The ground-state energy for $^8$He as a function of $\hbar\Omega$ and $N_{\rm max}$. The details are the same as in Fig.~\ref{He6_hw}.}
\end{center}
\end{figure}

\begin{figure}[t]
\begin{center}
\includegraphics[scale=0.35]{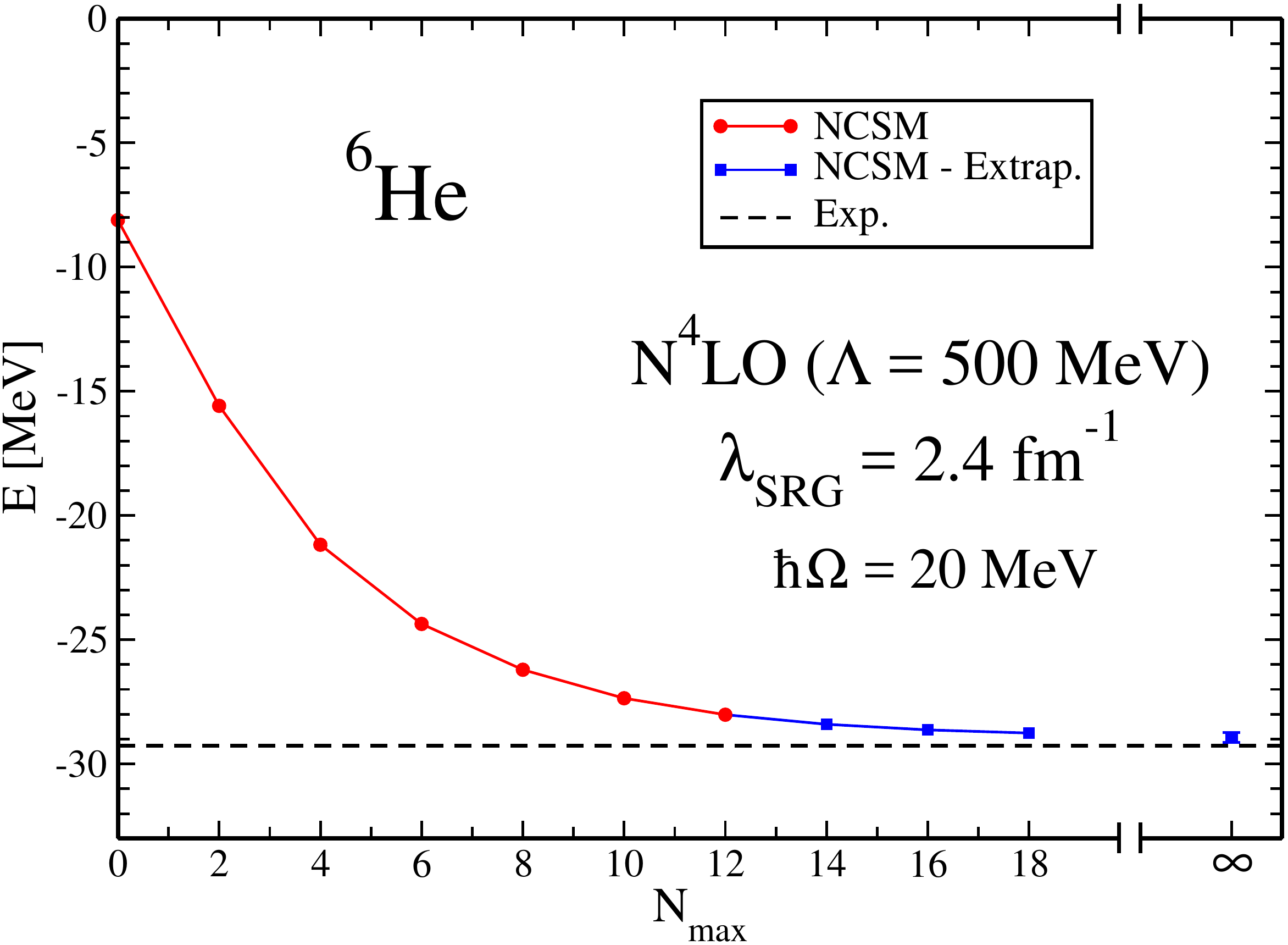}
\caption{\label{He6_nmax}  (Color online) Ground-state energy of $^6$He as function of $\nmax$ calculated with $\hb = 20$ MeV within the NCSM and using the SRG-evolved
N$^4$LO $NN$ potential~\cite{PhysRevC.91.014002,PhysRevC.96.024004} with $\lambda_{\rm SRG} = 2.4$ ${\rm fm}^{-1}$. The circles represent the calculated results while the squares are the energies obtained from the exponential extrapolation.}
\end{center}
\end{figure}

\begin{figure}[t]
\begin{center}
\includegraphics[scale=0.35]{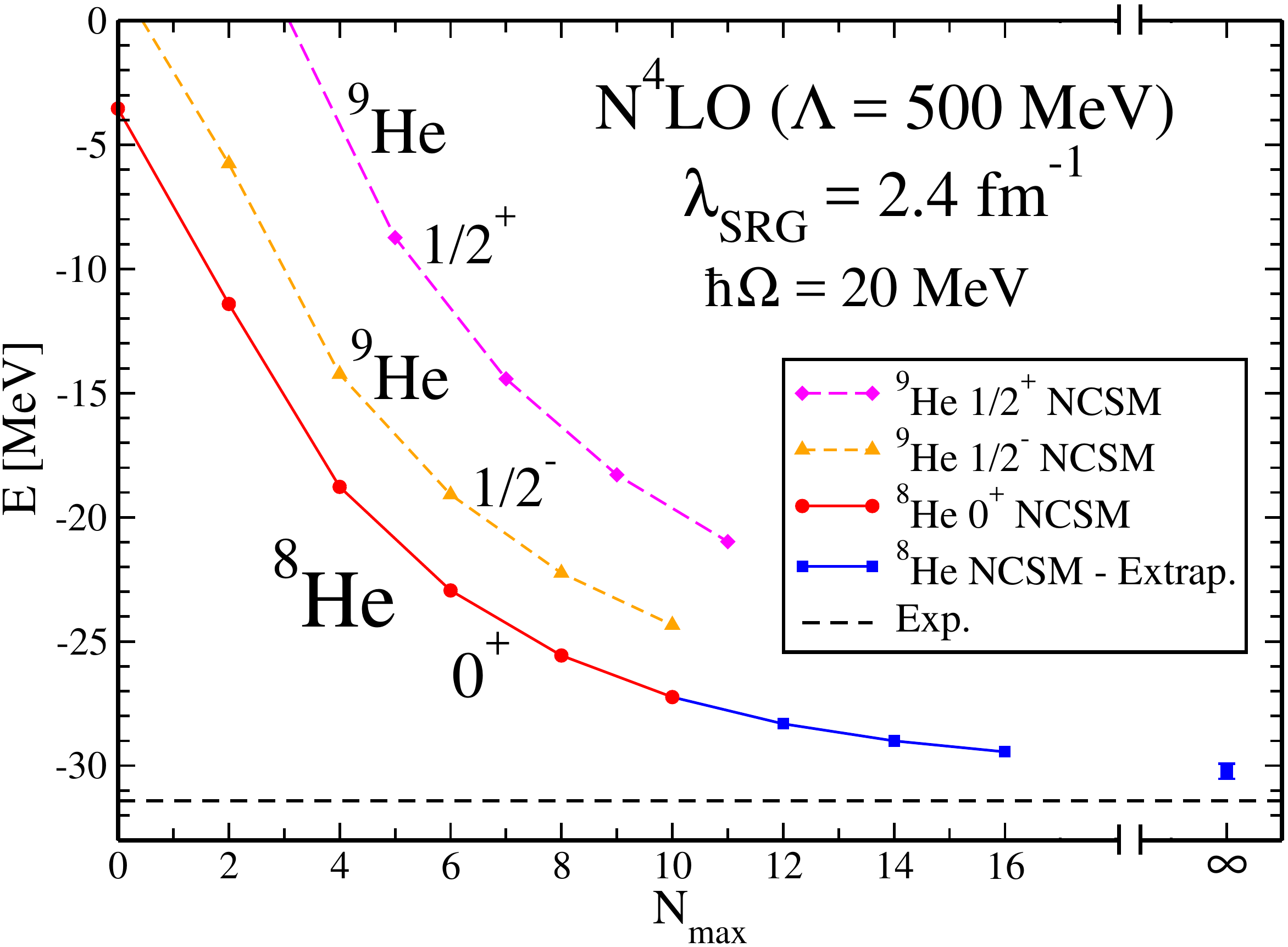}
\caption{\label{He8_nmax}  (Color online) The ground-state energy for $^8$He and the $1/2^-$ and $1/2^+$ eigenenergies of $^9$He as a function of $N_{\rm max}$. The details are the same as in Fig.~\ref{He6_nmax}.}
\end{center}
\end{figure}

\begin{table}[t]
\begin{center}
\begin{ruledtabular}
\begin{tabular}{c|ccc}
$E_{\mathrm{g.s.}}$ MeV & $^4$He & $^6$He & $^8$He \\
\hline
NCSM & -28.36 & -28.94(20) & -30.23(30) \\
Exp. & -28.30 & -29.27 & -31.41 \\
\end{tabular}
\end{ruledtabular}
\caption{Ground-state energies of $^{4,6,8}$He in MeV. NCSM calculations are performed using the SRG-evolved N$^4$LO $NN$
potential~\cite{PhysRevC.91.014002,PhysRevC.96.024004} with $\lambda_{\rm SRG} = 2.4$ ${\rm fm}^{-1}$. All results are obtained using the HO frequency $\hb = 20$ MeV.
The $^4$He energy is computed up to $\nmax = 20$ and is converged to a keV precision, while the energies for $^{6,8}$He are extrapolated using Eq.~(\ref{energy_extrapolation}).}  
\label{He_isotopes_energies}
\end{center} 
\end{table}

As stated in subsection~\ref{sec_interaction}, in the present work we used the new $NN$ chiral potential at N$^4$LO~\cite{PhysRevC.91.014002,PhysRevC.96.024004} with a cutoff $\Lambda = 500$ MeV that we evolved via the SRG, discarding both the induced and the initial chiral three-nucleon forces. In general, the more the potential is evolved the faster the many-body calculations converge, but induced $3N$ forces become larger and larger, such that the net effect of the $3N$ forces (initial plus induced) is no longer negligible. Our strategy to select the value of the SRG evolution parameter $\lambda_{\rm SRG}$ was to reproduce closely the binding energy of $^4$He and obtain realistic ones for $^{6,8}$He with the least amount of evolution. For our purposes it is important to show that our interaction predicts $^8$He bound with respect to $^6$He$+2 n$ and $^6$He bound with respect to $^4$He$+2 n$. For this reason, all our calculations have been performed with an SRG evolved $NN$ potential with $\lambda_{\rm SRG}=2.4$ ${\rm fm}^{-1}$, which has been identified as a satisfactory value.

\subsection{NCSM calculations for He isotopes}
\label{sec_He_NCSM}

We begin the discussion of our calculations with the NCSM results for $^{4,6}$He and $^{8,9}$He, with the latter very important for the subsequent NCSMC study of $^9$He. Let us note that we have developed a three-cluster version of the NCSMC applicable to $^6$He in particular~\cite{PhysRevLett.117.222501,Quaglioni:2017vpa} that provides a superior description of this nucleus compared to a simple NCSM calculation. However, since we are interested here only in the ground-state energy of $^6$He the NCSM, upon extrapolation to the infinite model space, is sufficient.

In Fig.~\ref{He6_hw} and in Fig.~\ref{He8_hw} we present the $^6$He and $^8$He ground-state energies as functions of $\hb$ and for different values of $\nmax$. In both cases, the computed energies display a convergence toward the experimental value represented by the dashed line; a rapid convergence is particularly
evident for $^6$He that can be calculated up to $\nmax = 12$. For both these nuclei, the variational NCSM calculations obtained with the largest $\nmax$ value exhibit a minimum in correspondence of $\hb=20$ MeV, which was then chosen for our subsequent $^9$He investigation. Due to the convergence pattern, it is possible to extrapolate the energies for the higher $\nmax$ values using the exponential function
\begin{equation}\label{energy_extrapolation}
E (\nmax) = E_{\infty} + a e^{- b \nmax} \, ,
\end{equation}
where $a$, $b$, and $E_{\infty}$ are free parameters and $E_{\infty}$ represent the extrapolated energy in the limit of $\nmax \rightarrow \infty$. The extrapolated energies at $\hb{=}20$ MeV are displayed with shaded bands because they include the theoretical error that was obtained as the difference between the fit done using the three points obtained with the last three $\nmax$ values and a second fit in which we used the last four values, and it was estimated of the order of $0.3$ MeV. 

In Fig.~\ref{He6_nmax} and in Fig.~\ref{He8_nmax} we show the calculated and extrapolated energies as functions of $\nmax$ computed with $\hb = 20$ MeV, where the point at infinity corresponds to the $E_{\infty}$ parameter of Eq.~(\ref{energy_extrapolation}). These results are summarized in Tab.~\ref{He_isotopes_energies}, where we also report the ground-state energy of $^4$He computed within the NCSM. In this case the calculation was done up to $\nmax = 20$ and the result is fully converged to a keV precision and close to the experimental value.

In Fig.~\ref{He8_nmax}, we also present the $^9$He NCSM eigenenergies of the $1/2^-_1$ and $1/2^+_1$ states up to $\nmax=10$ and 11, respectively, that serve as inputs into our NCSMC calculations of $^9$He. 

We also investigated the convergence of the $^8$He $2^+$ excited-state energy relevant for our NCSMC study of $^9$He. Using $\hb=20$~MeV, we find a change of the $E_x(2^+_1)$ from 4.67~MeV at $\nmax=6$ to 4.22~MeV at $\nmax=10$. This is a reasonable yet somewhat slower convergence rate compared to a typical well-bound-state calculation that can be attributed to the fact that the calculated $2^+$ state corresponds to an experimentally unbound state.

\subsection{${}^9\mbox{He}$ NCSMC calculations}

\begin{figure}[t]
\begin{center}
\includegraphics[scale=0.35]{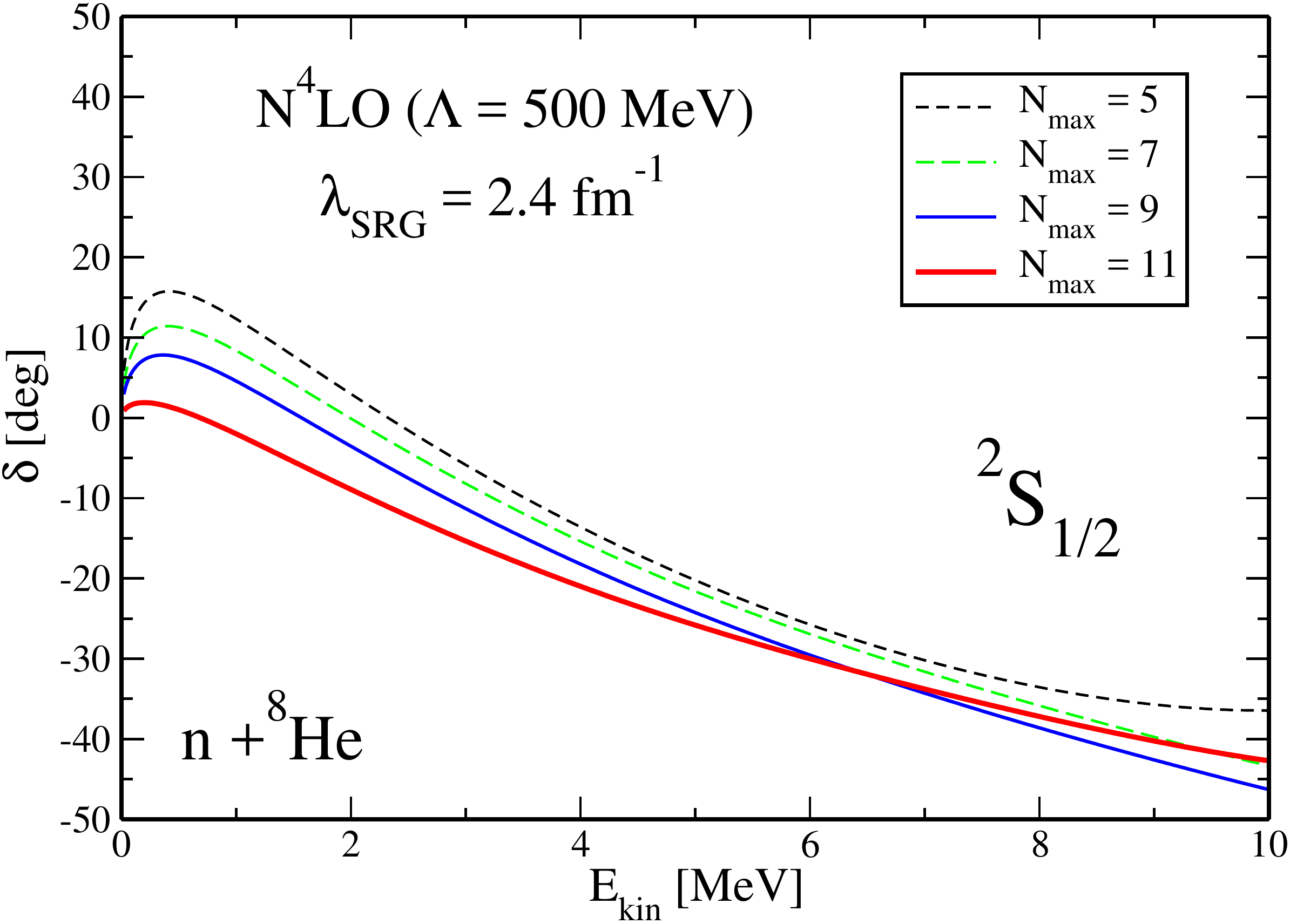}
\caption{\label{2shalf}  (Color online) Dependence of the NCSMC results from the $\nmax$ basis size of the $^2\mathrm{S}_{1/2}$ phase shift as a function of the kinetic energy
in the center of mass. The SRG-evolved N$^4$LO $NN$ potential~\cite{PhysRevC.91.014002,PhysRevC.96.024004} with $\lambda_{\rm SRG} = 2.4$ ${\rm fm}^{-1}$ and the
HO frequency of $\hb = 20$ MeV were used.}
\end{center}
\end{figure}

\begin{figure}[t]
\begin{center}
\includegraphics[scale=0.35]{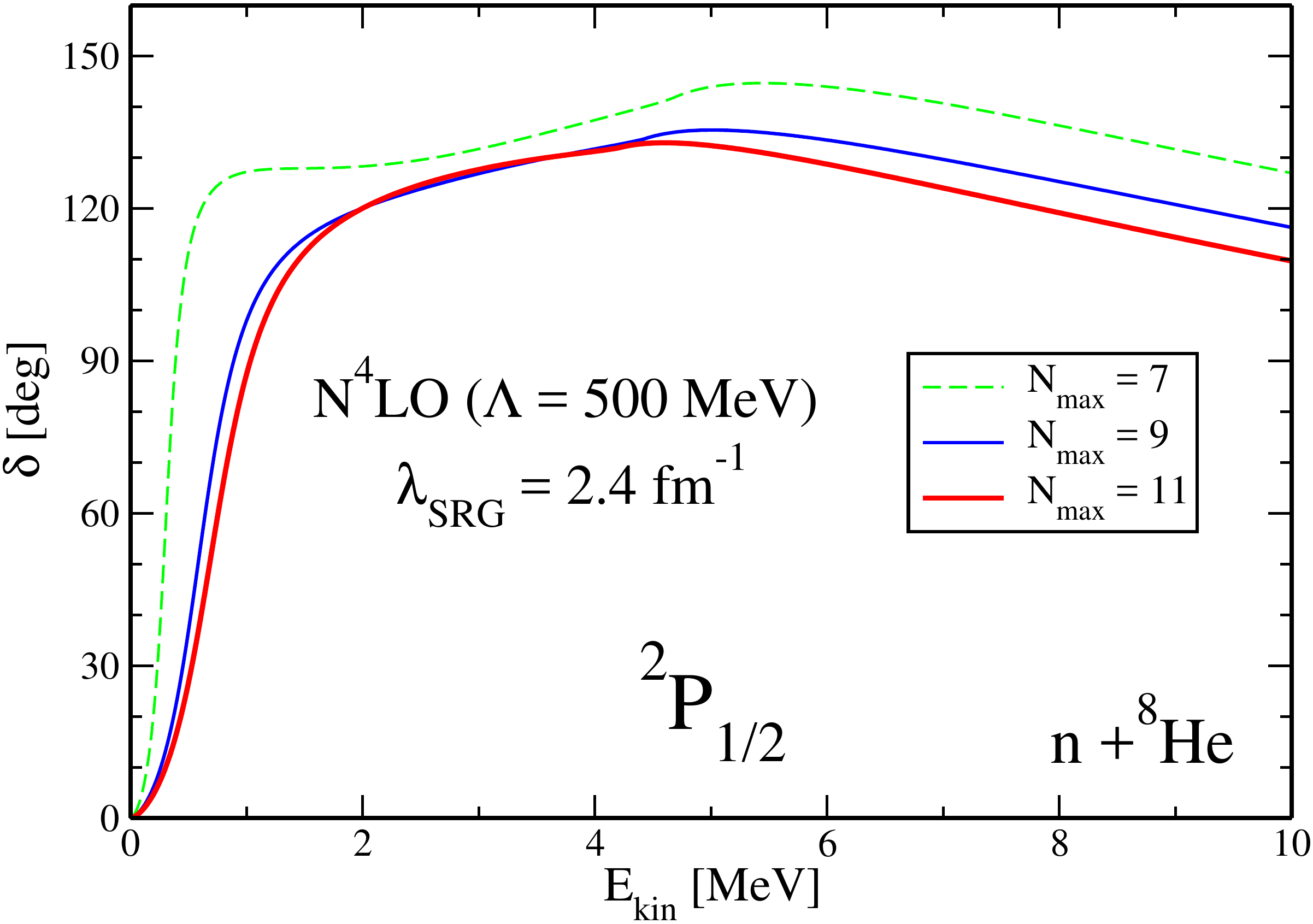}
\caption{\label{2phalf}  (Color online) The  $^2\mathrm{P}_{1/2}$ phase shift in analogy to Fig.~\ref{2shalf}.}
\end{center}
\end{figure}

\begin{figure}[t]
\begin{center}
\includegraphics[scale=0.35]{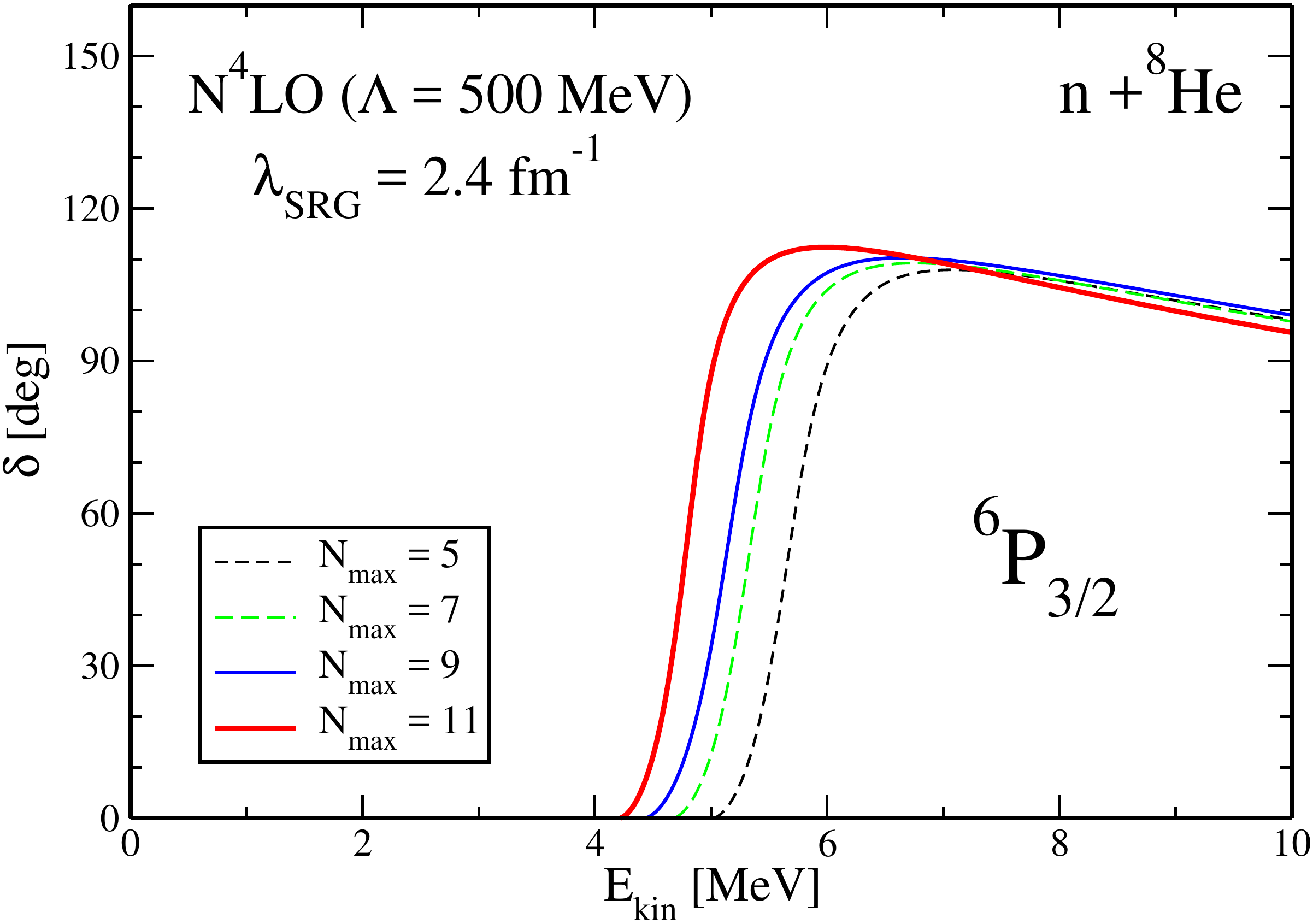}
\caption{\label{6pthreehalf}  (Color online) The $^6\mathrm{P}_{3/2}$ phase shift in analogy to Fig.~\ref{2shalf}.}
\end{center}
\end{figure}

\begin{figure}[t]
\begin{center}
\includegraphics[scale=0.35]{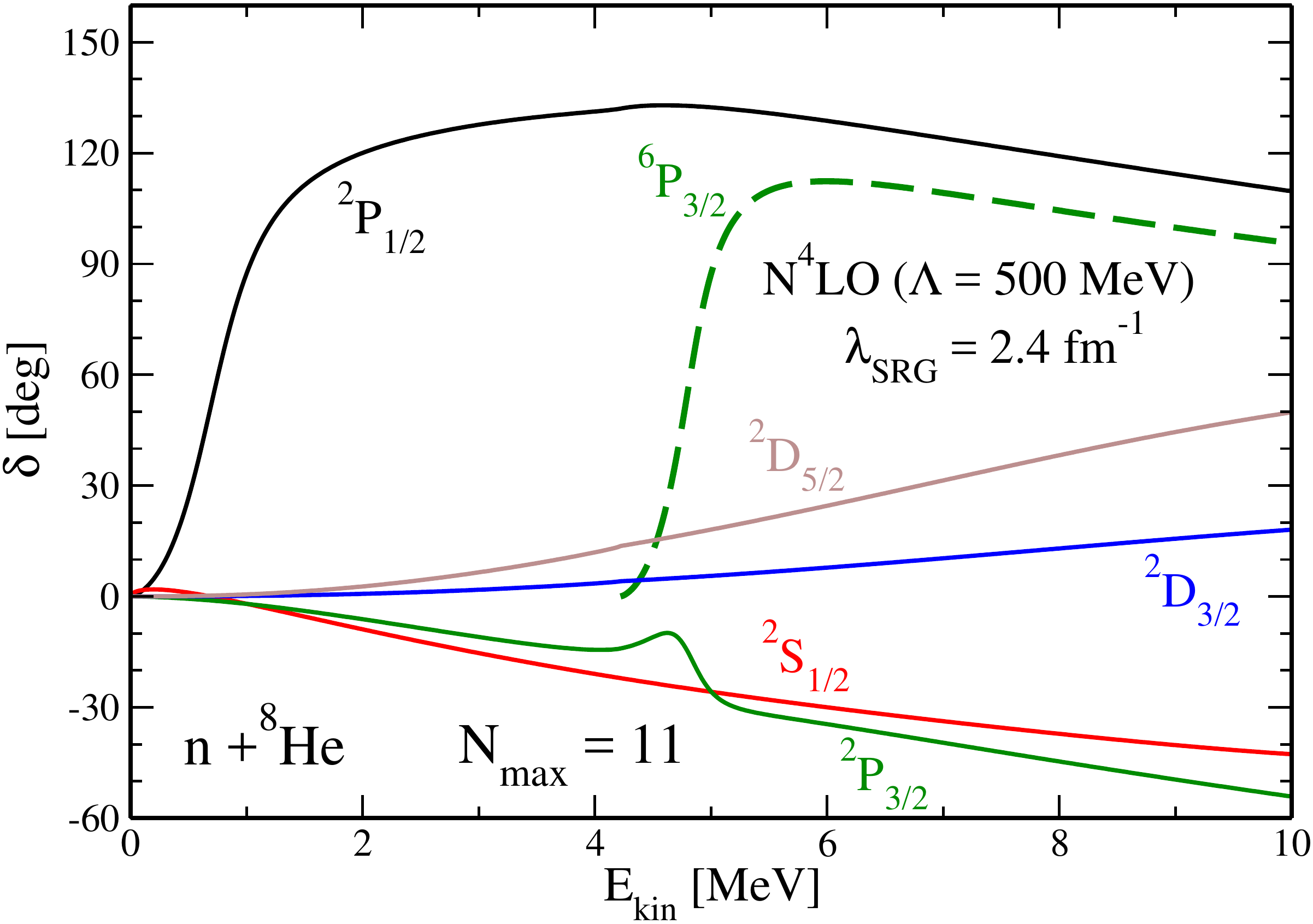}
\caption{\label{phase_shifts}  (Color online) NCSMC $n+^8$He diagonal phase shifts as a function of the kinetic energy in the center of mass computed at $\nmax = 11$.
The SRG-evolved N$^4$LO $NN$ potential~\cite{PhysRevC.91.014002,PhysRevC.96.024004} with $\lambda_{\rm SRG} = 2.4$ ${\rm fm}^{-1}$ and the HO frequency
of $\hb = 20$ MeV were used.}
\end{center}
\end{figure}

We now present the NCSMC results for the $^9$He nucleus. As discussed in subsection~\ref{sec_He_NCSM} and according to Eq.~(\ref{ncsmc_wavefunc}), we first computed the $^9$He and $^8$He eigenenergies and wave functions within the NCSM by diagonalizing the Hamiltonian of Eq.~(\ref{H}). The NCSMC calculation of the $^9$He system was performed within a model space up to $\nmax = 11 \, (10)$ for positive (negative) parity and including the six lowest
positive-parity $(1/2_1^+ , 5/2_1^+ , 3/2_1^+ , 5/2_2^+ , 1/2_2^+ , 3/2_2^+)$ and the four lowest negative-parity $(1/2^- , 3/2_1^- , 3/2_2^- , 3/2_3^-)$ NCSM eigenstates of $^9$He, while the binary-cluster sector was computed including the two lowest eigenstates of $^8$He, {\it i.e.} $(0^+ ,2^+)$.

We start by analyzing the convergence pattern ($\nmax$ dependence) of the three most important $n+^8$He phase shifts, {\it i.e.}
$^2\mathrm{S}_{1/2}$, $^2\mathrm{P}_{1/2}$, and $^6\mathrm{P}_{3/2}$. Here we denote the channels using the standard notation $^{2 s + 1}\ell_{J}$, where the quantum numbers $s$, $\ell$, and $J$ (compare Eq.~(\ref{eq:rgm-state})) represent the channel spin, the relative orbital momentum and the total angular momentum, respectively, of $^8$He and $n$. The $^2\mathrm{S}_{1/2}$ and $^2\mathrm{P}_{1/2}$ phase shifts correspond to the
experimentally debated ${1/2}^+$ and ${1/2}^-$ states while the third one corresponds to an excited state.

In Fig.~\ref{2shalf} we display the NCSMC result for the $^2\mathrm{S}_{1/2}$ phase shift as a function of the kinetic energy for different values of the $\nmax$
parameter. At $\nmax = 5$ the phase shift is positive with a maximum in correspondence of approximately $0.4$ MeV and changes sign at an energy of about $2.4$ MeV, becoming negative. Increasing the $\nmax$ value the maximum of the phase shift starts to decrease until it finally approaches a small positive value for $\nmax = 11$ becoming negative immediately after. This indicates a very weak attraction in this channel with a negative scattering length aproaching zero. Thus in this channel we do not find any resonance and our results suggest that the ${1/2}^+$ state is not the ground state of $^9$He, in agreement with the findings of the experiments of Refs.~\cite{Seth87,Bohlen88,VONOERTZEN1995c129,Bohlen99,PhysRevC.67.041603}.

In Fig.~\ref{2phalf} we show the convergence pattern obtained for the $^2\mathrm{P}_{1/2}$ phase shift. In this case we do not show the $\nmax = 5$ results, because we obtain a bound state in this very small basis space for this channel. For higher values of $\nmax$, the phase shifts present a good convergence and display a fairly narrow resonance, which bears the quantum numbers corresponding to the experimentally observed ${1/2}^-$ state.

In Fig.~\ref{6pthreehalf} we show the convergence pattern for the $^6\mathrm{P}_{3/2}$ phase shift. Here the increase of the $\nmax$ value produces a shift of
the curves towards smaller kinetic energies corresponding in part to the fall-off of the $^8$He $2^+$ state excitation energy, which presents a somewhat slower convergence than the $^8$He ground state. The $^6\mathrm{P}_{3/2}$ phase shifts are also resonant, corresponding to a ${3/2}^-$ state, which
is thus taken as the first excited state of $^9$He. It is important to notice that this state is built on the first $2^+$ excited state of $^8$He; a simpler calculation with only the $^8$He ground state would not produce this resonance. As a final comment we mention that our calculations
do not include the $^6\mathrm{He} + 2 n$ channel that opens at 2.14 MeV excitation energy of $^8$He~\cite{TILLEY2004155}. This introduces some uncertainties in our results especially for the ${3/2}^-$ resonance, which appears at energy where this channel is open. We note that while we are able to perform NCSMC calculations with three-body cluster states~\cite{PhysRevLett.117.222501,Quaglioni:2017vpa}, the $^9$He investigation with the $^6\mathrm{He} + 2 n$ channel open corresponds to a four-body cluster ($^6\mathrm{He} + 3 n$) that is beyond our computational capablity at present.

A qualitative idea of the energy spectrum of the $^9$He nucleus, which summarizes the current analysis, can be inferred from Fig.~\ref{phase_shifts} where we display the phase
shifts including higher partial waves computed within the NCSMC at $\nmax = 11$. We only found two resonances in the $^2\mathrm{P}_{1/2}$
and $^6\mathrm{P}_{3/2}$ channels, corresponding to the ${1/2}^-$ and ${3/2}^-$ states, respectively. In all other channels we did not find any resonance, especially in the
$^2\mathrm{S}_{1/2}$ channel, which represents the experimentally much debated ${1/2}^+$ state.

\begin{figure}[t]
\begin{center}
\includegraphics[scale=0.35]{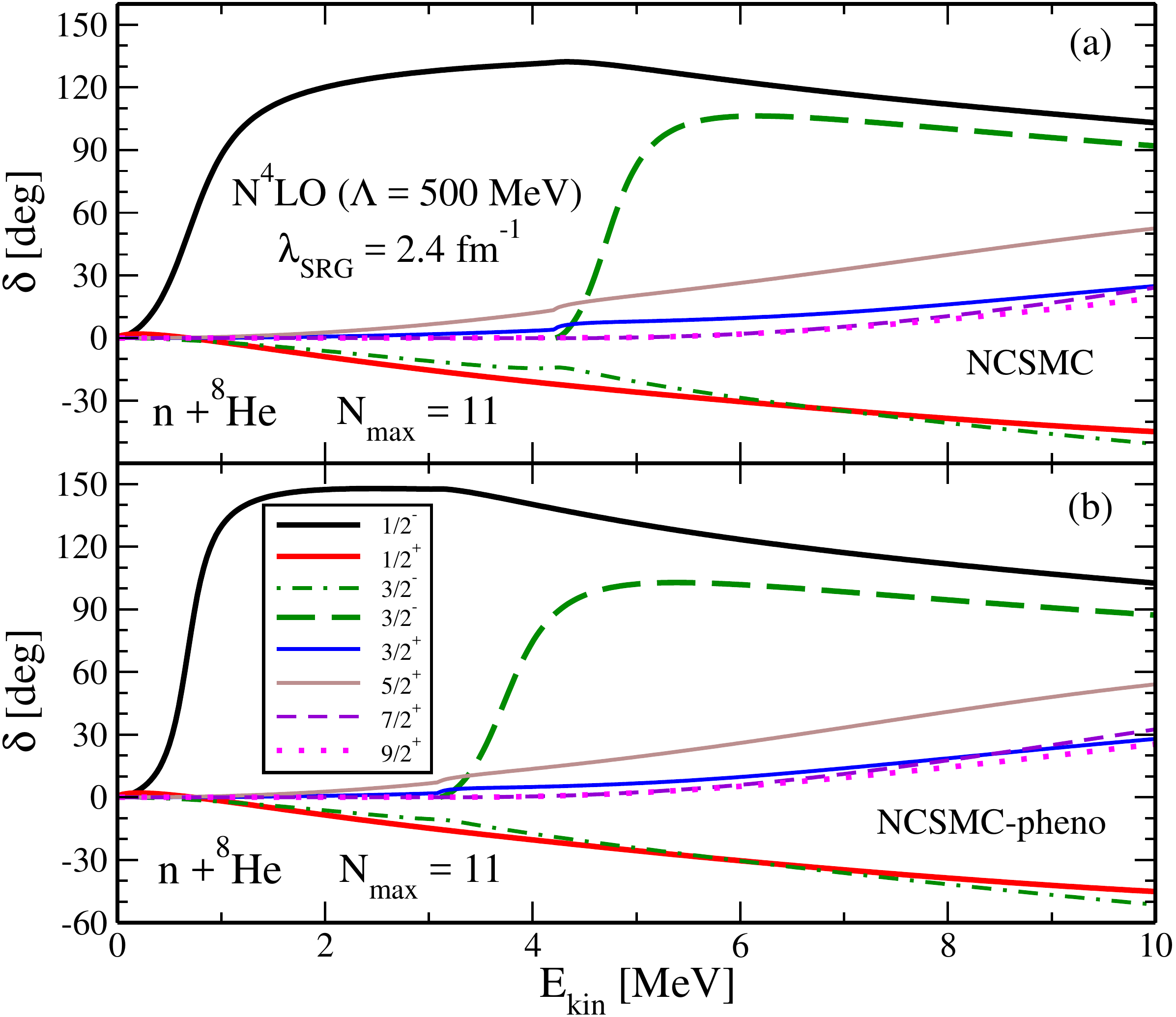}
\caption{\label{compeigen} (Color online) {\bf Panel (a):} NCSMC $n+^8$He eigenphase shifts as a function of the kinetic energy in the center of mass.
{\bf Panel (b):} NCSMC-pheno $n+^8$He eigenphase shifts as a function of the kinetic energy in the center of mass and setting the energy of the first excited state
of $^8$He at the experimental energy of $3.1$ MeV. In both cases the results were obtained at $\nmax = 11$ using the SRG-evolved N$^4$LO $NN$
potential~\cite{PhysRevC.91.014002,PhysRevC.96.024004} with $\lambda_{\rm SRG} = 2.4$ ${\rm fm}^{-1}$ and the HO frequency of $\hb = 20$ MeV were used.}
\end{center}
\end{figure}

\begin{table}[t]
\begin{center}
\begin{ruledtabular}
\begin{tabular}{c|c|c}
$J^{\pi}$ & NCSMC \hspace{0.25cm} & NCSMC-pheno \\
\hline
${1/2}^-$ & $E_R = 0.69$ \hspace{0.3cm} $\Gamma = 0.83$ \hspace{0.25cm} & $E_R = 0.68$ \hspace{0.3cm} $\Gamma = 0.37$ \\
${3/2}^-$ & $E_R = 4.70$ \hspace{0.3cm} $\Gamma = 0.74$ \hspace{0.25cm}  & $E_R = 3.72$ \hspace{0.3cm} $\Gamma = 0.95$ \\
\end{tabular}
\end{ruledtabular}
\caption{Theoretical values for the resonance centroids and widths in MeV for the ${1/2}^-$ ground state and the ${3/2}^-$ excited state of $^9$He. Calculations are carried out
as described in Fig.~\ref{compeigen} and in the text.}  
\label{resonances}
\end{center} 
\end{table}

Finally, in Fig.~\ref{compeigen} we present the results for the eigenphase shifts obtained by diagonalizing the scattering matrix. While the phase shifts allow us
to have an insight of the physics in the partial wave channels and identify the resonances, the eigenphase shifts take into account the coupling of different partial waves that are used for a quantitative analysis of these resonances. For example, they are used to compute the resonance centroid and width. In panel (a) of Fig.~\ref{compeigen} we show the
eigenphase shifts obtained within the NCSMC for various values of total angular momentum, while in panel (b) we show the eigenphase shifts obtained within the NCSMC-pheno~\cite{DOHETERALY2016430},
which means that the calculation was performed within the NCSMC but we set phenomenologically the NCSM eigenenergy of the $2^+$ state in $^8$He (which is an input to the NCSMC) to the experimental value~\cite{TILLEY2004155} of $3.1$ MeV. This produces a sizeable effect only for the two resonances corresponding to the ${1/2}^-$
and ${3/2}^-$ (dashed line) states; qualitatively we can see that the former becomes narrower while the latter becomes broader. This is confirmed from the calculation of the
resonance centroid $E_R$ and width $\Gamma$. In the present work the centroid of the two resonant states was computed performing the first derivative of the
eigenphase shift with respect the kinetic energy $E_{kin}$ in the center-of-mass frame and then taking the value of the kinetic energy for which the derivative has a maximum.
Instead, the calculation of the width was performed as specified in Ref.~\cite{thompson2009nuclear} and according to
\begin{equation}
\Gamma = \frac{2}{\mathrm{d} \delta (E_{kin})/\mathrm{d} E_{kin}} \Big|_{E_{kin} = E_R} \, ,
\end{equation}
with the eigenphase shift expressed in radians. The results of our analysis are summarized in Tab.~\ref{resonances}. For the ${1/2}^-$ state the value of the centroid remains
constant while the width is significantly reduced in the NCSMC-pheno calculation and it is closer to the experimental value of $0.1$ MeV. On the contrary, for
the ${3/2}^-$ state the NCSMC-pheno calculation gives a larger width and a different value of the centroid, which is about $1$ MeV smaller than that one obtained
within the NCSMC.

\section{Conclusions}
\label{sec_conclusions}
In this work we used the {\it ab initio} NCSMC approach to study the $^9$He resonances by analyzing the $n+^8$He scattering process. The exotic $^9$He is a very interesting nucleus due to its extreme $N/Z$ ratio and because it
is part of a series of $N=7$ isotones where, given the systematics, the ground state could be expected to be a positive-parity state. Despite having been extensively experimentally investigated, its structure is currently a matter of debate.

The NCSMC is a method capable of describing bound and unbound states in a unified way by combining an $A$-body square-integrable (which contains the many-body correlations) and a continuous basis (which enables the description of long-range interactions between cluster-type states). The NCSMC calculations do not involve any adjustable parameters except for those used to generate the $NN$ interaction, which is the input of our approach.

Our calculations were performed with the SRG-evolved $NN$ interaction derived from the new chiral potential at N$^4$LO~\cite{PhysRevC.91.014002,PhysRevC.96.024004} discarding the three-body terms.
This choice was motivated by the large basis needed to obtain reliable results for the $^9$He system, that makes the calculation with three-body forces computationally prohibitive at present.
We softened the $NN$ potential via the SRG transformation using $\lambda_{\rm SRG} = 2.4$ $\mathrm{fm}^{-1}$ for the evolution parameter. With this choice, the predicted binding energies of $^{4,6,8}$He are close to the experimental values.

Our analysis identified two resonances corresponding to spin-parity states of ${1/2}^-$ and ${3/2}^-$ respectively. The former is identified as the ground state of $^9$He, while the latter is built on the $2^+$ state of $^8$He and represents the first excited state of $^9$He. In particular we did not find any resonance corresponding to a ${1/2}^+$ state; according to our calculations $^9$He breaks the parity inversion observed in $^{11}$Be and in $^{10}$Li. 

In the future we plan to study the $^9$He nucleus including the three-body interactions. We note that we already performed exploratory calculations with the chiral N$^2$LO three-body forces for smaller values of the $\nmax$ parameters with results qualitatively consistent with those obtained with the SRG-evolved N$^4$LO $NN$ at the corresponding $N_{\rm max}$. Unfortunately, it is not currently possible to perform a calculation with the complete $3N$ force at $\nmax \geq 10$ due to the tremendous computational effort. The only possibility to achieve this goal is to adopt a truncation scheme, such as the normal ordering~\cite{PhysRevC.93.031301}, which aims to introduce a controlled approximation for the $3N$ terms and it is currently under development.
Finally, we also plan to study the $p+^8$He scattering with $^9$Li as the composite system including the $n+^8$Li charge exchange channel, which permits access to the $T=5/2$ isobaric analog states of $^9$He.

\acknowledgments

This work was supported by the NSERC Grant No. SAPIN-2016-00033 and by the U.S. Department of Energy, Office of Science, Office of Nuclear Physics, under Work Proposals No. SCW1158 and SCW0498. TRIUMF receives federal funding via a contribution agreement with the National Research Council of Canada. This work was prepared in part by LLNL under Contract No. DE-AC52-07NA27344. Computing support came from an INCITE Award on the Titan supercomputer of the Oak Ridge Leadership Computing Facility (OLCF) at ORNL, from Calcul Quebec, Westgrid and Compute Canada, and from the LLNL institutional Computing Grand Challenge Program.

%\bibliography{biblio}

\begin{thebibliography}{97}%
\makeatletter
\providecommand \@ifxundefined [1]{%
 \@ifx{#1\undefined}
}%
\providecommand \@ifnum [1]{%
 \ifnum #1\expandafter \@firstoftwo
 \else \expandafter \@secondoftwo
 \fi
}%
\providecommand \@ifx [1]{%
 \ifx #1\expandafter \@firstoftwo
 \else \expandafter \@secondoftwo
 \fi
}%
\providecommand \natexlab [1]{#1}%
\providecommand \enquote  [1]{``#1''}%
\providecommand \bibnamefont  [1]{#1}%
\providecommand \bibfnamefont [1]{#1}%
\providecommand \citenamefont [1]{#1}%
\providecommand \href@noop [0]{\@secondoftwo}%
\providecommand \href [0]{\begingroup \@sanitize@url \@href}%
\providecommand \@href[1]{\@@startlink{#1}\@@href}%
\providecommand \@@href[1]{\endgroup#1\@@endlink}%
\providecommand \@sanitize@url [0]{\catcode `\\12\catcode `\$12\catcode
  `\&12\catcode `\#12\catcode `\^12\catcode `\_12\catcode `\%12\relax}%
\providecommand \@@startlink[1]{}%
\providecommand \@@endlink[0]{}%
\providecommand \url  [0]{\begingroup\@sanitize@url \@url }%
\providecommand \@url [1]{\endgroup\@href {#1}{\urlprefix }}%
\providecommand \urlprefix  [0]{URL }%
\providecommand \Eprint [0]{\href }%
\providecommand \doibase [0]{http://dx.doi.org/}%
\providecommand \selectlanguage [0]{\@gobble}%
\providecommand \bibinfo  [0]{\@secondoftwo}%
\providecommand \bibfield  [0]{\@secondoftwo}%
\providecommand \translation [1]{[#1]}%
\providecommand \BibitemOpen [0]{}%
\providecommand \bibitemStop [0]{}%
\providecommand \bibitemNoStop [0]{.\EOS\space}%
\providecommand \EOS [0]{\spacefactor3000\relax}%
\providecommand \BibitemShut  [1]{\csname bibitem#1\endcsname}%
\let\auto@bib@innerbib\@empty
%</preamble>
\bibitem [{fri()}]{frib}%
  \BibitemOpen
  \href@noop {} {}\bibinfo {howpublished}
  {\url{http://www.frib.msu.edu/}}\BibitemShut {NoStop}%
\bibitem [{\citenamefont {Talmi}\ and\ \citenamefont
  {Unna}(1960)}]{PhysRevLett.4.469}%
  \BibitemOpen
  \bibfield  {author} {\bibinfo {author} {\bibfnamefont {I.}~\bibnamefont
  {Talmi}}\ and\ \bibinfo {author} {\bibfnamefont {I.}~\bibnamefont {Unna}},\
  }\href {\doibase 10.1103/PhysRevLett.4.469} {\bibfield  {journal} {\bibinfo
  {journal} {Phys. Rev. Lett.}\ }\textbf {\bibinfo {volume} {4}},\ \bibinfo
  {pages} {469} (\bibinfo {year} {1960})}\BibitemShut {NoStop}%
\bibitem [{\citenamefont {Otsuka}\ \emph {et~al.}(1993)\citenamefont {Otsuka},
  \citenamefont {Fukunishi},\ and\ \citenamefont
  {Sagawa}}]{PhysRevLett.70.1385}%
  \BibitemOpen
  \bibfield  {author} {\bibinfo {author} {\bibfnamefont {T.}~\bibnamefont
  {Otsuka}}, \bibinfo {author} {\bibfnamefont {N.}~\bibnamefont {Fukunishi}}, \
  and\ \bibinfo {author} {\bibfnamefont {H.}~\bibnamefont {Sagawa}},\ }\href
  {\doibase 10.1103/PhysRevLett.70.1385} {\bibfield  {journal} {\bibinfo
  {journal} {Phys. Rev. Lett.}\ }\textbf {\bibinfo {volume} {70}},\ \bibinfo
  {pages} {1385} (\bibinfo {year} {1993})}\BibitemShut {NoStop}%
\bibitem [{\citenamefont {Sagawa}\ \emph {et~al.}(1993)\citenamefont {Sagawa},
  \citenamefont {Brown},\ and\ \citenamefont {Esbensen}}]{SAGAWA19931}%
  \BibitemOpen
  \bibfield  {author} {\bibinfo {author} {\bibfnamefont {H.}~\bibnamefont
  {Sagawa}}, \bibinfo {author} {\bibfnamefont {B.}~\bibnamefont {Brown}}, \
  and\ \bibinfo {author} {\bibfnamefont {H.}~\bibnamefont {Esbensen}},\ }\href
  {\doibase https://doi.org/10.1016/0370-2693(93)91493-7} {\bibfield  {journal}
  {\bibinfo  {journal} {Physics Letters B}\ }\textbf {\bibinfo {volume}
  {309}},\ \bibinfo {pages} {1 } (\bibinfo {year} {1993})}\BibitemShut
  {NoStop}%
\bibitem [{\citenamefont {Mau}(1995)}]{VINHMAU199533}%
  \BibitemOpen
  \bibfield  {author} {\bibinfo {author} {\bibfnamefont {N.~V.}\ \bibnamefont
  {Mau}},\ }\href {\doibase https://doi.org/10.1016/0375-9474(95)00298-F}
  {\bibfield  {journal} {\bibinfo  {journal} {Nuclear Physics A}\ }\textbf
  {\bibinfo {volume} {592}},\ \bibinfo {pages} {33 } (\bibinfo {year}
  {1995})}\BibitemShut {NoStop}%
\bibitem [{\citenamefont {Gori}\ \emph {et~al.}(2004)\citenamefont {Gori},
  \citenamefont {Barranco}, \citenamefont {Vigezzi},\ and\ \citenamefont
  {Broglia}}]{PhysRevC.69.041302}%
  \BibitemOpen
  \bibfield  {author} {\bibinfo {author} {\bibfnamefont {G.}~\bibnamefont
  {Gori}}, \bibinfo {author} {\bibfnamefont {F.}~\bibnamefont {Barranco}},
  \bibinfo {author} {\bibfnamefont {E.}~\bibnamefont {Vigezzi}}, \ and\
  \bibinfo {author} {\bibfnamefont {R.~A.}\ \bibnamefont {Broglia}},\ }\href
  {\doibase 10.1103/PhysRevC.69.041302} {\bibfield  {journal} {\bibinfo
  {journal} {Phys. Rev. C}\ }\textbf {\bibinfo {volume} {69}},\ \bibinfo
  {pages} {041302} (\bibinfo {year} {2004})}\BibitemShut {NoStop}%
\bibitem [{\citenamefont {Nunes}\ \emph {et~al.}(1996)\citenamefont {Nunes},
  \citenamefont {Thompson},\ and\ \citenamefont {Johnson}}]{NUNES1996171}%
  \BibitemOpen
  \bibfield  {author} {\bibinfo {author} {\bibfnamefont {F.}~\bibnamefont
  {Nunes}}, \bibinfo {author} {\bibfnamefont {I.}~\bibnamefont {Thompson}}, \
  and\ \bibinfo {author} {\bibfnamefont {R.}~\bibnamefont {Johnson}},\ }\href
  {\doibase https://doi.org/10.1016/0375-9474(95)00398-3} {\bibfield  {journal}
  {\bibinfo  {journal} {Nuclear Physics A}\ }\textbf {\bibinfo {volume}
  {596}},\ \bibinfo {pages} {171 } (\bibinfo {year} {1996})}\BibitemShut
  {NoStop}%
\bibitem [{\citenamefont {Fossez}\ \emph {et~al.}(2016)\citenamefont {Fossez},
  \citenamefont {Nazarewicz}, \citenamefont {Jaganathen}, \citenamefont
  {Michel},\ and\ \citenamefont {P\l{}oszajczak}}]{PhysRevC.93.011305}%
  \BibitemOpen
  \bibfield  {author} {\bibinfo {author} {\bibfnamefont {K.}~\bibnamefont
  {Fossez}}, \bibinfo {author} {\bibfnamefont {W.}~\bibnamefont {Nazarewicz}},
  \bibinfo {author} {\bibfnamefont {Y.}~\bibnamefont {Jaganathen}}, \bibinfo
  {author} {\bibfnamefont {N.}~\bibnamefont {Michel}}, \ and\ \bibinfo {author}
  {\bibfnamefont {M.}~\bibnamefont {P\l{}oszajczak}},\ }\href {\doibase
  10.1103/PhysRevC.93.011305} {\bibfield  {journal} {\bibinfo  {journal} {Phys.
  Rev. C}\ }\textbf {\bibinfo {volume} {93}},\ \bibinfo {pages} {011305}
  (\bibinfo {year} {2016})}\BibitemShut {NoStop}%
\bibitem [{\citenamefont {Hamamoto}\ and\ \citenamefont
  {Shimoura}(2007)}]{HamamotoBe11}%
  \BibitemOpen
  \bibfield  {author} {\bibinfo {author} {\bibfnamefont {I.}~\bibnamefont
  {Hamamoto}}\ and\ \bibinfo {author} {\bibfnamefont {S.}~\bibnamefont
  {Shimoura}},\ }\href {http://stacks.iop.org/0954-3899/34/i=12/a=015}
  {\bibfield  {journal} {\bibinfo  {journal} {Journal of Physics G: Nuclear and
  Particle Physics}\ }\textbf {\bibinfo {volume} {34}},\ \bibinfo {pages}
  {2715} (\bibinfo {year} {2007})}\BibitemShut {NoStop}%
\bibitem [{\citenamefont {Kanada-En'yo}\ and\ \citenamefont
  {Horiuchi}(2002)}]{PhysRevC.66.024305}%
  \BibitemOpen
  \bibfield  {author} {\bibinfo {author} {\bibfnamefont {Y.}~\bibnamefont
  {Kanada-En'yo}}\ and\ \bibinfo {author} {\bibfnamefont {H.}~\bibnamefont
  {Horiuchi}},\ }\href {\doibase 10.1103/PhysRevC.66.024305} {\bibfield
  {journal} {\bibinfo  {journal} {Phys. Rev. C}\ }\textbf {\bibinfo {volume}
  {66}},\ \bibinfo {pages} {024305} (\bibinfo {year} {2002})}\BibitemShut
  {NoStop}%
\bibitem [{\citenamefont {Quaglioni}\ and\ \citenamefont
  {Navr\'atil}(2008)}]{PhysRevLett.101.092501}%
  \BibitemOpen
  \bibfield  {author} {\bibinfo {author} {\bibfnamefont {S.}~\bibnamefont
  {Quaglioni}}\ and\ \bibinfo {author} {\bibfnamefont {P.}~\bibnamefont
  {Navr\'atil}},\ }\href {\doibase 10.1103/PhysRevLett.101.092501} {\bibfield
  {journal} {\bibinfo  {journal} {Phys. Rev. Lett.}\ }\textbf {\bibinfo
  {volume} {101}},\ \bibinfo {pages} {092501} (\bibinfo {year}
  {2008})}\BibitemShut {NoStop}%
\bibitem [{\citenamefont {Calci}\ \emph {et~al.}(2016)\citenamefont {Calci},
  \citenamefont {Navr\'atil}, \citenamefont {Roth}, \citenamefont
  {Dohet-Eraly}, \citenamefont {Quaglioni},\ and\ \citenamefont
  {Hupin}}]{PhysRevLett.117.242501}%
  \BibitemOpen
  \bibfield  {author} {\bibinfo {author} {\bibfnamefont {A.}~\bibnamefont
  {Calci}}, \bibinfo {author} {\bibfnamefont {P.}~\bibnamefont {Navr\'atil}},
  \bibinfo {author} {\bibfnamefont {R.}~\bibnamefont {Roth}}, \bibinfo {author}
  {\bibfnamefont {J.}~\bibnamefont {Dohet-Eraly}}, \bibinfo {author}
  {\bibfnamefont {S.}~\bibnamefont {Quaglioni}}, \ and\ \bibinfo {author}
  {\bibfnamefont {G.}~\bibnamefont {Hupin}},\ }\href {\doibase
  10.1103/PhysRevLett.117.242501} {\bibfield  {journal} {\bibinfo  {journal}
  {Phys. Rev. Lett.}\ }\textbf {\bibinfo {volume} {117}},\ \bibinfo {pages}
  {242501} (\bibinfo {year} {2016})}\BibitemShut {NoStop}%
\bibitem [{\citenamefont {Krieger}\ \emph {et~al.}(2012)\citenamefont
  {Krieger}, \citenamefont {Blaum}, \citenamefont {Bissell}, \citenamefont
  {Fr\"ommgen}, \citenamefont {Geppert}, \citenamefont {Hammen}, \citenamefont
  {Kreim}, \citenamefont {Kowalska}, \citenamefont {Kr\"amer}, \citenamefont
  {Neff}, \citenamefont {Neugart}, \citenamefont {Neyens}, \citenamefont
  {N\"ortersh\"auser}, \citenamefont {Novotny}, \citenamefont {S\'anchez},\
  and\ \citenamefont {Yordanov}}]{PhysRevLett.108.142501}%
  \BibitemOpen
  \bibfield  {author} {\bibinfo {author} {\bibfnamefont {A.}~\bibnamefont
  {Krieger}}, \bibinfo {author} {\bibfnamefont {K.}~\bibnamefont {Blaum}},
  \bibinfo {author} {\bibfnamefont {M.~L.}\ \bibnamefont {Bissell}}, \bibinfo
  {author} {\bibfnamefont {N.}~\bibnamefont {Fr\"ommgen}}, \bibinfo {author}
  {\bibfnamefont {C.}~\bibnamefont {Geppert}}, \bibinfo {author} {\bibfnamefont
  {M.}~\bibnamefont {Hammen}}, \bibinfo {author} {\bibfnamefont
  {K.}~\bibnamefont {Kreim}}, \bibinfo {author} {\bibfnamefont
  {M.}~\bibnamefont {Kowalska}}, \bibinfo {author} {\bibfnamefont
  {J.}~\bibnamefont {Kr\"amer}}, \bibinfo {author} {\bibfnamefont
  {T.}~\bibnamefont {Neff}}, \bibinfo {author} {\bibfnamefont {R.}~\bibnamefont
  {Neugart}}, \bibinfo {author} {\bibfnamefont {G.}~\bibnamefont {Neyens}},
  \bibinfo {author} {\bibfnamefont {W.}~\bibnamefont {N\"ortersh\"auser}},
  \bibinfo {author} {\bibfnamefont {C.}~\bibnamefont {Novotny}}, \bibinfo
  {author} {\bibfnamefont {R.}~\bibnamefont {S\'anchez}}, \ and\ \bibinfo
  {author} {\bibfnamefont {D.~T.}\ \bibnamefont {Yordanov}},\ }\href {\doibase
  10.1103/PhysRevLett.108.142501} {\bibfield  {journal} {\bibinfo  {journal}
  {Phys. Rev. Lett.}\ }\textbf {\bibinfo {volume} {108}},\ \bibinfo {pages}
  {142501} (\bibinfo {year} {2012})}\BibitemShut {NoStop}%
\bibitem [{\citenamefont {Timofeyuk}\ and\ \citenamefont
  {Johnson}(1999)}]{PhysRevC.59.1545}%
  \BibitemOpen
  \bibfield  {author} {\bibinfo {author} {\bibfnamefont {N.~K.}\ \bibnamefont
  {Timofeyuk}}\ and\ \bibinfo {author} {\bibfnamefont {R.~C.}\ \bibnamefont
  {Johnson}},\ }\href {\doibase 10.1103/PhysRevC.59.1545} {\bibfield  {journal}
  {\bibinfo  {journal} {Phys. Rev. C}\ }\textbf {\bibinfo {volume} {59}},\
  \bibinfo {pages} {1545} (\bibinfo {year} {1999})}\BibitemShut {NoStop}%
\bibitem [{\citenamefont {Keeley}\ \emph {et~al.}(2004)\citenamefont {Keeley},
  \citenamefont {Alamanos},\ and\ \citenamefont {Lapoux}}]{PhysRevC.69.064604}%
  \BibitemOpen
  \bibfield  {author} {\bibinfo {author} {\bibfnamefont {N.}~\bibnamefont
  {Keeley}}, \bibinfo {author} {\bibfnamefont {N.}~\bibnamefont {Alamanos}}, \
  and\ \bibinfo {author} {\bibfnamefont {V.}~\bibnamefont {Lapoux}},\ }\href
  {\doibase 10.1103/PhysRevC.69.064604} {\bibfield  {journal} {\bibinfo
  {journal} {Phys. Rev. C}\ }\textbf {\bibinfo {volume} {69}},\ \bibinfo
  {pages} {064604} (\bibinfo {year} {2004})}\BibitemShut {NoStop}%
\bibitem [{\citenamefont {Deltuva}(2009)}]{PhysRevC.79.054603}%
  \BibitemOpen
  \bibfield  {author} {\bibinfo {author} {\bibfnamefont {A.}~\bibnamefont
  {Deltuva}},\ }\href {\doibase 10.1103/PhysRevC.79.054603} {\bibfield
  {journal} {\bibinfo  {journal} {Phys. Rev. C}\ }\textbf {\bibinfo {volume}
  {79}},\ \bibinfo {pages} {054603} (\bibinfo {year} {2009})}\BibitemShut
  {NoStop}%
\bibitem [{\citenamefont {Deltuva}(2013)}]{PhysRevC.88.011601}%
  \BibitemOpen
  \bibfield  {author} {\bibinfo {author} {\bibfnamefont {A.}~\bibnamefont
  {Deltuva}},\ }\href {\doibase 10.1103/PhysRevC.88.011601} {\bibfield
  {journal} {\bibinfo  {journal} {Phys. Rev. C}\ }\textbf {\bibinfo {volume}
  {88}},\ \bibinfo {pages} {011601} (\bibinfo {year} {2013})}\BibitemShut
  {NoStop}%
\bibitem [{\citenamefont {Lay}\ \emph {et~al.}(2014)\citenamefont {Lay},
  \citenamefont {Moro}, \citenamefont {Arias},\ and\ \citenamefont
  {Kanada-En'yo}}]{PhysRevC.89.014333}%
  \BibitemOpen
  \bibfield  {author} {\bibinfo {author} {\bibfnamefont {J.~A.}\ \bibnamefont
  {Lay}}, \bibinfo {author} {\bibfnamefont {A.~M.}\ \bibnamefont {Moro}},
  \bibinfo {author} {\bibfnamefont {J.~M.}\ \bibnamefont {Arias}}, \ and\
  \bibinfo {author} {\bibfnamefont {Y.}~\bibnamefont {Kanada-En'yo}},\ }\href
  {\doibase 10.1103/PhysRevC.89.014333} {\bibfield  {journal} {\bibinfo
  {journal} {Phys. Rev. C}\ }\textbf {\bibinfo {volume} {89}},\ \bibinfo
  {pages} {014333} (\bibinfo {year} {2014})}\BibitemShut {NoStop}%
\bibitem [{\citenamefont {de~Diego}\ \emph {et~al.}(2014)\citenamefont
  {de~Diego}, \citenamefont {Arias}, \citenamefont {Lay},\ and\ \citenamefont
  {Moro}}]{PhysRevC.89.064609}%
  \BibitemOpen
  \bibfield  {author} {\bibinfo {author} {\bibfnamefont {R.}~\bibnamefont
  {de~Diego}}, \bibinfo {author} {\bibfnamefont {J.~M.}\ \bibnamefont {Arias}},
  \bibinfo {author} {\bibfnamefont {J.~A.}\ \bibnamefont {Lay}}, \ and\
  \bibinfo {author} {\bibfnamefont {A.~M.}\ \bibnamefont {Moro}},\ }\href
  {\doibase 10.1103/PhysRevC.89.064609} {\bibfield  {journal} {\bibinfo
  {journal} {Phys. Rev. C}\ }\textbf {\bibinfo {volume} {89}},\ \bibinfo
  {pages} {064609} (\bibinfo {year} {2014})}\BibitemShut {NoStop}%
\bibitem [{\citenamefont {Barranco}\ \emph {et~al.}(2017)\citenamefont
  {Barranco}, \citenamefont {Potel}, \citenamefont {Broglia},\ and\
  \citenamefont {Vigezzi}}]{PhysRevLett.119.082501}%
  \BibitemOpen
  \bibfield  {author} {\bibinfo {author} {\bibfnamefont {F.}~\bibnamefont
  {Barranco}}, \bibinfo {author} {\bibfnamefont {G.}~\bibnamefont {Potel}},
  \bibinfo {author} {\bibfnamefont {R.~A.}\ \bibnamefont {Broglia}}, \ and\
  \bibinfo {author} {\bibfnamefont {E.}~\bibnamefont {Vigezzi}},\ }\href
  {\doibase 10.1103/PhysRevLett.119.082501} {\bibfield  {journal} {\bibinfo
  {journal} {Phys. Rev. Lett.}\ }\textbf {\bibinfo {volume} {119}},\ \bibinfo
  {pages} {082501} (\bibinfo {year} {2017})}\BibitemShut {NoStop}%
\bibitem [{\citenamefont {Auton}(1970)}]{AUTON1970305}%
  \BibitemOpen
  \bibfield  {author} {\bibinfo {author} {\bibfnamefont {D.}~\bibnamefont
  {Auton}},\ }\href {\doibase https://doi.org/10.1016/0375-9474(70)90115-6}
  {\bibfield  {journal} {\bibinfo  {journal} {Nuclear Physics A}\ }\textbf
  {\bibinfo {volume} {157}},\ \bibinfo {pages} {305 } (\bibinfo {year}
  {1970})}\BibitemShut {NoStop}%
\bibitem [{\citenamefont {Zwieglinski}\ \emph {et~al.}(1979)\citenamefont
  {Zwieglinski}, \citenamefont {Benenson}, \citenamefont {Robertson},\ and\
  \citenamefont {Coker}}]{ZWIEGLINSKI1979124}%
  \BibitemOpen
  \bibfield  {author} {\bibinfo {author} {\bibfnamefont {B.}~\bibnamefont
  {Zwieglinski}}, \bibinfo {author} {\bibfnamefont {W.}~\bibnamefont
  {Benenson}}, \bibinfo {author} {\bibfnamefont {R.}~\bibnamefont {Robertson}},
  \ and\ \bibinfo {author} {\bibfnamefont {W.}~\bibnamefont {Coker}},\ }\href
  {\doibase https://doi.org/10.1016/0375-9474(79)90637-7} {\bibfield  {journal}
  {\bibinfo  {journal} {Nuclear Physics A}\ }\textbf {\bibinfo {volume}
  {315}},\ \bibinfo {pages} {124 } (\bibinfo {year} {1979})}\BibitemShut
  {NoStop}%
\bibitem [{\citenamefont {Fortier}\ \emph {et~al.}(1999)\citenamefont
  {Fortier}, \citenamefont {Pita}, \citenamefont {Winfield}, \citenamefont
  {Catford}, \citenamefont {Orr}, \citenamefont {de~Wiele}, \citenamefont
  {Blumenfeld}, \citenamefont {Chapman}, \citenamefont {Chappell},
  \citenamefont {Clarke}, \citenamefont {Curtis}, \citenamefont {Freer},
  \citenamefont {Gal\`es}, \citenamefont {Jones}, \citenamefont
  {Langevin-Joliot}, \citenamefont {Laurent}, \citenamefont {Lhenry},
  \citenamefont {Maison}, \citenamefont {Roussel-Chomaz}, \citenamefont
  {Shawcross}, \citenamefont {Smith}, \citenamefont {Spohr}, \citenamefont
  {Suomij$\ddot{\mathrm{a}}$rvi},\ and\ \citenamefont
  {de~Vismes}}]{FORTIER199922}%
  \BibitemOpen
  \bibfield  {author} {\bibinfo {author} {\bibfnamefont {S.}~\bibnamefont
  {Fortier}}, \bibinfo {author} {\bibfnamefont {S.}~\bibnamefont {Pita}},
  \bibinfo {author} {\bibfnamefont {J.}~\bibnamefont {Winfield}}, \bibinfo
  {author} {\bibfnamefont {W.}~\bibnamefont {Catford}}, \bibinfo {author}
  {\bibfnamefont {N.}~\bibnamefont {Orr}}, \bibinfo {author} {\bibfnamefont
  {J.~V.}\ \bibnamefont {de~Wiele}}, \bibinfo {author} {\bibfnamefont
  {Y.}~\bibnamefont {Blumenfeld}}, \bibinfo {author} {\bibfnamefont
  {R.}~\bibnamefont {Chapman}}, \bibinfo {author} {\bibfnamefont
  {S.}~\bibnamefont {Chappell}}, \bibinfo {author} {\bibfnamefont
  {N.}~\bibnamefont {Clarke}}, \bibinfo {author} {\bibfnamefont
  {N.}~\bibnamefont {Curtis}}, \bibinfo {author} {\bibfnamefont
  {M.}~\bibnamefont {Freer}}, \bibinfo {author} {\bibfnamefont
  {S.}~\bibnamefont {Gal\`es}}, \bibinfo {author} {\bibfnamefont
  {K.}~\bibnamefont {Jones}}, \bibinfo {author} {\bibfnamefont
  {H.}~\bibnamefont {Langevin-Joliot}}, \bibinfo {author} {\bibfnamefont
  {H.}~\bibnamefont {Laurent}}, \bibinfo {author} {\bibfnamefont
  {I.}~\bibnamefont {Lhenry}}, \bibinfo {author} {\bibfnamefont
  {J.}~\bibnamefont {Maison}}, \bibinfo {author} {\bibfnamefont
  {P.}~\bibnamefont {Roussel-Chomaz}}, \bibinfo {author} {\bibfnamefont
  {M.}~\bibnamefont {Shawcross}}, \bibinfo {author} {\bibfnamefont
  {M.}~\bibnamefont {Smith}}, \bibinfo {author} {\bibfnamefont
  {K.}~\bibnamefont {Spohr}}, \bibinfo {author} {\bibfnamefont
  {T.}~\bibnamefont {Suomij$\ddot{\mathrm{a}}$rvi}}, \ and\ \bibinfo {author}
  {\bibfnamefont {A.}~\bibnamefont {de~Vismes}},\ }\href {\doibase
  https://doi.org/10.1016/S0370-2693(99)00825-4} {\bibfield  {journal}
  {\bibinfo  {journal} {Physics Letters B}\ }\textbf {\bibinfo {volume}
  {461}},\ \bibinfo {pages} {22 } (\bibinfo {year} {1999})}\BibitemShut
  {NoStop}%
\bibitem [{\citenamefont {Aumann}\ \emph {et~al.}(2000)\citenamefont {Aumann},
  \citenamefont {Navin}, \citenamefont {Balamuth}, \citenamefont {Bazin},
  \citenamefont {Blank}, \citenamefont {Brown}, \citenamefont {Bush},
  \citenamefont {Caggiano}, \citenamefont {Davids}, \citenamefont {Glasmacher},
  \citenamefont {Guimar\~aes}, \citenamefont {Hansen}, \citenamefont
  {Ibbotson}, \citenamefont {Karnes}, \citenamefont {Kolata}, \citenamefont
  {Maddalena}, \citenamefont {Pritychenko}, \citenamefont {Scheit},
  \citenamefont {Sherrill},\ and\ \citenamefont {Tostevin}}]{Be11Exp}%
  \BibitemOpen
  \bibfield  {author} {\bibinfo {author} {\bibfnamefont {T.}~\bibnamefont
  {Aumann}}, \bibinfo {author} {\bibfnamefont {A.}~\bibnamefont {Navin}},
  \bibinfo {author} {\bibfnamefont {D.~P.}\ \bibnamefont {Balamuth}}, \bibinfo
  {author} {\bibfnamefont {D.}~\bibnamefont {Bazin}}, \bibinfo {author}
  {\bibfnamefont {B.}~\bibnamefont {Blank}}, \bibinfo {author} {\bibfnamefont
  {B.~A.}\ \bibnamefont {Brown}}, \bibinfo {author} {\bibfnamefont {J.~E.}\
  \bibnamefont {Bush}}, \bibinfo {author} {\bibfnamefont {J.~A.}\ \bibnamefont
  {Caggiano}}, \bibinfo {author} {\bibfnamefont {B.}~\bibnamefont {Davids}},
  \bibinfo {author} {\bibfnamefont {T.}~\bibnamefont {Glasmacher}}, \bibinfo
  {author} {\bibfnamefont {V.}~\bibnamefont {Guimar\~aes}}, \bibinfo {author}
  {\bibfnamefont {P.~G.}\ \bibnamefont {Hansen}}, \bibinfo {author}
  {\bibfnamefont {R.~W.}\ \bibnamefont {Ibbotson}}, \bibinfo {author}
  {\bibfnamefont {D.}~\bibnamefont {Karnes}}, \bibinfo {author} {\bibfnamefont
  {J.~J.}\ \bibnamefont {Kolata}}, \bibinfo {author} {\bibfnamefont
  {V.}~\bibnamefont {Maddalena}}, \bibinfo {author} {\bibfnamefont
  {B.}~\bibnamefont {Pritychenko}}, \bibinfo {author} {\bibfnamefont
  {H.}~\bibnamefont {Scheit}}, \bibinfo {author} {\bibfnamefont {B.~M.}\
  \bibnamefont {Sherrill}}, \ and\ \bibinfo {author} {\bibfnamefont {J.~A.}\
  \bibnamefont {Tostevin}},\ }\href {\doibase 10.1103/PhysRevLett.84.35}
  {\bibfield  {journal} {\bibinfo  {journal} {Phys. Rev. Lett.}\ }\textbf
  {\bibinfo {volume} {84}},\ \bibinfo {pages} {35} (\bibinfo {year}
  {2000})}\BibitemShut {NoStop}%
\bibitem [{\citenamefont {Iwasaki}\ \emph {et~al.}(2000)\citenamefont
  {Iwasaki}, \citenamefont {Motobayashi}, \citenamefont {Akiyoshi},
  \citenamefont {Ando}, \citenamefont {Fukuda}, \citenamefont {Fujiwara},
  \citenamefont {Fülöp}, \citenamefont {Hahn}, \citenamefont {Higurashi},
  \citenamefont {Hirai}, \citenamefont {Hisanaga}, \citenamefont {Iwasa},
  \citenamefont {Kijima}, \citenamefont {Minemura}, \citenamefont {Nakamura},
  \citenamefont {Notani}, \citenamefont {Ozawa}, \citenamefont {Sakurai},
  \citenamefont {Shimoura}, \citenamefont {Takeuchi}, \citenamefont
  {Teranishi}, \citenamefont {Yanagisawa},\ and\ \citenamefont
  {Ishihara}}]{IWASAKI20007}%
  \BibitemOpen
  \bibfield  {author} {\bibinfo {author} {\bibfnamefont {H.}~\bibnamefont
  {Iwasaki}}, \bibinfo {author} {\bibfnamefont {T.}~\bibnamefont
  {Motobayashi}}, \bibinfo {author} {\bibfnamefont {H.}~\bibnamefont
  {Akiyoshi}}, \bibinfo {author} {\bibfnamefont {Y.}~\bibnamefont {Ando}},
  \bibinfo {author} {\bibfnamefont {N.}~\bibnamefont {Fukuda}}, \bibinfo
  {author} {\bibfnamefont {H.}~\bibnamefont {Fujiwara}}, \bibinfo {author}
  {\bibfnamefont {Z.}~\bibnamefont {Fülöp}}, \bibinfo {author} {\bibfnamefont
  {K.}~\bibnamefont {Hahn}}, \bibinfo {author} {\bibfnamefont {Y.}~\bibnamefont
  {Higurashi}}, \bibinfo {author} {\bibfnamefont {M.}~\bibnamefont {Hirai}},
  \bibinfo {author} {\bibfnamefont {I.}~\bibnamefont {Hisanaga}}, \bibinfo
  {author} {\bibfnamefont {N.}~\bibnamefont {Iwasa}}, \bibinfo {author}
  {\bibfnamefont {T.}~\bibnamefont {Kijima}}, \bibinfo {author} {\bibfnamefont
  {T.}~\bibnamefont {Minemura}}, \bibinfo {author} {\bibfnamefont
  {T.}~\bibnamefont {Nakamura}}, \bibinfo {author} {\bibfnamefont
  {M.}~\bibnamefont {Notani}}, \bibinfo {author} {\bibfnamefont
  {S.}~\bibnamefont {Ozawa}}, \bibinfo {author} {\bibfnamefont
  {H.}~\bibnamefont {Sakurai}}, \bibinfo {author} {\bibfnamefont
  {S.}~\bibnamefont {Shimoura}}, \bibinfo {author} {\bibfnamefont
  {S.}~\bibnamefont {Takeuchi}}, \bibinfo {author} {\bibfnamefont
  {T.}~\bibnamefont {Teranishi}}, \bibinfo {author} {\bibfnamefont
  {Y.}~\bibnamefont {Yanagisawa}}, \ and\ \bibinfo {author} {\bibfnamefont
  {M.}~\bibnamefont {Ishihara}},\ }\href {\doibase
  https://doi.org/10.1016/S0370-2693(00)00428-7} {\bibfield  {journal}
  {\bibinfo  {journal} {Physics Letters B}\ }\textbf {\bibinfo {volume}
  {481}},\ \bibinfo {pages} {7 } (\bibinfo {year} {2000})}\BibitemShut
  {NoStop}%
\bibitem [{\citenamefont {Winfield}\ \emph {et~al.}(2001)\citenamefont
  {Winfield}, \citenamefont {Fortier}, \citenamefont {Catford}, \citenamefont
  {Pita}, \citenamefont {Orr}, \citenamefont {de~Wiele}, \citenamefont
  {Blumenfeld}, \citenamefont {Chapman}, \citenamefont {Chappell},
  \citenamefont {Clarke}, \citenamefont {Curtis}, \citenamefont {Freer},
  \citenamefont {Gal\`es}, \citenamefont {Langevin-Joliot}, \citenamefont
  {Laurent}, \citenamefont {Lhenry}, \citenamefont {Maison}, \citenamefont
  {Roussel-Chomaz}, \citenamefont {Shawcross}, \citenamefont {Spohr},
  \citenamefont {Suomij$\ddot{\mathrm{a}}$rvi},\ and\ \citenamefont
  {de~Vismes}}]{WINFIELD200148}%
  \BibitemOpen
  \bibfield  {author} {\bibinfo {author} {\bibfnamefont {J.}~\bibnamefont
  {Winfield}}, \bibinfo {author} {\bibfnamefont {S.}~\bibnamefont {Fortier}},
  \bibinfo {author} {\bibfnamefont {W.}~\bibnamefont {Catford}}, \bibinfo
  {author} {\bibfnamefont {S.}~\bibnamefont {Pita}}, \bibinfo {author}
  {\bibfnamefont {N.}~\bibnamefont {Orr}}, \bibinfo {author} {\bibfnamefont
  {J.~V.}\ \bibnamefont {de~Wiele}}, \bibinfo {author} {\bibfnamefont
  {Y.}~\bibnamefont {Blumenfeld}}, \bibinfo {author} {\bibfnamefont
  {R.}~\bibnamefont {Chapman}}, \bibinfo {author} {\bibfnamefont
  {S.}~\bibnamefont {Chappell}}, \bibinfo {author} {\bibfnamefont
  {N.}~\bibnamefont {Clarke}}, \bibinfo {author} {\bibfnamefont
  {N.}~\bibnamefont {Curtis}}, \bibinfo {author} {\bibfnamefont
  {M.}~\bibnamefont {Freer}}, \bibinfo {author} {\bibfnamefont
  {S.}~\bibnamefont {Gal\`es}}, \bibinfo {author} {\bibfnamefont
  {H.}~\bibnamefont {Langevin-Joliot}}, \bibinfo {author} {\bibfnamefont
  {H.}~\bibnamefont {Laurent}}, \bibinfo {author} {\bibfnamefont
  {I.}~\bibnamefont {Lhenry}}, \bibinfo {author} {\bibfnamefont
  {J.}~\bibnamefont {Maison}}, \bibinfo {author} {\bibfnamefont
  {P.}~\bibnamefont {Roussel-Chomaz}}, \bibinfo {author} {\bibfnamefont
  {M.}~\bibnamefont {Shawcross}}, \bibinfo {author} {\bibfnamefont
  {K.}~\bibnamefont {Spohr}}, \bibinfo {author} {\bibfnamefont
  {T.}~\bibnamefont {Suomij$\ddot{\mathrm{a}}$rvi}}, \ and\ \bibinfo {author}
  {\bibfnamefont {A.}~\bibnamefont {de~Vismes}},\ }\href {\doibase
  https://doi.org/10.1016/S0375-9474(00)00463-2} {\bibfield  {journal}
  {\bibinfo  {journal} {Nuclear Physics A}\ }\textbf {\bibinfo {volume}
  {683}},\ \bibinfo {pages} {48 } (\bibinfo {year} {2001})}\BibitemShut
  {NoStop}%
\bibitem [{\citenamefont {N\"ortersh\"auser}\ \emph {et~al.}(2009)\citenamefont
  {N\"ortersh\"auser}, \citenamefont {Tiedemann}, \citenamefont
  {\ifmmode~\check{Z}\else \v{Z}\fi{}\'akov\'a}, \citenamefont {Andjelkovic},
  \citenamefont {Blaum}, \citenamefont {Bissell}, \citenamefont {Cazan},
  \citenamefont {Drake}, \citenamefont {Geppert}, \citenamefont {Kowalska},
  \citenamefont {Kr\"amer}, \citenamefont {Krieger}, \citenamefont {Neugart},
  \citenamefont {S\'anchez}, \citenamefont {Schmidt-Kaler}, \citenamefont
  {Yan}, \citenamefont {Yordanov},\ and\ \citenamefont
  {Zimmermann}}]{PhysRevLett.102.062503}%
  \BibitemOpen
  \bibfield  {author} {\bibinfo {author} {\bibfnamefont {W.}~\bibnamefont
  {N\"ortersh\"auser}}, \bibinfo {author} {\bibfnamefont {D.}~\bibnamefont
  {Tiedemann}}, \bibinfo {author} {\bibfnamefont {M.}~\bibnamefont
  {\ifmmode~\check{Z}\else \v{Z}\fi{}\'akov\'a}}, \bibinfo {author}
  {\bibfnamefont {Z.}~\bibnamefont {Andjelkovic}}, \bibinfo {author}
  {\bibfnamefont {K.}~\bibnamefont {Blaum}}, \bibinfo {author} {\bibfnamefont
  {M.~L.}\ \bibnamefont {Bissell}}, \bibinfo {author} {\bibfnamefont
  {R.}~\bibnamefont {Cazan}}, \bibinfo {author} {\bibfnamefont {G.~W.~F.}\
  \bibnamefont {Drake}}, \bibinfo {author} {\bibfnamefont {C.}~\bibnamefont
  {Geppert}}, \bibinfo {author} {\bibfnamefont {M.}~\bibnamefont {Kowalska}},
  \bibinfo {author} {\bibfnamefont {J.}~\bibnamefont {Kr\"amer}}, \bibinfo
  {author} {\bibfnamefont {A.}~\bibnamefont {Krieger}}, \bibinfo {author}
  {\bibfnamefont {R.}~\bibnamefont {Neugart}}, \bibinfo {author} {\bibfnamefont
  {R.}~\bibnamefont {S\'anchez}}, \bibinfo {author} {\bibfnamefont
  {F.}~\bibnamefont {Schmidt-Kaler}}, \bibinfo {author} {\bibfnamefont {Z.-C.}\
  \bibnamefont {Yan}}, \bibinfo {author} {\bibfnamefont {D.~T.}\ \bibnamefont
  {Yordanov}}, \ and\ \bibinfo {author} {\bibfnamefont {C.}~\bibnamefont
  {Zimmermann}},\ }\href {\doibase 10.1103/PhysRevLett.102.062503} {\bibfield
  {journal} {\bibinfo  {journal} {Phys. Rev. Lett.}\ }\textbf {\bibinfo
  {volume} {102}},\ \bibinfo {pages} {062503} (\bibinfo {year}
  {2009})}\BibitemShut {NoStop}%
\bibitem [{\citenamefont {Schmitt}\ \emph {et~al.}(2013)\citenamefont
  {Schmitt}, \citenamefont {Jones}, \citenamefont {Ahn}, \citenamefont
  {Bardayan}, \citenamefont {Bey}, \citenamefont {Blackmon}, \citenamefont
  {Brown}, \citenamefont {Chae}, \citenamefont {Chipps}, \citenamefont
  {Cizewski}, \citenamefont {Hahn}, \citenamefont {Kolata}, \citenamefont
  {Kozub}, \citenamefont {Liang}, \citenamefont {Matei}, \citenamefont {Matos},
  \citenamefont {Matyas}, \citenamefont {Moazen}, \citenamefont {Nesaraja},
  \citenamefont {Nunes}, \citenamefont {O'Malley}, \citenamefont {Pain},
  \citenamefont {Peters}, \citenamefont {Pittman}, \citenamefont {Roberts},
  \citenamefont {Shapira}, \citenamefont {Shriner}, \citenamefont {Smith},
  \citenamefont {Spassova}, \citenamefont {Stracener}, \citenamefont
  {Upadhyay}, \citenamefont {Villano},\ and\ \citenamefont
  {Wilson}}]{PhysRevC.88.064612}%
  \BibitemOpen
  \bibfield  {author} {\bibinfo {author} {\bibfnamefont {K.~T.}\ \bibnamefont
  {Schmitt}}, \bibinfo {author} {\bibfnamefont {K.~L.}\ \bibnamefont {Jones}},
  \bibinfo {author} {\bibfnamefont {S.}~\bibnamefont {Ahn}}, \bibinfo {author}
  {\bibfnamefont {D.~W.}\ \bibnamefont {Bardayan}}, \bibinfo {author}
  {\bibfnamefont {A.}~\bibnamefont {Bey}}, \bibinfo {author} {\bibfnamefont
  {J.~C.}\ \bibnamefont {Blackmon}}, \bibinfo {author} {\bibfnamefont {S.~M.}\
  \bibnamefont {Brown}}, \bibinfo {author} {\bibfnamefont {K.~Y.}\ \bibnamefont
  {Chae}}, \bibinfo {author} {\bibfnamefont {K.~A.}\ \bibnamefont {Chipps}},
  \bibinfo {author} {\bibfnamefont {J.~A.}\ \bibnamefont {Cizewski}}, \bibinfo
  {author} {\bibfnamefont {K.~I.}\ \bibnamefont {Hahn}}, \bibinfo {author}
  {\bibfnamefont {J.~J.}\ \bibnamefont {Kolata}}, \bibinfo {author}
  {\bibfnamefont {R.~L.}\ \bibnamefont {Kozub}}, \bibinfo {author}
  {\bibfnamefont {J.~F.}\ \bibnamefont {Liang}}, \bibinfo {author}
  {\bibfnamefont {C.}~\bibnamefont {Matei}}, \bibinfo {author} {\bibfnamefont
  {M.}~\bibnamefont {Matos}}, \bibinfo {author} {\bibfnamefont
  {D.}~\bibnamefont {Matyas}}, \bibinfo {author} {\bibfnamefont
  {B.}~\bibnamefont {Moazen}}, \bibinfo {author} {\bibfnamefont {C.~D.}\
  \bibnamefont {Nesaraja}}, \bibinfo {author} {\bibfnamefont {F.~M.}\
  \bibnamefont {Nunes}}, \bibinfo {author} {\bibfnamefont {P.~D.}\ \bibnamefont
  {O'Malley}}, \bibinfo {author} {\bibfnamefont {S.~D.}\ \bibnamefont {Pain}},
  \bibinfo {author} {\bibfnamefont {W.~A.}\ \bibnamefont {Peters}}, \bibinfo
  {author} {\bibfnamefont {S.~T.}\ \bibnamefont {Pittman}}, \bibinfo {author}
  {\bibfnamefont {A.}~\bibnamefont {Roberts}}, \bibinfo {author} {\bibfnamefont
  {D.}~\bibnamefont {Shapira}}, \bibinfo {author} {\bibfnamefont {J.~F.}\
  \bibnamefont {Shriner}}, \bibinfo {author} {\bibfnamefont {M.~S.}\
  \bibnamefont {Smith}}, \bibinfo {author} {\bibfnamefont {I.}~\bibnamefont
  {Spassova}}, \bibinfo {author} {\bibfnamefont {D.~W.}\ \bibnamefont
  {Stracener}}, \bibinfo {author} {\bibfnamefont {N.~J.}\ \bibnamefont
  {Upadhyay}}, \bibinfo {author} {\bibfnamefont {A.~N.}\ \bibnamefont
  {Villano}}, \ and\ \bibinfo {author} {\bibfnamefont {G.~L.}\ \bibnamefont
  {Wilson}},\ }\href {\doibase 10.1103/PhysRevC.88.064612} {\bibfield
  {journal} {\bibinfo  {journal} {Phys. Rev. C}\ }\textbf {\bibinfo {volume}
  {88}},\ \bibinfo {pages} {064612} (\bibinfo {year} {2013})}\BibitemShut
  {NoStop}%
\bibitem [{\citenamefont {Kwan}\ \emph {et~al.}(2014)\citenamefont {Kwan},
  \citenamefont {Wu}, \citenamefont {Summers}, \citenamefont {Hackman},
  \citenamefont {Drake}, \citenamefont {Andreoiu}, \citenamefont {Ashley},
  \citenamefont {Ball}, \citenamefont {Bender}, \citenamefont {Boston},
  \citenamefont {Boston}, \citenamefont {Chester}, \citenamefont {Close},
  \citenamefont {Cline}, \citenamefont {Cross}, \citenamefont {Dunlop},
  \citenamefont {Finlay}, \citenamefont {Garnsworthy}, \citenamefont {Hayes},
  \citenamefont {Laffoley}, \citenamefont {Nano}, \citenamefont {Navr\`atil},
  \citenamefont {Pearson}, \citenamefont {Pore}, \citenamefont {Quaglioni},
  \citenamefont {Svensson}, \citenamefont {Starosta}, \citenamefont {Thompson},
  \citenamefont {Voss}, \citenamefont {Williams},\ and\ \citenamefont
  {Wang}}]{KWAN2014210}%
  \BibitemOpen
  \bibfield  {author} {\bibinfo {author} {\bibfnamefont {E.}~\bibnamefont
  {Kwan}}, \bibinfo {author} {\bibfnamefont {C.}~\bibnamefont {Wu}}, \bibinfo
  {author} {\bibfnamefont {N.}~\bibnamefont {Summers}}, \bibinfo {author}
  {\bibfnamefont {G.}~\bibnamefont {Hackman}}, \bibinfo {author} {\bibfnamefont
  {T.}~\bibnamefont {Drake}}, \bibinfo {author} {\bibfnamefont
  {C.}~\bibnamefont {Andreoiu}}, \bibinfo {author} {\bibfnamefont
  {R.}~\bibnamefont {Ashley}}, \bibinfo {author} {\bibfnamefont
  {G.}~\bibnamefont {Ball}}, \bibinfo {author} {\bibfnamefont {P.}~\bibnamefont
  {Bender}}, \bibinfo {author} {\bibfnamefont {A.}~\bibnamefont {Boston}},
  \bibinfo {author} {\bibfnamefont {H.}~\bibnamefont {Boston}}, \bibinfo
  {author} {\bibfnamefont {A.}~\bibnamefont {Chester}}, \bibinfo {author}
  {\bibfnamefont {A.}~\bibnamefont {Close}}, \bibinfo {author} {\bibfnamefont
  {D.}~\bibnamefont {Cline}}, \bibinfo {author} {\bibfnamefont
  {D.}~\bibnamefont {Cross}}, \bibinfo {author} {\bibfnamefont
  {R.}~\bibnamefont {Dunlop}}, \bibinfo {author} {\bibfnamefont
  {A.}~\bibnamefont {Finlay}}, \bibinfo {author} {\bibfnamefont
  {A.}~\bibnamefont {Garnsworthy}}, \bibinfo {author} {\bibfnamefont
  {A.}~\bibnamefont {Hayes}}, \bibinfo {author} {\bibfnamefont
  {A.}~\bibnamefont {Laffoley}}, \bibinfo {author} {\bibfnamefont
  {T.}~\bibnamefont {Nano}}, \bibinfo {author} {\bibfnamefont {P.}~\bibnamefont
  {Navr\`atil}}, \bibinfo {author} {\bibfnamefont {C.}~\bibnamefont {Pearson}},
  \bibinfo {author} {\bibfnamefont {J.}~\bibnamefont {Pore}}, \bibinfo {author}
  {\bibfnamefont {S.}~\bibnamefont {Quaglioni}}, \bibinfo {author}
  {\bibfnamefont {C.}~\bibnamefont {Svensson}}, \bibinfo {author}
  {\bibfnamefont {K.}~\bibnamefont {Starosta}}, \bibinfo {author}
  {\bibfnamefont {I.}~\bibnamefont {Thompson}}, \bibinfo {author}
  {\bibfnamefont {P.}~\bibnamefont {Voss}}, \bibinfo {author} {\bibfnamefont
  {S.}~\bibnamefont {Williams}}, \ and\ \bibinfo {author} {\bibfnamefont
  {Z.}~\bibnamefont {Wang}},\ }\href {\doibase
  https://doi.org/10.1016/j.physletb.2014.03.049} {\bibfield  {journal}
  {\bibinfo  {journal} {Physics Letters B}\ }\textbf {\bibinfo {volume}
  {732}},\ \bibinfo {pages} {210 } (\bibinfo {year} {2014})}\BibitemShut
  {NoStop}%
\bibitem [{\citenamefont {Thoennessen}\ \emph {et~al.}(1999)\citenamefont
  {Thoennessen}, \citenamefont {Yokoyama}, \citenamefont {Azhari},
  \citenamefont {Baumann}, \citenamefont {Brown}, \citenamefont {Galonsky},
  \citenamefont {Hansen}, \citenamefont {Kelley}, \citenamefont {Kryger},
  \citenamefont {Ramakrishnan},\ and\ \citenamefont {Thirolf}}]{Thoen99}%
  \BibitemOpen
  \bibfield  {author} {\bibinfo {author} {\bibfnamefont {M.}~\bibnamefont
  {Thoennessen}}, \bibinfo {author} {\bibfnamefont {S.}~\bibnamefont
  {Yokoyama}}, \bibinfo {author} {\bibfnamefont {A.}~\bibnamefont {Azhari}},
  \bibinfo {author} {\bibfnamefont {T.}~\bibnamefont {Baumann}}, \bibinfo
  {author} {\bibfnamefont {J.~A.}\ \bibnamefont {Brown}}, \bibinfo {author}
  {\bibfnamefont {A.}~\bibnamefont {Galonsky}}, \bibinfo {author}
  {\bibfnamefont {P.~G.}\ \bibnamefont {Hansen}}, \bibinfo {author}
  {\bibfnamefont {J.~H.}\ \bibnamefont {Kelley}}, \bibinfo {author}
  {\bibfnamefont {R.~A.}\ \bibnamefont {Kryger}}, \bibinfo {author}
  {\bibfnamefont {E.}~\bibnamefont {Ramakrishnan}}, \ and\ \bibinfo {author}
  {\bibfnamefont {P.}~\bibnamefont {Thirolf}},\ }\href {\doibase
  10.1103/PhysRevC.59.111} {\bibfield  {journal} {\bibinfo  {journal} {Phys.
  Rev. C}\ }\textbf {\bibinfo {volume} {59}},\ \bibinfo {pages} {111} (\bibinfo
  {year} {1999})}\BibitemShut {NoStop}%
\bibitem [{\citenamefont {Simon}\ \emph {et~al.}(2004)\citenamefont {Simon},
  \citenamefont {Aumann}, \citenamefont {Borge}, \citenamefont {Chulkov},
  \citenamefont {Elze}, \citenamefont {Emling}, \citenamefont {Forss\'en},
  \citenamefont {Geissel}, \citenamefont {Hellstr$\ddot{\mathrm{o}}$m},
  \citenamefont {Jonson}, \citenamefont {Kratz}, \citenamefont {Kulessa},
  \citenamefont {Leifels}, \citenamefont {Markenroth}, \citenamefont {Meister},
  \citenamefont {M$\ddot{\mathrm{u}}$nzenberg}, \citenamefont {Nickel},
  \citenamefont {Nilsson}, \citenamefont {Nyman}, \citenamefont {Pribora},
  \citenamefont {Richter}, \citenamefont {Riisager}, \citenamefont
  {Scheidenberger}, \citenamefont {Schrieder}, \citenamefont {Tengblad},\ and\
  \citenamefont {Zhukov}}]{SIMON2004323}%
  \BibitemOpen
  \bibfield  {author} {\bibinfo {author} {\bibfnamefont {H.}~\bibnamefont
  {Simon}}, \bibinfo {author} {\bibfnamefont {T.}~\bibnamefont {Aumann}},
  \bibinfo {author} {\bibfnamefont {M.}~\bibnamefont {Borge}}, \bibinfo
  {author} {\bibfnamefont {L.}~\bibnamefont {Chulkov}}, \bibinfo {author}
  {\bibfnamefont {T.}~\bibnamefont {Elze}}, \bibinfo {author} {\bibfnamefont
  {H.}~\bibnamefont {Emling}}, \bibinfo {author} {\bibfnamefont
  {C.}~\bibnamefont {Forss\'en}}, \bibinfo {author} {\bibfnamefont
  {H.}~\bibnamefont {Geissel}}, \bibinfo {author} {\bibfnamefont
  {M.}~\bibnamefont {Hellstr$\ddot{\mathrm{o}}$m}}, \bibinfo {author}
  {\bibfnamefont {B.}~\bibnamefont {Jonson}}, \bibinfo {author} {\bibfnamefont
  {J.}~\bibnamefont {Kratz}}, \bibinfo {author} {\bibfnamefont
  {R.}~\bibnamefont {Kulessa}}, \bibinfo {author} {\bibfnamefont
  {Y.}~\bibnamefont {Leifels}}, \bibinfo {author} {\bibfnamefont
  {K.}~\bibnamefont {Markenroth}}, \bibinfo {author} {\bibfnamefont
  {M.}~\bibnamefont {Meister}}, \bibinfo {author} {\bibfnamefont
  {G.}~\bibnamefont {M$\ddot{\mathrm{u}}$nzenberg}}, \bibinfo {author}
  {\bibfnamefont {F.}~\bibnamefont {Nickel}}, \bibinfo {author} {\bibfnamefont
  {T.}~\bibnamefont {Nilsson}}, \bibinfo {author} {\bibfnamefont
  {G.}~\bibnamefont {Nyman}}, \bibinfo {author} {\bibfnamefont
  {V.}~\bibnamefont {Pribora}}, \bibinfo {author} {\bibfnamefont
  {A.}~\bibnamefont {Richter}}, \bibinfo {author} {\bibfnamefont
  {K.}~\bibnamefont {Riisager}}, \bibinfo {author} {\bibfnamefont
  {C.}~\bibnamefont {Scheidenberger}}, \bibinfo {author} {\bibfnamefont
  {G.}~\bibnamefont {Schrieder}}, \bibinfo {author} {\bibfnamefont
  {O.}~\bibnamefont {Tengblad}}, \ and\ \bibinfo {author} {\bibfnamefont
  {M.}~\bibnamefont {Zhukov}},\ }\href {\doibase
  https://doi.org/10.1016/j.nuclphysa.2004.01.058} {\bibfield  {journal}
  {\bibinfo  {journal} {Nuclear Physics A}\ }\textbf {\bibinfo {volume}
  {734}},\ \bibinfo {pages} {323 } (\bibinfo {year} {2004})}\BibitemShut
  {NoStop}%
\bibitem [{\citenamefont {Jeppesen}\ \emph {et~al.}(2006)\citenamefont
  {Jeppesen}, \citenamefont {Moro}, \citenamefont {Bergmann}, \citenamefont
  {Borge}, \citenamefont {Cederk$\ddot{\mathrm{a}}$ll}, \citenamefont {Fraile},
  \citenamefont {Fynbo}, \citenamefont {G\'omez-Camacho}, \citenamefont
  {Johansson}, \citenamefont {Jonson}, \citenamefont {Meister}, \citenamefont
  {Nilsson}, \citenamefont {Nyman}, \citenamefont {Pantea}, \citenamefont
  {Riisager}, \citenamefont {Richter}, \citenamefont {Schrieder}, \citenamefont
  {Sieber}, \citenamefont {Tengblad}, \citenamefont {Tengborn}, \citenamefont
  {Turri\'on},\ and\ \citenamefont {Wenander}}]{JEPPESEN2006449}%
  \BibitemOpen
  \bibfield  {author} {\bibinfo {author} {\bibfnamefont {H.}~\bibnamefont
  {Jeppesen}}, \bibinfo {author} {\bibfnamefont {A.}~\bibnamefont {Moro}},
  \bibinfo {author} {\bibfnamefont {U.}~\bibnamefont {Bergmann}}, \bibinfo
  {author} {\bibfnamefont {M.}~\bibnamefont {Borge}}, \bibinfo {author}
  {\bibfnamefont {J.}~\bibnamefont {Cederk$\ddot{\mathrm{a}}$ll}}, \bibinfo
  {author} {\bibfnamefont {L.}~\bibnamefont {Fraile}}, \bibinfo {author}
  {\bibfnamefont {H.}~\bibnamefont {Fynbo}}, \bibinfo {author} {\bibfnamefont
  {J.}~\bibnamefont {G\'omez-Camacho}}, \bibinfo {author} {\bibfnamefont
  {H.}~\bibnamefont {Johansson}}, \bibinfo {author} {\bibfnamefont
  {B.}~\bibnamefont {Jonson}}, \bibinfo {author} {\bibfnamefont
  {M.}~\bibnamefont {Meister}}, \bibinfo {author} {\bibfnamefont
  {T.}~\bibnamefont {Nilsson}}, \bibinfo {author} {\bibfnamefont
  {G.}~\bibnamefont {Nyman}}, \bibinfo {author} {\bibfnamefont
  {M.}~\bibnamefont {Pantea}}, \bibinfo {author} {\bibfnamefont
  {K.}~\bibnamefont {Riisager}}, \bibinfo {author} {\bibfnamefont
  {A.}~\bibnamefont {Richter}}, \bibinfo {author} {\bibfnamefont
  {G.}~\bibnamefont {Schrieder}}, \bibinfo {author} {\bibfnamefont
  {T.}~\bibnamefont {Sieber}}, \bibinfo {author} {\bibfnamefont
  {O.}~\bibnamefont {Tengblad}}, \bibinfo {author} {\bibfnamefont
  {E.}~\bibnamefont {Tengborn}}, \bibinfo {author} {\bibfnamefont
  {M.}~\bibnamefont {Turri\'on}}, \ and\ \bibinfo {author} {\bibfnamefont
  {F.}~\bibnamefont {Wenander}},\ }\href {\doibase
  https://doi.org/10.1016/j.physletb.2006.09.060} {\bibfield  {journal}
  {\bibinfo  {journal} {Physics Letters B}\ }\textbf {\bibinfo {volume}
  {642}},\ \bibinfo {pages} {449 } (\bibinfo {year} {2006})}\BibitemShut
  {NoStop}%
\bibitem [{\citenamefont {Aksyutina}\ \emph {et~al.}()\citenamefont
  {Aksyutina}, \citenamefont {Johansson}, \citenamefont {Adrich}, \citenamefont
  {Aksouh}, \citenamefont {Aumann}, \citenamefont {Boretzky}, \citenamefont
  {Borge}, \citenamefont {Chatillon}, \citenamefont {Chulkov}, \citenamefont
  {Cortina-Gil}, \citenamefont {Pramanik}, \citenamefont {Emling},
  \citenamefont {Forss\'en}, \citenamefont {Fynbo}, \citenamefont {Geissel},
  \citenamefont {Hellstr$\ddot{\mathrm{o}}$m}, \citenamefont {Ickert},
  \citenamefont {Jones}, \citenamefont {Jonson}, \citenamefont {Kliemkiewicz},
  \citenamefont {Kratz}, \citenamefont {Kulessa}, \citenamefont {Lantz},
  \citenamefont {LeBleis}, \citenamefont {Lindahl}, \citenamefont {Mahata},
  \citenamefont {Matos}, \citenamefont {Meister}, \citenamefont
  {M$\ddot{\mathrm{u}}$nzenberg}, \citenamefont {Nilsson}, \citenamefont
  {Nyman}, \citenamefont {Palit}, \citenamefont {Pantea}, \citenamefont
  {Paschalis}, \citenamefont {Prokopowicz}, \citenamefont {Reifarth},
  \citenamefont {Richter}, \citenamefont {Riisager}, \citenamefont {Schrieder},
  \citenamefont {Simon}, \citenamefont {S$\ddot{\mathrm{u}}$mmerer},
  \citenamefont {Tengblad}, \citenamefont {Walus}, \citenamefont {Weick},\ and\
  \citenamefont {Zhukov}}]{AKSYUTINA430}%
  \BibitemOpen
  \bibfield  {author} {\bibinfo {author} {\bibfnamefont {Y.}~\bibnamefont
  {Aksyutina}}, \bibinfo {author} {\bibfnamefont {H.}~\bibnamefont
  {Johansson}}, \bibinfo {author} {\bibfnamefont {P.}~\bibnamefont {Adrich}},
  \bibinfo {author} {\bibfnamefont {F.}~\bibnamefont {Aksouh}}, \bibinfo
  {author} {\bibfnamefont {T.}~\bibnamefont {Aumann}}, \bibinfo {author}
  {\bibfnamefont {K.}~\bibnamefont {Boretzky}}, \bibinfo {author}
  {\bibfnamefont {M.}~\bibnamefont {Borge}}, \bibinfo {author} {\bibfnamefont
  {A.}~\bibnamefont {Chatillon}}, \bibinfo {author} {\bibfnamefont
  {L.}~\bibnamefont {Chulkov}}, \bibinfo {author} {\bibfnamefont
  {D.}~\bibnamefont {Cortina-Gil}}, \bibinfo {author} {\bibfnamefont {U.~D.}\
  \bibnamefont {Pramanik}}, \bibinfo {author} {\bibfnamefont {H.}~\bibnamefont
  {Emling}}, \bibinfo {author} {\bibfnamefont {C.}~\bibnamefont {Forss\'en}},
  \bibinfo {author} {\bibfnamefont {H.}~\bibnamefont {Fynbo}}, \bibinfo
  {author} {\bibfnamefont {H.}~\bibnamefont {Geissel}}, \bibinfo {author}
  {\bibfnamefont {M.}~\bibnamefont {Hellstr$\ddot{\mathrm{o}}$m}}, \bibinfo
  {author} {\bibfnamefont {G.}~\bibnamefont {Ickert}}, \bibinfo {author}
  {\bibfnamefont {K.}~\bibnamefont {Jones}}, \bibinfo {author} {\bibfnamefont
  {B.}~\bibnamefont {Jonson}}, \bibinfo {author} {\bibfnamefont
  {A.}~\bibnamefont {Kliemkiewicz}}, \bibinfo {author} {\bibfnamefont
  {J.}~\bibnamefont {Kratz}}, \bibinfo {author} {\bibfnamefont
  {R.}~\bibnamefont {Kulessa}}, \bibinfo {author} {\bibfnamefont
  {M.}~\bibnamefont {Lantz}}, \bibinfo {author} {\bibfnamefont
  {T.}~\bibnamefont {LeBleis}}, \bibinfo {author} {\bibfnamefont
  {A.}~\bibnamefont {Lindahl}}, \bibinfo {author} {\bibfnamefont
  {K.}~\bibnamefont {Mahata}}, \bibinfo {author} {\bibfnamefont
  {M.}~\bibnamefont {Matos}}, \bibinfo {author} {\bibfnamefont
  {M.}~\bibnamefont {Meister}}, \bibinfo {author} {\bibfnamefont
  {G.}~\bibnamefont {M$\ddot{\mathrm{u}}$nzenberg}}, \bibinfo {author}
  {\bibfnamefont {T.}~\bibnamefont {Nilsson}}, \bibinfo {author} {\bibfnamefont
  {G.}~\bibnamefont {Nyman}}, \bibinfo {author} {\bibfnamefont
  {R.}~\bibnamefont {Palit}}, \bibinfo {author} {\bibfnamefont
  {M.}~\bibnamefont {Pantea}}, \bibinfo {author} {\bibfnamefont
  {S.}~\bibnamefont {Paschalis}}, \bibinfo {author} {\bibfnamefont
  {W.}~\bibnamefont {Prokopowicz}}, \bibinfo {author} {\bibfnamefont
  {R.}~\bibnamefont {Reifarth}}, \bibinfo {author} {\bibfnamefont
  {A.}~\bibnamefont {Richter}}, \bibinfo {author} {\bibfnamefont
  {K.}~\bibnamefont {Riisager}}, \bibinfo {author} {\bibfnamefont
  {G.}~\bibnamefont {Schrieder}}, \bibinfo {author} {\bibfnamefont
  {H.}~\bibnamefont {Simon}}, \bibinfo {author} {\bibfnamefont
  {K.}~\bibnamefont {S$\ddot{\mathrm{u}}$mmerer}}, \bibinfo {author}
  {\bibfnamefont {O.}~\bibnamefont {Tengblad}}, \bibinfo {author}
  {\bibfnamefont {W.}~\bibnamefont {Walus}}, \bibinfo {author} {\bibfnamefont
  {H.}~\bibnamefont {Weick}}, \ and\ \bibinfo {author} {\bibfnamefont
  {M.}~\bibnamefont {Zhukov}},\ }\href {\doibase
  https://doi.org/10.1016/j.physletb.2008.07.093} {\bibfield  {journal}
  {\bibinfo  {journal} {Physics Letters B}\ }\textbf {\bibinfo {volume}
  {666}},\ \bibinfo {pages} {430 }}\BibitemShut {NoStop}%
\bibitem [{\citenamefont {Falou}\ \emph
  {et~al.}(2011{\natexlab{a}})\citenamefont {Falou}, \citenamefont {Leprince},\
  and\ \citenamefont {Orr}}]{Falou2011}%
  \BibitemOpen
  \bibfield  {author} {\bibinfo {author} {\bibfnamefont {H.~A.}\ \bibnamefont
  {Falou}}, \bibinfo {author} {\bibfnamefont {A.}~\bibnamefont {Leprince}}, \
  and\ \bibinfo {author} {\bibfnamefont {N.}~\bibnamefont {Orr}},\ }\href
  {http://stacks.iop.org/1742-6596/312/i=9/a=092012} {\bibfield  {journal}
  {\bibinfo  {journal} {Journal of Physics: Conference Series}\ }\textbf
  {\bibinfo {volume} {312}},\ \bibinfo {pages} {092012} (\bibinfo {year}
  {2011}{\natexlab{a}})}\BibitemShut {NoStop}%
\bibitem [{\citenamefont {Seth}\ \emph {et~al.}(1987)\citenamefont {Seth},
  \citenamefont {Artuso}, \citenamefont {Barlow}, \citenamefont {Iversen},
  \citenamefont {Kaletka}, \citenamefont {Nann}, \citenamefont {Parker},\ and\
  \citenamefont {Soundranayagam}}]{Seth87}%
  \BibitemOpen
  \bibfield  {author} {\bibinfo {author} {\bibfnamefont {K.~K.}\ \bibnamefont
  {Seth}}, \bibinfo {author} {\bibfnamefont {M.}~\bibnamefont {Artuso}},
  \bibinfo {author} {\bibfnamefont {D.}~\bibnamefont {Barlow}}, \bibinfo
  {author} {\bibfnamefont {S.}~\bibnamefont {Iversen}}, \bibinfo {author}
  {\bibfnamefont {M.}~\bibnamefont {Kaletka}}, \bibinfo {author} {\bibfnamefont
  {H.}~\bibnamefont {Nann}}, \bibinfo {author} {\bibfnamefont {B.}~\bibnamefont
  {Parker}}, \ and\ \bibinfo {author} {\bibfnamefont {R.}~\bibnamefont
  {Soundranayagam}},\ }\href {\doibase 10.1103/PhysRevLett.58.1930} {\bibfield
  {journal} {\bibinfo  {journal} {Phys. Rev. Lett.}\ }\textbf {\bibinfo
  {volume} {58}},\ \bibinfo {pages} {1930} (\bibinfo {year}
  {1987})}\BibitemShut {NoStop}%
\bibitem [{\citenamefont {Bohlen}\ \emph {et~al.}(1988)\citenamefont {Bohlen},
  \citenamefont {Gebauer}, \citenamefont {Kolbert}, \citenamefont {von
  Oertzen}, \citenamefont {Stiliaris}, \citenamefont {Wilpert},\ and\
  \citenamefont {Wilpert}}]{Bohlen88}%
  \BibitemOpen
  \bibfield  {author} {\bibinfo {author} {\bibfnamefont {H.~G.}\ \bibnamefont
  {Bohlen}}, \bibinfo {author} {\bibfnamefont {B.}~\bibnamefont {Gebauer}},
  \bibinfo {author} {\bibfnamefont {D.}~\bibnamefont {Kolbert}}, \bibinfo
  {author} {\bibfnamefont {W.}~\bibnamefont {von Oertzen}}, \bibinfo {author}
  {\bibfnamefont {E.}~\bibnamefont {Stiliaris}}, \bibinfo {author}
  {\bibfnamefont {M.}~\bibnamefont {Wilpert}}, \ and\ \bibinfo {author}
  {\bibfnamefont {T.}~\bibnamefont {Wilpert}},\ }\href
  {http://dx.doi.org/10.1007/BF01293402} {\bibfield  {journal} {\bibinfo
  {journal} {Zeitschrift für Physik A Atomic Nuclei}\ }\textbf {\bibinfo
  {volume} {330}},\ \bibinfo {pages} {227} (\bibinfo {year} {1988})},\ \bibinfo
  {note} {10.1007/BF01293402}\BibitemShut {NoStop}%
\bibitem [{\citenamefont {von Oertzen}\ \emph {et~al.}(1995)\citenamefont {von
  Oertzen}, \citenamefont {Bohlen}, \citenamefont {Gebauer}, \citenamefont {von
  Lucke-Petsch}, \citenamefont {Ostrowski}, \citenamefont {Seyfert},
  \citenamefont {Stolla}, \citenamefont {Wilpert}, \citenamefont {Wilpert},
  \citenamefont {Alexandrov}, \citenamefont {Korsheninnikov}, \citenamefont
  {Mukha}, \citenamefont {Ogloblin}, \citenamefont {Kalpakchieva},
  \citenamefont {Penionzhkevich}, \citenamefont {Piskor}, \citenamefont
  {Grimes},\ and\ \citenamefont {Massey}}]{VONOERTZEN1995c129}%
  \BibitemOpen
  \bibfield  {author} {\bibinfo {author} {\bibfnamefont {W.}~\bibnamefont {von
  Oertzen}}, \bibinfo {author} {\bibfnamefont {H.}~\bibnamefont {Bohlen}},
  \bibinfo {author} {\bibfnamefont {B.}~\bibnamefont {Gebauer}}, \bibinfo
  {author} {\bibfnamefont {M.}~\bibnamefont {von Lucke-Petsch}}, \bibinfo
  {author} {\bibfnamefont {A.}~\bibnamefont {Ostrowski}}, \bibinfo {author}
  {\bibfnamefont {C.}~\bibnamefont {Seyfert}}, \bibinfo {author} {\bibfnamefont
  {T.}~\bibnamefont {Stolla}}, \bibinfo {author} {\bibfnamefont
  {M.}~\bibnamefont {Wilpert}}, \bibinfo {author} {\bibfnamefont
  {T.}~\bibnamefont {Wilpert}}, \bibinfo {author} {\bibfnamefont
  {D.}~\bibnamefont {Alexandrov}}, \bibinfo {author} {\bibfnamefont
  {A.}~\bibnamefont {Korsheninnikov}}, \bibinfo {author} {\bibfnamefont
  {I.}~\bibnamefont {Mukha}}, \bibinfo {author} {\bibfnamefont
  {A.}~\bibnamefont {Ogloblin}}, \bibinfo {author} {\bibfnamefont
  {R.}~\bibnamefont {Kalpakchieva}}, \bibinfo {author} {\bibfnamefont
  {Y.}~\bibnamefont {Penionzhkevich}}, \bibinfo {author} {\bibfnamefont
  {S.}~\bibnamefont {Piskor}}, \bibinfo {author} {\bibfnamefont
  {S.}~\bibnamefont {Grimes}}, \ and\ \bibinfo {author} {\bibfnamefont
  {T.}~\bibnamefont {Massey}},\ }\href {\doibase
  https://doi.org/10.1016/0375-9474(95)00111-D} {\bibfield  {journal} {\bibinfo
   {journal} {Nuclear Physics A}\ }\textbf {\bibinfo {volume} {588}},\ \bibinfo
  {pages} {c129 } (\bibinfo {year} {1995})},\ \bibinfo {note} {proceedings of
  the Fifth International Symposium on Physics of Unstable Nuclei}\BibitemShut
  {NoStop}%
\bibitem [{\citenamefont {Bohlen}\ \emph {et~al.}(1999)\citenamefont {Bohlen},
  \citenamefont {Blazevi$\breve{c}$}, \citenamefont {Gebauer}, \citenamefont
  {Oertzen}, \citenamefont {Thummerer}, \citenamefont {Kalpakchieva},
  \citenamefont {Grimes},\ and\ \citenamefont {Massey}}]{Bohlen99}%
  \BibitemOpen
  \bibfield  {author} {\bibinfo {author} {\bibfnamefont {H.}~\bibnamefont
  {Bohlen}}, \bibinfo {author} {\bibfnamefont {A.}~\bibnamefont
  {Blazevi$\breve{c}$}}, \bibinfo {author} {\bibfnamefont {B.}~\bibnamefont
  {Gebauer}}, \bibinfo {author} {\bibfnamefont {W.~V.}\ \bibnamefont
  {Oertzen}}, \bibinfo {author} {\bibfnamefont {S.}~\bibnamefont {Thummerer}},
  \bibinfo {author} {\bibfnamefont {R.}~\bibnamefont {Kalpakchieva}}, \bibinfo
  {author} {\bibfnamefont {S.}~\bibnamefont {Grimes}}, \ and\ \bibinfo {author}
  {\bibfnamefont {T.}~\bibnamefont {Massey}},\ }\href {\doibase
  https://doi.org/10.1016/S0146-6410(99)00056-3} {\bibfield  {journal}
  {\bibinfo  {journal} {Progress in Particle and Nuclear Physics}\ }\textbf
  {\bibinfo {volume} {42}},\ \bibinfo {pages} {17 } (\bibinfo {year} {1999})},\
  \bibinfo {note} {{H}eavy Ion Collisions from Nuclear to Quark
  Matter}\BibitemShut {NoStop}%
\bibitem [{\citenamefont {Rogachev}\ \emph {et~al.}(2003)\citenamefont
  {Rogachev}, \citenamefont {Goldberg}, \citenamefont {Kolata}, \citenamefont
  {Chubarian}, \citenamefont {Aleksandrov}, \citenamefont {Fomichev},
  \citenamefont {Golovkov}, \citenamefont {Oganessian}, \citenamefont {Rodin},
  \citenamefont {Skorodumov}, \citenamefont {Slepnev}, \citenamefont
  {Ter-Akopian}, \citenamefont {Trzaska},\ and\ \citenamefont
  {Wolski}}]{PhysRevC.67.041603}%
  \BibitemOpen
  \bibfield  {author} {\bibinfo {author} {\bibfnamefont {G.~V.}\ \bibnamefont
  {Rogachev}}, \bibinfo {author} {\bibfnamefont {V.~Z.}\ \bibnamefont
  {Goldberg}}, \bibinfo {author} {\bibfnamefont {J.~J.}\ \bibnamefont
  {Kolata}}, \bibinfo {author} {\bibfnamefont {G.}~\bibnamefont {Chubarian}},
  \bibinfo {author} {\bibfnamefont {D.}~\bibnamefont {Aleksandrov}}, \bibinfo
  {author} {\bibfnamefont {A.}~\bibnamefont {Fomichev}}, \bibinfo {author}
  {\bibfnamefont {M.~S.}\ \bibnamefont {Golovkov}}, \bibinfo {author}
  {\bibfnamefont {Y.~T.}\ \bibnamefont {Oganessian}}, \bibinfo {author}
  {\bibfnamefont {A.}~\bibnamefont {Rodin}}, \bibinfo {author} {\bibfnamefont
  {B.}~\bibnamefont {Skorodumov}}, \bibinfo {author} {\bibfnamefont {R.~S.}\
  \bibnamefont {Slepnev}}, \bibinfo {author} {\bibfnamefont {G.}~\bibnamefont
  {Ter-Akopian}}, \bibinfo {author} {\bibfnamefont {W.~H.}\ \bibnamefont
  {Trzaska}}, \ and\ \bibinfo {author} {\bibfnamefont {R.}~\bibnamefont
  {Wolski}},\ }\href {\doibase 10.1103/PhysRevC.67.041603} {\bibfield
  {journal} {\bibinfo  {journal} {Phys. Rev. C}\ }\textbf {\bibinfo {volume}
  {67}},\ \bibinfo {pages} {041603} (\bibinfo {year} {2003})}\BibitemShut
  {NoStop}%
\bibitem [{\citenamefont {Fortier}\ \emph {et~al.}(2007)\citenamefont
  {Fortier}, \citenamefont {Tryggestad}, \citenamefont {Rich}, \citenamefont
  {Beaumel}, \citenamefont {Becheva}, \citenamefont {Blumenfeld}, \citenamefont
  {Delaunay}, \citenamefont {Drouart}, \citenamefont {Fomichev}, \citenamefont
  {Frascaria}, \citenamefont {Gales}, \citenamefont {Gaudefroy}, \citenamefont
  {Gillibert}, \citenamefont {Guillot}, \citenamefont {Hammache}, \citenamefont
  {Kemper}, \citenamefont {Khan}, \citenamefont {Lapoux}, \citenamefont {Lima},
  \citenamefont {Nalpas}, \citenamefont {Obertelli}, \citenamefont {Pollacco},
  \citenamefont {Skaza}, \citenamefont {Pramanik}, \citenamefont
  {Roussel-Chomaz}, \citenamefont {Santonocito}, \citenamefont {Scarpaci},
  \citenamefont {Sorlin}, \citenamefont {Stepantsov}, \citenamefont {Akopian},\
  and\ \citenamefont {Wolski}}]{fortier2007}%
  \BibitemOpen
  \bibfield  {author} {\bibinfo {author} {\bibfnamefont {S.}~\bibnamefont
  {Fortier}}, \bibinfo {author} {\bibfnamefont {E.}~\bibnamefont {Tryggestad}},
  \bibinfo {author} {\bibfnamefont {E.}~\bibnamefont {Rich}}, \bibinfo {author}
  {\bibfnamefont {D.}~\bibnamefont {Beaumel}}, \bibinfo {author} {\bibfnamefont
  {E.}~\bibnamefont {Becheva}}, \bibinfo {author} {\bibfnamefont
  {Y.}~\bibnamefont {Blumenfeld}}, \bibinfo {author} {\bibfnamefont
  {F.}~\bibnamefont {Delaunay}}, \bibinfo {author} {\bibfnamefont
  {A.}~\bibnamefont {Drouart}}, \bibinfo {author} {\bibfnamefont
  {A.}~\bibnamefont {Fomichev}}, \bibinfo {author} {\bibfnamefont
  {N.}~\bibnamefont {Frascaria}}, \bibinfo {author} {\bibfnamefont
  {S.}~\bibnamefont {Gales}}, \bibinfo {author} {\bibfnamefont
  {L.}~\bibnamefont {Gaudefroy}}, \bibinfo {author} {\bibfnamefont
  {A.}~\bibnamefont {Gillibert}}, \bibinfo {author} {\bibfnamefont
  {J.}~\bibnamefont {Guillot}}, \bibinfo {author} {\bibfnamefont
  {F.}~\bibnamefont {Hammache}}, \bibinfo {author} {\bibfnamefont {K.~W.}\
  \bibnamefont {Kemper}}, \bibinfo {author} {\bibfnamefont {E.}~\bibnamefont
  {Khan}}, \bibinfo {author} {\bibfnamefont {V.}~\bibnamefont {Lapoux}},
  \bibinfo {author} {\bibfnamefont {V.}~\bibnamefont {Lima}}, \bibinfo {author}
  {\bibfnamefont {L.}~\bibnamefont {Nalpas}}, \bibinfo {author} {\bibfnamefont
  {A.}~\bibnamefont {Obertelli}}, \bibinfo {author} {\bibfnamefont {E.~C.}\
  \bibnamefont {Pollacco}}, \bibinfo {author} {\bibfnamefont {F.}~\bibnamefont
  {Skaza}}, \bibinfo {author} {\bibfnamefont {U.~D.}\ \bibnamefont {Pramanik}},
  \bibinfo {author} {\bibfnamefont {P.}~\bibnamefont {Roussel-Chomaz}},
  \bibinfo {author} {\bibfnamefont {D.}~\bibnamefont {Santonocito}}, \bibinfo
  {author} {\bibfnamefont {J.~A.}\ \bibnamefont {Scarpaci}}, \bibinfo {author}
  {\bibfnamefont {O.}~\bibnamefont {Sorlin}}, \bibinfo {author} {\bibfnamefont
  {S.~V.}\ \bibnamefont {Stepantsov}}, \bibinfo {author} {\bibfnamefont
  {G.~M.~T.}\ \bibnamefont {Akopian}}, \ and\ \bibinfo {author} {\bibfnamefont
  {R.}~\bibnamefont {Wolski}},\ }\href {\doibase 10.1063/1.2746575} {\bibfield
  {journal} {\bibinfo  {journal} {AIP Conference Proceedings}\ }\textbf
  {\bibinfo {volume} {912}},\ \bibinfo {pages} {3} (\bibinfo {year} {2007})},\
  \Eprint
  {http://arxiv.org/abs/http://aip.scitation.org/doi/pdf/10.1063/1.2746575}
  {http://aip.scitation.org/doi/pdf/10.1063/1.2746575} \BibitemShut {NoStop}%
\bibitem [{\citenamefont {Al~Falou}(2007)}]{alfalou2007}%
  \BibitemOpen
  \bibfield  {author} {\bibinfo {author} {\bibfnamefont {H.}~\bibnamefont
  {Al~Falou}},\ }\emph {\bibinfo {title} {{Study of the unbound nuclei
  $^{7,9}$He and $^{10}$Li}}},\ \href
  {https://tel.archives-ouvertes.fr/tel-00212214} {\bibinfo {type} {Theses}},\
  \bibinfo  {school} {{Universit{\'e} de Caen}} (\bibinfo {year}
  {2007})\BibitemShut {NoStop}%
\bibitem [{\citenamefont {Falou}\ \emph
  {et~al.}(2011{\natexlab{b}})\citenamefont {Falou}, \citenamefont {Leprince},\
  and\ \citenamefont {Orr}}]{alfalou2011}%
  \BibitemOpen
  \bibfield  {author} {\bibinfo {author} {\bibfnamefont {H.~A.}\ \bibnamefont
  {Falou}}, \bibinfo {author} {\bibfnamefont {A.}~\bibnamefont {Leprince}}, \
  and\ \bibinfo {author} {\bibfnamefont {N.}~\bibnamefont {Orr}},\ }\href
  {http://stacks.iop.org/1742-6596/312/i=9/a=092012} {\bibfield  {journal}
  {\bibinfo  {journal} {Journal of Physics: Conference Series}\ }\textbf
  {\bibinfo {volume} {312}},\ \bibinfo {pages} {092012} (\bibinfo {year}
  {2011}{\natexlab{b}})}\BibitemShut {NoStop}%
\bibitem [{\citenamefont {Johansson}\ \emph {et~al.}(2010)\citenamefont
  {Johansson}, \citenamefont {Aksyutina}, \citenamefont {Aumann}, \citenamefont
  {Boretzky}, \citenamefont {Borge}, \citenamefont {Chatillon}, \citenamefont
  {Chulkov}, \citenamefont {Cortina-Gil}, \citenamefont {Pramanik},
  \citenamefont {Emling}, \citenamefont {Forss\'en}, \citenamefont {Fynbo},
  \citenamefont {Geissel}, \citenamefont {Ickert}, \citenamefont {Jonson},
  \citenamefont {Kulessa}, \citenamefont {Langer}, \citenamefont {Lantz},
  \citenamefont {LeBleis}, \citenamefont {Mahata}, \citenamefont {Meister},
  \citenamefont {M$\ddot{\mathrm{u}}$nzenberg}, \citenamefont {Nilsson},
  \citenamefont {Nyman}, \citenamefont {Palit}, \citenamefont {Paschalis},
  \citenamefont {Prokopowicz}, \citenamefont {Reifarth}, \citenamefont
  {Richter}, \citenamefont {Riisager}, \citenamefont {Schrieder}, \citenamefont
  {Simon}, \citenamefont {S$\ddot{\mathrm{u}}$mmerer}, \citenamefont
  {Tengblad}, \citenamefont {Weick},\ and\ \citenamefont
  {Zhukov}}]{JOHANSSON201015}%
  \BibitemOpen
  \bibfield  {author} {\bibinfo {author} {\bibfnamefont {H.}~\bibnamefont
  {Johansson}}, \bibinfo {author} {\bibfnamefont {Y.}~\bibnamefont
  {Aksyutina}}, \bibinfo {author} {\bibfnamefont {T.}~\bibnamefont {Aumann}},
  \bibinfo {author} {\bibfnamefont {K.}~\bibnamefont {Boretzky}}, \bibinfo
  {author} {\bibfnamefont {M.}~\bibnamefont {Borge}}, \bibinfo {author}
  {\bibfnamefont {A.}~\bibnamefont {Chatillon}}, \bibinfo {author}
  {\bibfnamefont {L.}~\bibnamefont {Chulkov}}, \bibinfo {author} {\bibfnamefont
  {D.}~\bibnamefont {Cortina-Gil}}, \bibinfo {author} {\bibfnamefont {U.~D.}\
  \bibnamefont {Pramanik}}, \bibinfo {author} {\bibfnamefont {H.}~\bibnamefont
  {Emling}}, \bibinfo {author} {\bibfnamefont {C.}~\bibnamefont {Forss\'en}},
  \bibinfo {author} {\bibfnamefont {H.}~\bibnamefont {Fynbo}}, \bibinfo
  {author} {\bibfnamefont {H.}~\bibnamefont {Geissel}}, \bibinfo {author}
  {\bibfnamefont {G.}~\bibnamefont {Ickert}}, \bibinfo {author} {\bibfnamefont
  {B.}~\bibnamefont {Jonson}}, \bibinfo {author} {\bibfnamefont
  {R.}~\bibnamefont {Kulessa}}, \bibinfo {author} {\bibfnamefont
  {C.}~\bibnamefont {Langer}}, \bibinfo {author} {\bibfnamefont
  {M.}~\bibnamefont {Lantz}}, \bibinfo {author} {\bibfnamefont
  {T.}~\bibnamefont {LeBleis}}, \bibinfo {author} {\bibfnamefont
  {K.}~\bibnamefont {Mahata}}, \bibinfo {author} {\bibfnamefont
  {M.}~\bibnamefont {Meister}}, \bibinfo {author} {\bibfnamefont
  {G.}~\bibnamefont {M$\ddot{\mathrm{u}}$nzenberg}}, \bibinfo {author}
  {\bibfnamefont {T.}~\bibnamefont {Nilsson}}, \bibinfo {author} {\bibfnamefont
  {G.}~\bibnamefont {Nyman}}, \bibinfo {author} {\bibfnamefont
  {R.}~\bibnamefont {Palit}}, \bibinfo {author} {\bibfnamefont
  {S.}~\bibnamefont {Paschalis}}, \bibinfo {author} {\bibfnamefont
  {W.}~\bibnamefont {Prokopowicz}}, \bibinfo {author} {\bibfnamefont
  {R.}~\bibnamefont {Reifarth}}, \bibinfo {author} {\bibfnamefont
  {A.}~\bibnamefont {Richter}}, \bibinfo {author} {\bibfnamefont
  {K.}~\bibnamefont {Riisager}}, \bibinfo {author} {\bibfnamefont
  {G.}~\bibnamefont {Schrieder}}, \bibinfo {author} {\bibfnamefont
  {H.}~\bibnamefont {Simon}}, \bibinfo {author} {\bibfnamefont
  {K.}~\bibnamefont {S$\ddot{\mathrm{u}}$mmerer}}, \bibinfo {author}
  {\bibfnamefont {O.}~\bibnamefont {Tengblad}}, \bibinfo {author}
  {\bibfnamefont {H.}~\bibnamefont {Weick}}, \ and\ \bibinfo {author}
  {\bibfnamefont {M.}~\bibnamefont {Zhukov}},\ }\href {\doibase
  https://doi.org/10.1016/j.nuclphysa.2010.04.006} {\bibfield  {journal}
  {\bibinfo  {journal} {Nuclear Physics A}\ }\textbf {\bibinfo {volume}
  {842}},\ \bibinfo {pages} {15 } (\bibinfo {year} {2010})}\BibitemShut
  {NoStop}%
\bibitem [{\citenamefont {Golovkov}\ \emph {et~al.}(2007)\citenamefont
  {Golovkov}, \citenamefont {Grigorenko}, \citenamefont {Fomichev},
  \citenamefont {Gorshkov}, \citenamefont {Gorshkov}, \citenamefont {Krupko},
  \citenamefont {Oganessian}, \citenamefont {Rodin}, \citenamefont {Sidorchuk},
  \citenamefont {Slepnev}, \citenamefont {Stepantsov}, \citenamefont
  {Ter-Akopian}, \citenamefont {Wolski}, \citenamefont {Korsheninnikov},
  \citenamefont {Nikolskii}, \citenamefont {Kuzmin}, \citenamefont {Novatskii},
  \citenamefont {Stepanov}, \citenamefont {Roussel-Chomaz},\ and\ \citenamefont
  {Mittig}}]{PhysRevC.76.021605}%
  \BibitemOpen
  \bibfield  {author} {\bibinfo {author} {\bibfnamefont {M.~S.}\ \bibnamefont
  {Golovkov}}, \bibinfo {author} {\bibfnamefont {L.~V.}\ \bibnamefont
  {Grigorenko}}, \bibinfo {author} {\bibfnamefont {A.~S.}\ \bibnamefont
  {Fomichev}}, \bibinfo {author} {\bibfnamefont {A.~V.}\ \bibnamefont
  {Gorshkov}}, \bibinfo {author} {\bibfnamefont {V.~A.}\ \bibnamefont
  {Gorshkov}}, \bibinfo {author} {\bibfnamefont {S.~A.}\ \bibnamefont
  {Krupko}}, \bibinfo {author} {\bibfnamefont {Y.~T.}\ \bibnamefont
  {Oganessian}}, \bibinfo {author} {\bibfnamefont {A.~M.}\ \bibnamefont
  {Rodin}}, \bibinfo {author} {\bibfnamefont {S.~I.}\ \bibnamefont
  {Sidorchuk}}, \bibinfo {author} {\bibfnamefont {R.~S.}\ \bibnamefont
  {Slepnev}}, \bibinfo {author} {\bibfnamefont {S.~V.}\ \bibnamefont
  {Stepantsov}}, \bibinfo {author} {\bibfnamefont {G.~M.}\ \bibnamefont
  {Ter-Akopian}}, \bibinfo {author} {\bibfnamefont {R.}~\bibnamefont {Wolski}},
  \bibinfo {author} {\bibfnamefont {A.~A.}\ \bibnamefont {Korsheninnikov}},
  \bibinfo {author} {\bibfnamefont {E.~Y.}\ \bibnamefont {Nikolskii}}, \bibinfo
  {author} {\bibfnamefont {V.~A.}\ \bibnamefont {Kuzmin}}, \bibinfo {author}
  {\bibfnamefont {B.~G.}\ \bibnamefont {Novatskii}}, \bibinfo {author}
  {\bibfnamefont {D.~N.}\ \bibnamefont {Stepanov}}, \bibinfo {author}
  {\bibfnamefont {P.}~\bibnamefont {Roussel-Chomaz}}, \ and\ \bibinfo {author}
  {\bibfnamefont {W.}~\bibnamefont {Mittig}},\ }\href {\doibase
  10.1103/PhysRevC.76.021605} {\bibfield  {journal} {\bibinfo  {journal} {Phys.
  Rev. C}\ }\textbf {\bibinfo {volume} {76}},\ \bibinfo {pages} {021605}
  (\bibinfo {year} {2007})}\BibitemShut {NoStop}%
\bibitem [{\citenamefont {Chen}\ \emph {et~al.}(2001)\citenamefont {Chen},
  \citenamefont {Blank}, \citenamefont {Brown}, \citenamefont {Chartier},
  \citenamefont {Galonsky}, \citenamefont {Hansen},\ and\ \citenamefont
  {Thoennessen}}]{chen01}%
  \BibitemOpen
  \bibfield  {author} {\bibinfo {author} {\bibfnamefont {L.}~\bibnamefont
  {Chen}}, \bibinfo {author} {\bibfnamefont {B.}~\bibnamefont {Blank}},
  \bibinfo {author} {\bibfnamefont {B.}~\bibnamefont {Brown}}, \bibinfo
  {author} {\bibfnamefont {M.}~\bibnamefont {Chartier}}, \bibinfo {author}
  {\bibfnamefont {A.}~\bibnamefont {Galonsky}}, \bibinfo {author}
  {\bibfnamefont {P.}~\bibnamefont {Hansen}}, \ and\ \bibinfo {author}
  {\bibfnamefont {M.}~\bibnamefont {Thoennessen}},\ }\href {\doibase
  https://doi.org/10.1016/S0370-2693(01)00313-6} {\bibfield  {journal}
  {\bibinfo  {journal} {Physics Letters B}\ }\textbf {\bibinfo {volume}
  {505}},\ \bibinfo {pages} {21 } (\bibinfo {year} {2001})}\BibitemShut
  {NoStop}%
\bibitem [{\citenamefont {Al~Kalanee}\ \emph {et~al.}(2013)\citenamefont
  {Al~Kalanee}, \citenamefont {Gibelin}, \citenamefont {Roussel-Chomaz},
  \citenamefont {Keeley}, \citenamefont {Beaumel}, \citenamefont {Blumenfeld},
  \citenamefont {Fern\'andez-Dom\'{\i}nguez}, \citenamefont {Force},
  \citenamefont {Gaudefroy}, \citenamefont {Gillibert}, \citenamefont
  {Guillot}, \citenamefont {Iwasaki}, \citenamefont {Krupko}, \citenamefont
  {Lapoux}, \citenamefont {Mittig}, \citenamefont {Mougeot}, \citenamefont
  {Nalpas}, \citenamefont {Pollacco}, \citenamefont {Rusek}, \citenamefont
  {Roger}, \citenamefont {Savajols}, \citenamefont {de~S\'er\'eville},
  \citenamefont {Sidorchuk}, \citenamefont {Suzuki}, \citenamefont {Strojek},\
  and\ \citenamefont {Orr}}]{PhysRevC.88.034301}%
  \BibitemOpen
  \bibfield  {author} {\bibinfo {author} {\bibfnamefont {T.}~\bibnamefont
  {Al~Kalanee}}, \bibinfo {author} {\bibfnamefont {J.}~\bibnamefont {Gibelin}},
  \bibinfo {author} {\bibfnamefont {P.}~\bibnamefont {Roussel-Chomaz}},
  \bibinfo {author} {\bibfnamefont {N.}~\bibnamefont {Keeley}}, \bibinfo
  {author} {\bibfnamefont {D.}~\bibnamefont {Beaumel}}, \bibinfo {author}
  {\bibfnamefont {Y.}~\bibnamefont {Blumenfeld}}, \bibinfo {author}
  {\bibfnamefont {B.}~\bibnamefont {Fern\'andez-Dom\'{\i}nguez}}, \bibinfo
  {author} {\bibfnamefont {C.}~\bibnamefont {Force}}, \bibinfo {author}
  {\bibfnamefont {L.}~\bibnamefont {Gaudefroy}}, \bibinfo {author}
  {\bibfnamefont {A.}~\bibnamefont {Gillibert}}, \bibinfo {author}
  {\bibfnamefont {J.}~\bibnamefont {Guillot}}, \bibinfo {author} {\bibfnamefont
  {H.}~\bibnamefont {Iwasaki}}, \bibinfo {author} {\bibfnamefont
  {S.}~\bibnamefont {Krupko}}, \bibinfo {author} {\bibfnamefont
  {V.}~\bibnamefont {Lapoux}}, \bibinfo {author} {\bibfnamefont
  {W.}~\bibnamefont {Mittig}}, \bibinfo {author} {\bibfnamefont
  {X.}~\bibnamefont {Mougeot}}, \bibinfo {author} {\bibfnamefont
  {L.}~\bibnamefont {Nalpas}}, \bibinfo {author} {\bibfnamefont
  {E.}~\bibnamefont {Pollacco}}, \bibinfo {author} {\bibfnamefont
  {K.}~\bibnamefont {Rusek}}, \bibinfo {author} {\bibfnamefont
  {T.}~\bibnamefont {Roger}}, \bibinfo {author} {\bibfnamefont
  {H.}~\bibnamefont {Savajols}}, \bibinfo {author} {\bibfnamefont
  {N.}~\bibnamefont {de~S\'er\'eville}}, \bibinfo {author} {\bibfnamefont
  {S.}~\bibnamefont {Sidorchuk}}, \bibinfo {author} {\bibfnamefont
  {D.}~\bibnamefont {Suzuki}}, \bibinfo {author} {\bibfnamefont
  {I.}~\bibnamefont {Strojek}}, \ and\ \bibinfo {author} {\bibfnamefont
  {N.~A.}\ \bibnamefont {Orr}},\ }\href {\doibase 10.1103/PhysRevC.88.034301}
  {\bibfield  {journal} {\bibinfo  {journal} {Phys. Rev. C}\ }\textbf {\bibinfo
  {volume} {88}},\ \bibinfo {pages} {034301} (\bibinfo {year}
  {2013})}\BibitemShut {NoStop}%
\bibitem [{\citenamefont {Uberseder}\ \emph {et~al.}(2016)\citenamefont
  {Uberseder}, \citenamefont {Rogachev}, \citenamefont {Goldberg},
  \citenamefont {Koshchiy}, \citenamefont {Roeder}, \citenamefont {Alcorta},
  \citenamefont {Chubarian}, \citenamefont {Davids}, \citenamefont {Fu},
  \citenamefont {Hooker}, \citenamefont {Jayatissa}, \citenamefont
  {Melconian},\ and\ \citenamefont {Tribble}}]{UBERSEDER2016323}%
  \BibitemOpen
  \bibfield  {author} {\bibinfo {author} {\bibfnamefont {E.}~\bibnamefont
  {Uberseder}}, \bibinfo {author} {\bibfnamefont {G.}~\bibnamefont {Rogachev}},
  \bibinfo {author} {\bibfnamefont {V.}~\bibnamefont {Goldberg}}, \bibinfo
  {author} {\bibfnamefont {E.}~\bibnamefont {Koshchiy}}, \bibinfo {author}
  {\bibfnamefont {B.}~\bibnamefont {Roeder}}, \bibinfo {author} {\bibfnamefont
  {M.}~\bibnamefont {Alcorta}}, \bibinfo {author} {\bibfnamefont
  {G.}~\bibnamefont {Chubarian}}, \bibinfo {author} {\bibfnamefont
  {B.}~\bibnamefont {Davids}}, \bibinfo {author} {\bibfnamefont
  {C.}~\bibnamefont {Fu}}, \bibinfo {author} {\bibfnamefont {J.}~\bibnamefont
  {Hooker}}, \bibinfo {author} {\bibfnamefont {H.}~\bibnamefont {Jayatissa}},
  \bibinfo {author} {\bibfnamefont {D.}~\bibnamefont {Melconian}}, \ and\
  \bibinfo {author} {\bibfnamefont {R.}~\bibnamefont {Tribble}},\ }\href
  {\doibase https://doi.org/10.1016/j.physletb.2016.01.014} {\bibfield
  {journal} {\bibinfo  {journal} {Physics Letters B}\ }\textbf {\bibinfo
  {volume} {754}},\ \bibinfo {pages} {323 } (\bibinfo {year}
  {2016})}\BibitemShut {NoStop}%
\bibitem [{\citenamefont {Poppelier}\ \emph {et~al.}(1985)\citenamefont
  {Poppelier}, \citenamefont {Wood},\ and\ \citenamefont
  {Glaudemans}}]{POPPELIER1985120}%
  \BibitemOpen
  \bibfield  {author} {\bibinfo {author} {\bibfnamefont {N.}~\bibnamefont
  {Poppelier}}, \bibinfo {author} {\bibfnamefont {L.}~\bibnamefont {Wood}}, \
  and\ \bibinfo {author} {\bibfnamefont {P.}~\bibnamefont {Glaudemans}},\
  }\href {\doibase https://doi.org/10.1016/0370-2693(85)91529-1} {\bibfield
  {journal} {\bibinfo  {journal} {Physics Letters B}\ }\textbf {\bibinfo
  {volume} {157}},\ \bibinfo {pages} {120 } (\bibinfo {year}
  {1985})}\BibitemShut {NoStop}%
\bibitem [{\citenamefont {Warburton}\ and\ \citenamefont
  {Brown}(1992)}]{PhysRevC.46.923}%
  \BibitemOpen
  \bibfield  {author} {\bibinfo {author} {\bibfnamefont {E.~K.}\ \bibnamefont
  {Warburton}}\ and\ \bibinfo {author} {\bibfnamefont {B.~A.}\ \bibnamefont
  {Brown}},\ }\href {\doibase 10.1103/PhysRevC.46.923} {\bibfield  {journal}
  {\bibinfo  {journal} {Phys. Rev. C}\ }\textbf {\bibinfo {volume} {46}},\
  \bibinfo {pages} {923} (\bibinfo {year} {1992})}\BibitemShut {NoStop}%
\bibitem [{\citenamefont {Hisashi}\ and\ \citenamefont
  {Hiroyuki}(1993)}]{HISASHI199316}%
  \BibitemOpen
  \bibfield  {author} {\bibinfo {author} {\bibfnamefont {K.}~\bibnamefont
  {Hisashi}}\ and\ \bibinfo {author} {\bibfnamefont {S.}~\bibnamefont
  {Hiroyuki}},\ }\href {\doibase https://doi.org/10.1016/0375-9474(93)90300-M}
  {\bibfield  {journal} {\bibinfo  {journal} {Nuclear Physics A}\ }\textbf
  {\bibinfo {volume} {551}},\ \bibinfo {pages} {16 } (\bibinfo {year}
  {1993})}\BibitemShut {NoStop}%
\bibitem [{\citenamefont {Poppelier}\ \emph {et~al.}(1993)\citenamefont
  {Poppelier}, \citenamefont {Wolters},\ and\ \citenamefont
  {Glaudemans}}]{Poppelier1993}%
  \BibitemOpen
  \bibfield  {author} {\bibinfo {author} {\bibfnamefont {N.~A. F.~M.}\
  \bibnamefont {Poppelier}}, \bibinfo {author} {\bibfnamefont {A.~A.}\
  \bibnamefont {Wolters}}, \ and\ \bibinfo {author} {\bibfnamefont {P.~W.~M.}\
  \bibnamefont {Glaudemans}},\ }\href {\doibase 10.1007/BF01290776} {\bibfield
  {journal} {\bibinfo  {journal} {Zeitschrift f$\ddot{\mathrm{u}}$r Physik A
  Hadrons and Nuclei}\ }\textbf {\bibinfo {volume} {346}},\ \bibinfo {pages}
  {11} (\bibinfo {year} {1993})}\BibitemShut {NoStop}%
\bibitem [{\citenamefont {Ogloblin}(1995)}]{Ogloblin1995}%
  \BibitemOpen
  \bibfield  {author} {\bibinfo {author} {\bibfnamefont {A.~A.}\ \bibnamefont
  {Ogloblin}},\ }\href {\doibase 10.1007/BF01291136} {\bibfield  {journal}
  {\bibinfo  {journal} {Zeitschrift f$\ddot{\mathrm{u}}$r Physik A Hadrons and
  Nuclei}\ }\textbf {\bibinfo {volume} {351}},\ \bibinfo {pages} {355}
  (\bibinfo {year} {1995})}\BibitemShut {NoStop}%
\bibitem [{\citenamefont {Otsuka}\ \emph {et~al.}(2001)\citenamefont {Otsuka},
  \citenamefont {Fujimoto}, \citenamefont {Utsuno}, \citenamefont {Brown},
  \citenamefont {Honma},\ and\ \citenamefont
  {Mizusaki}}]{PhysRevLett.87.082502}%
  \BibitemOpen
  \bibfield  {author} {\bibinfo {author} {\bibfnamefont {T.}~\bibnamefont
  {Otsuka}}, \bibinfo {author} {\bibfnamefont {R.}~\bibnamefont {Fujimoto}},
  \bibinfo {author} {\bibfnamefont {Y.}~\bibnamefont {Utsuno}}, \bibinfo
  {author} {\bibfnamefont {B.~A.}\ \bibnamefont {Brown}}, \bibinfo {author}
  {\bibfnamefont {M.}~\bibnamefont {Honma}}, \ and\ \bibinfo {author}
  {\bibfnamefont {T.}~\bibnamefont {Mizusaki}},\ }\href {\doibase
  10.1103/PhysRevLett.87.082502} {\bibfield  {journal} {\bibinfo  {journal}
  {Phys. Rev. Lett.}\ }\textbf {\bibinfo {volume} {87}},\ \bibinfo {pages}
  {082502} (\bibinfo {year} {2001})}\BibitemShut {NoStop}%
\bibitem [{\citenamefont {Pieper}\ \emph {et~al.}(2002)\citenamefont {Pieper},
  \citenamefont {Varga},\ and\ \citenamefont {Wiringa}}]{PhysRevC.66.044310}%
  \BibitemOpen
  \bibfield  {author} {\bibinfo {author} {\bibfnamefont {S.~C.}\ \bibnamefont
  {Pieper}}, \bibinfo {author} {\bibfnamefont {K.}~\bibnamefont {Varga}}, \
  and\ \bibinfo {author} {\bibfnamefont {R.~B.}\ \bibnamefont {Wiringa}},\
  }\href {\doibase 10.1103/PhysRevC.66.044310} {\bibfield  {journal} {\bibinfo
  {journal} {Phys. Rev. C}\ }\textbf {\bibinfo {volume} {66}},\ \bibinfo
  {pages} {044310} (\bibinfo {year} {2002})}\BibitemShut {NoStop}%
\bibitem [{\citenamefont {Barker}(2004)}]{BARKER200442}%
  \BibitemOpen
  \bibfield  {author} {\bibinfo {author} {\bibfnamefont {F.}~\bibnamefont
  {Barker}},\ }\href {\doibase https://doi.org/10.1016/j.nuclphysa.2004.06.001}
  {\bibfield  {journal} {\bibinfo  {journal} {Nuclear Physics A}\ }\textbf
  {\bibinfo {volume} {741}},\ \bibinfo {pages} {42 } (\bibinfo {year}
  {2004})}\BibitemShut {NoStop}%
\bibitem [{\citenamefont {Volya}\ and\ \citenamefont
  {Zelevinsky}(2005)}]{PhysRevLett.94.052501}%
  \BibitemOpen
  \bibfield  {author} {\bibinfo {author} {\bibfnamefont {A.}~\bibnamefont
  {Volya}}\ and\ \bibinfo {author} {\bibfnamefont {V.}~\bibnamefont
  {Zelevinsky}},\ }\href {\doibase 10.1103/PhysRevLett.94.052501} {\bibfield
  {journal} {\bibinfo  {journal} {Phys. Rev. Lett.}\ }\textbf {\bibinfo
  {volume} {94}},\ \bibinfo {pages} {052501} (\bibinfo {year}
  {2005})}\BibitemShut {NoStop}%
\bibitem [{\citenamefont {Lisetskiy}\ \emph {et~al.}(2008)\citenamefont
  {Lisetskiy}, \citenamefont {Barrett}, \citenamefont {Kruse}, \citenamefont
  {Navr\'atil}, \citenamefont {Stetcu},\ and\ \citenamefont
  {Vary}}]{PhysRevC.78.044302}%
  \BibitemOpen
  \bibfield  {author} {\bibinfo {author} {\bibfnamefont {A.~F.}\ \bibnamefont
  {Lisetskiy}}, \bibinfo {author} {\bibfnamefont {B.~R.}\ \bibnamefont
  {Barrett}}, \bibinfo {author} {\bibfnamefont {M.~K.~G.}\ \bibnamefont
  {Kruse}}, \bibinfo {author} {\bibfnamefont {P.}~\bibnamefont {Navr\'atil}},
  \bibinfo {author} {\bibfnamefont {I.}~\bibnamefont {Stetcu}}, \ and\ \bibinfo
  {author} {\bibfnamefont {J.~P.}\ \bibnamefont {Vary}},\ }\href {\doibase
  10.1103/PhysRevC.78.044302} {\bibfield  {journal} {\bibinfo  {journal} {Phys.
  Rev. C}\ }\textbf {\bibinfo {volume} {78}},\ \bibinfo {pages} {044302}
  (\bibinfo {year} {2008})}\BibitemShut {NoStop}%
\bibitem [{\citenamefont {Nollett}(2012)}]{PhysRevC.86.044330}%
  \BibitemOpen
  \bibfield  {author} {\bibinfo {author} {\bibfnamefont {K.~M.}\ \bibnamefont
  {Nollett}},\ }\href {\doibase 10.1103/PhysRevC.86.044330} {\bibfield
  {journal} {\bibinfo  {journal} {Phys. Rev. C}\ }\textbf {\bibinfo {volume}
  {86}},\ \bibinfo {pages} {044330} (\bibinfo {year} {2012})}\BibitemShut
  {NoStop}%
\bibitem [{\citenamefont {Volya}\ and\ \citenamefont
  {Zelevinsky}(2014)}]{Volya2014}%
  \BibitemOpen
  \bibfield  {author} {\bibinfo {author} {\bibfnamefont {A.}~\bibnamefont
  {Volya}}\ and\ \bibinfo {author} {\bibfnamefont {V.}~\bibnamefont
  {Zelevinsky}},\ }\href {\doibase 10.1134/S1063778814070163} {\bibfield
  {journal} {\bibinfo  {journal} {Physics of Atomic Nuclei}\ }\textbf {\bibinfo
  {volume} {77}},\ \bibinfo {pages} {969} (\bibinfo {year} {2014})}\BibitemShut
  {NoStop}%
\bibitem [{\citenamefont {Fortune}(2015)}]{PhysRevC.91.034306}%
  \BibitemOpen
  \bibfield  {author} {\bibinfo {author} {\bibfnamefont {H.~T.}\ \bibnamefont
  {Fortune}},\ }\href {\doibase 10.1103/PhysRevC.91.034306} {\bibfield
  {journal} {\bibinfo  {journal} {Phys. Rev. C}\ }\textbf {\bibinfo {volume}
  {91}},\ \bibinfo {pages} {034306} (\bibinfo {year} {2015})}\BibitemShut
  {NoStop}%
\bibitem [{\citenamefont {{Jaganathen}}\ \emph {et~al.}(2017)\citenamefont
  {{Jaganathen}}, \citenamefont {{Betan}}, \citenamefont {{Michel}},
  \citenamefont {{Nazarewicz}},\ and\ \citenamefont
  {{Ploszajczak}}}]{2017arXiv171004727J}%
  \BibitemOpen
  \bibfield  {author} {\bibinfo {author} {\bibfnamefont {Y.}~\bibnamefont
  {{Jaganathen}}}, \bibinfo {author} {\bibfnamefont {R.~M.~I.}\ \bibnamefont
  {{Betan}}}, \bibinfo {author} {\bibfnamefont {N.}~\bibnamefont {{Michel}}},
  \bibinfo {author} {\bibfnamefont {W.}~\bibnamefont {{Nazarewicz}}}, \ and\
  \bibinfo {author} {\bibfnamefont {M.}~\bibnamefont {{Ploszajczak}}},\
  }\href@noop {} {\bibfield  {journal} {\bibinfo  {journal} {ArXiv e-prints}\ }
  (\bibinfo {year} {2017})},\ \Eprint {http://arxiv.org/abs/1710.04727}
  {arXiv:1710.04727 [nucl-th]} \BibitemShut {NoStop}%
\bibitem [{\citenamefont {Baroni}\ \emph
  {et~al.}(2013{\natexlab{a}})\citenamefont {Baroni}, \citenamefont
  {Navr\'atil},\ and\ \citenamefont {Quaglioni}}]{PhysRevLett.110.022505}%
  \BibitemOpen
  \bibfield  {author} {\bibinfo {author} {\bibfnamefont {S.}~\bibnamefont
  {Baroni}}, \bibinfo {author} {\bibfnamefont {P.}~\bibnamefont {Navr\'atil}},
  \ and\ \bibinfo {author} {\bibfnamefont {S.}~\bibnamefont {Quaglioni}},\
  }\href {\doibase 10.1103/PhysRevLett.110.022505} {\bibfield  {journal}
  {\bibinfo  {journal} {Phys. Rev. Lett.}\ }\textbf {\bibinfo {volume} {110}},\
  \bibinfo {pages} {022505} (\bibinfo {year} {2013}{\natexlab{a}})}\BibitemShut
  {NoStop}%
\bibitem [{\citenamefont {Baroni}\ \emph
  {et~al.}(2013{\natexlab{b}})\citenamefont {Baroni}, \citenamefont
  {Navr\'atil},\ and\ \citenamefont {Quaglioni}}]{PhysRevC.87.034326}%
  \BibitemOpen
  \bibfield  {author} {\bibinfo {author} {\bibfnamefont {S.}~\bibnamefont
  {Baroni}}, \bibinfo {author} {\bibfnamefont {P.}~\bibnamefont {Navr\'atil}},
  \ and\ \bibinfo {author} {\bibfnamefont {S.}~\bibnamefont {Quaglioni}},\
  }\href {\doibase 10.1103/PhysRevC.87.034326} {\bibfield  {journal} {\bibinfo
  {journal} {Phys. Rev. C}\ }\textbf {\bibinfo {volume} {87}},\ \bibinfo
  {pages} {034326} (\bibinfo {year} {2013}{\natexlab{b}})}\BibitemShut
  {NoStop}%
\bibitem [{\citenamefont {Navr\'atil}\ \emph {et~al.}(2016)\citenamefont
  {Navr\'atil}, \citenamefont {Quaglioni}, \citenamefont {Hupin}, \citenamefont
  {Romero-Redondo},\ and\ \citenamefont {Calci}}]{physcripnavratil}%
  \BibitemOpen
  \bibfield  {author} {\bibinfo {author} {\bibfnamefont {P.}~\bibnamefont
  {Navr\'atil}}, \bibinfo {author} {\bibfnamefont {S.}~\bibnamefont
  {Quaglioni}}, \bibinfo {author} {\bibfnamefont {G.}~\bibnamefont {Hupin}},
  \bibinfo {author} {\bibfnamefont {C.}~\bibnamefont {Romero-Redondo}}, \ and\
  \bibinfo {author} {\bibfnamefont {A.}~\bibnamefont {Calci}},\ }\href
  {http://stacks.iop.org/1402-4896/91/i=5/a=053002} {\bibfield  {journal}
  {\bibinfo  {journal} {Physica Scripta}\ }\textbf {\bibinfo {volume} {91}},\
  \bibinfo {pages} {053002} (\bibinfo {year} {2016})}\BibitemShut {NoStop}%
\bibitem [{\citenamefont {Weinberg}(1990)}]{Weinberg1990}%
  \BibitemOpen
  \bibfield  {author} {\bibinfo {author} {\bibfnamefont {S.}~\bibnamefont
  {Weinberg}},\ }\href {\doibase 10.1016/0370-2693(90)90938-3} {\bibfield
  {journal} {\bibinfo  {journal} {Phys. Lett. B}\ }\textbf {\bibinfo {volume}
  {251}},\ \bibinfo {pages} {288} (\bibinfo {year} {1990})}\BibitemShut
  {NoStop}%
\bibitem [{\citenamefont {Weinberg}(1991)}]{Weinberg1991}%
  \BibitemOpen
  \bibfield  {author} {\bibinfo {author} {\bibfnamefont {S.}~\bibnamefont
  {Weinberg}},\ }\href {\doibase 10.1016/0550-3213(91)90231-L} {\bibfield
  {journal} {\bibinfo  {journal} {Nucl. Phys. B}\ }\textbf {\bibinfo {volume}
  {363}},\ \bibinfo {pages} {3} (\bibinfo {year} {1991})}\BibitemShut {NoStop}%
\bibitem [{\citenamefont {Navr\'atil}\ \emph
  {et~al.}(2000{\natexlab{a}})\citenamefont {Navr\'atil}, \citenamefont
  {Vary},\ and\ \citenamefont {Barrett}}]{PhysRevLett.84.5728}%
  \BibitemOpen
  \bibfield  {author} {\bibinfo {author} {\bibfnamefont {P.}~\bibnamefont
  {Navr\'atil}}, \bibinfo {author} {\bibfnamefont {J.~P.}\ \bibnamefont
  {Vary}}, \ and\ \bibinfo {author} {\bibfnamefont {B.~R.}\ \bibnamefont
  {Barrett}},\ }\href {\doibase 10.1103/PhysRevLett.84.5728} {\bibfield
  {journal} {\bibinfo  {journal} {Phys. Rev. Lett.}\ }\textbf {\bibinfo
  {volume} {84}},\ \bibinfo {pages} {5728} (\bibinfo {year}
  {2000}{\natexlab{a}})}\BibitemShut {NoStop}%
\bibitem [{\citenamefont {Navr\'atil}\ \emph
  {et~al.}(2000{\natexlab{b}})\citenamefont {Navr\'atil}, \citenamefont
  {Vary},\ and\ \citenamefont {Barrett}}]{PhysRevC.62.054311}%
  \BibitemOpen
  \bibfield  {author} {\bibinfo {author} {\bibfnamefont {P.}~\bibnamefont
  {Navr\'atil}}, \bibinfo {author} {\bibfnamefont {J.~P.}\ \bibnamefont
  {Vary}}, \ and\ \bibinfo {author} {\bibfnamefont {B.~R.}\ \bibnamefont
  {Barrett}},\ }\href {\doibase 10.1103/PhysRevC.62.054311} {\bibfield
  {journal} {\bibinfo  {journal} {Phys. Rev. C}\ }\textbf {\bibinfo {volume}
  {62}},\ \bibinfo {pages} {054311} (\bibinfo {year}
  {2000}{\natexlab{b}})}\BibitemShut {NoStop}%
\bibitem [{\citenamefont {Barrett}\ \emph {et~al.}(2013)\citenamefont
  {Barrett}, \citenamefont {Navr\'atil},\ and\ \citenamefont
  {Vary}}]{BARRETT2013131}%
  \BibitemOpen
  \bibfield  {author} {\bibinfo {author} {\bibfnamefont {B.~R.}\ \bibnamefont
  {Barrett}}, \bibinfo {author} {\bibfnamefont {P.}~\bibnamefont {Navr\'atil}},
  \ and\ \bibinfo {author} {\bibfnamefont {J.~P.}\ \bibnamefont {Vary}},\
  }\href {\doibase https://doi.org/10.1016/j.ppnp.2012.10.003} {\bibfield
  {journal} {\bibinfo  {journal} {Progress in Particle and Nuclear Physics}\
  }\textbf {\bibinfo {volume} {69}},\ \bibinfo {pages} {131 } (\bibinfo {year}
  {2013})}\BibitemShut {NoStop}%
\bibitem [{\citenamefont {Tilley}\ \emph {et~al.}(2004)\citenamefont {Tilley},
  \citenamefont {Kelley}, \citenamefont {Godwin}, \citenamefont {Millener},
  \citenamefont {Purcell}, \citenamefont {Sheu},\ and\ \citenamefont
  {Weller}}]{TILLEY2004155}%
  \BibitemOpen
  \bibfield  {author} {\bibinfo {author} {\bibfnamefont {D.}~\bibnamefont
  {Tilley}}, \bibinfo {author} {\bibfnamefont {J.}~\bibnamefont {Kelley}},
  \bibinfo {author} {\bibfnamefont {J.}~\bibnamefont {Godwin}}, \bibinfo
  {author} {\bibfnamefont {D.}~\bibnamefont {Millener}}, \bibinfo {author}
  {\bibfnamefont {J.}~\bibnamefont {Purcell}}, \bibinfo {author} {\bibfnamefont
  {C.}~\bibnamefont {Sheu}}, \ and\ \bibinfo {author} {\bibfnamefont
  {H.}~\bibnamefont {Weller}},\ }\href {\doibase
  https://doi.org/10.1016/j.nuclphysa.2004.09.059} {\bibfield  {journal}
  {\bibinfo  {journal} {Nuclear Physics A}\ }\textbf {\bibinfo {volume}
  {745}},\ \bibinfo {pages} {155 } (\bibinfo {year} {2004})}\BibitemShut
  {NoStop}%
\bibitem [{\citenamefont {Wildermuth}\ and\ \citenamefont
  {Tang}(1977)}]{wildermuth1977unified}%
  \BibitemOpen
  \bibfield  {author} {\bibinfo {author} {\bibfnamefont {K.}~\bibnamefont
  {Wildermuth}}\ and\ \bibinfo {author} {\bibfnamefont {Y.}~\bibnamefont
  {Tang}},\ }\href {https://books.google.ca/books?id=Ti3wAAAAIAAJ} {\emph
  {\bibinfo {title} {A unified theory of the nucleus}}}\ (\bibinfo  {publisher}
  {Vieweg, Braunschweig},\ \bibinfo {year} {1977})\BibitemShut {NoStop}%
\bibitem [{\citenamefont {Tang}\ \emph {et~al.}(1978)\citenamefont {Tang},
  \citenamefont {LeMere},\ and\ \citenamefont {Thompsom}}]{TANG1978167}%
  \BibitemOpen
  \bibfield  {author} {\bibinfo {author} {\bibfnamefont {Y.}~\bibnamefont
  {Tang}}, \bibinfo {author} {\bibfnamefont {M.}~\bibnamefont {LeMere}}, \ and\
  \bibinfo {author} {\bibfnamefont {D.}~\bibnamefont {Thompsom}},\ }\href
  {\doibase https://doi.org/10.1016/0370-1573(78)90175-8} {\bibfield  {journal}
  {\bibinfo  {journal} {Physics Reports}\ }\textbf {\bibinfo {volume} {47}},\
  \bibinfo {pages} {167 } (\bibinfo {year} {1978})}\BibitemShut {NoStop}%
\bibitem [{\citenamefont {Fliessbach}\ and\ \citenamefont
  {Walliser}(1982)}]{FLIESSBACH198284}%
  \BibitemOpen
  \bibfield  {author} {\bibinfo {author} {\bibfnamefont {T.}~\bibnamefont
  {Fliessbach}}\ and\ \bibinfo {author} {\bibfnamefont {H.}~\bibnamefont
  {Walliser}},\ }\href {\doibase https://doi.org/10.1016/0375-9474(82)90322-0}
  {\bibfield  {journal} {\bibinfo  {journal} {Nuclear Physics A}\ }\textbf
  {\bibinfo {volume} {377}},\ \bibinfo {pages} {84 } (\bibinfo {year}
  {1982})}\BibitemShut {NoStop}%
\bibitem [{\citenamefont {Langanke}\ and\ \citenamefont
  {Friedrich}(1986)}]{langanke1986}%
  \BibitemOpen
  \bibfield  {author} {\bibinfo {author} {\bibfnamefont {K.}~\bibnamefont
  {Langanke}}\ and\ \bibinfo {author} {\bibfnamefont {H.}~\bibnamefont
  {Friedrich}},\ }\href@noop {} {\emph {\bibinfo {title} {Advances in Nuclear
  Physics}}},\ edited by J. W. Negele and E. Vogt\ (\bibinfo  {publisher}
  {Plenum, New York},\ \bibinfo {year} {1986})\BibitemShut {NoStop}%
\bibitem [{\citenamefont {Hofmann}\ and\ \citenamefont
  {Hale}(2008)}]{PhysRevC.77.044002}%
  \BibitemOpen
  \bibfield  {author} {\bibinfo {author} {\bibfnamefont {H.~M.}\ \bibnamefont
  {Hofmann}}\ and\ \bibinfo {author} {\bibfnamefont {G.~M.}\ \bibnamefont
  {Hale}},\ }\href {\doibase 10.1103/PhysRevC.77.044002} {\bibfield  {journal}
  {\bibinfo  {journal} {Phys. Rev. C}\ }\textbf {\bibinfo {volume} {77}},\
  \bibinfo {pages} {044002} (\bibinfo {year} {2008})}\BibitemShut {NoStop}%
\bibitem [{\citenamefont {Descouvemont}\ and\ \citenamefont
  {Baye}(2010)}]{Descouvemont2010}%
  \BibitemOpen
  \bibfield  {author} {\bibinfo {author} {\bibfnamefont {P.}~\bibnamefont
  {Descouvemont}}\ and\ \bibinfo {author} {\bibfnamefont {D.}~\bibnamefont
  {Baye}},\ }\href {\doibase 10.1088/0034-4885/73/3/036301} {\bibfield
  {journal} {\bibinfo  {journal} {Reports Prog. Phys.}\ }\textbf {\bibinfo
  {volume} {73}},\ \bibinfo {pages} {036301} (\bibinfo {year}
  {2010})}\BibitemShut {NoStop}%
\bibitem [{\citenamefont {Hesse}\ \emph {et~al.}(1998)\citenamefont {Hesse},
  \citenamefont {Sparenberg}, \citenamefont {{Van Raemdonck}},\ and\
  \citenamefont {Baye}}]{Hesse1998}%
  \BibitemOpen
  \bibfield  {author} {\bibinfo {author} {\bibfnamefont {M.}~\bibnamefont
  {Hesse}}, \bibinfo {author} {\bibfnamefont {J.-M.}\ \bibnamefont
  {Sparenberg}}, \bibinfo {author} {\bibfnamefont {F.}~\bibnamefont {{Van
  Raemdonck}}}, \ and\ \bibinfo {author} {\bibfnamefont {D.}~\bibnamefont
  {Baye}},\ }\href {\doibase 10.1016/S0375-9474(98)00435-7} {\bibfield
  {journal} {\bibinfo  {journal} {Nucl. Phys. A}\ }\textbf {\bibinfo {volume}
  {640}},\ \bibinfo {pages} {37} (\bibinfo {year} {1998})}\BibitemShut
  {NoStop}%
\bibitem [{\citenamefont {Hesse}\ \emph {et~al.}(2002)\citenamefont {Hesse},
  \citenamefont {Roland},\ and\ \citenamefont {Baye}}]{Hesse2002}%
  \BibitemOpen
  \bibfield  {author} {\bibinfo {author} {\bibfnamefont {M.}~\bibnamefont
  {Hesse}}, \bibinfo {author} {\bibfnamefont {J.}~\bibnamefont {Roland}}, \
  and\ \bibinfo {author} {\bibfnamefont {D.}~\bibnamefont {Baye}},\ }\href
  {\doibase 10.1016/S0375-9474(02)01040-0} {\bibfield  {journal} {\bibinfo
  {journal} {Nucl. Phys. A}\ }\textbf {\bibinfo {volume} {709}},\ \bibinfo
  {pages} {184} (\bibinfo {year} {2002})}\BibitemShut {NoStop}%
\bibitem [{\citenamefont {Ord\'o\~nez}\ \emph {et~al.}(1994)\citenamefont
  {Ord\'o\~nez}, \citenamefont {Ray},\ and\ \citenamefont {van
  Kolck}}]{OrRa94}%
  \BibitemOpen
  \bibfield  {author} {\bibinfo {author} {\bibfnamefont {C.}~\bibnamefont
  {Ord\'o\~nez}}, \bibinfo {author} {\bibfnamefont {L.}~\bibnamefont {Ray}}, \
  and\ \bibinfo {author} {\bibfnamefont {U.}~\bibnamefont {van Kolck}},\ }\href
  {\doibase 10.1103/PhysRevLett.72.1982} {\bibfield  {journal} {\bibinfo
  {journal} {Phys. Rev. Lett.}\ }\textbf {\bibinfo {volume} {72}},\ \bibinfo
  {pages} {1982} (\bibinfo {year} {1994})}\BibitemShut {NoStop}%
\bibitem [{\citenamefont {van Kolck}(1994)}]{VanKolck94}%
  \BibitemOpen
  \bibfield  {author} {\bibinfo {author} {\bibfnamefont {U.}~\bibnamefont {van
  Kolck}},\ }\href {\doibase 10.1103/PhysRevC.49.2932} {\bibfield  {journal}
  {\bibinfo  {journal} {Phys. Rev. C}\ }\textbf {\bibinfo {volume} {49}},\
  \bibinfo {pages} {2932} (\bibinfo {year} {1994})}\BibitemShut {NoStop}%
\bibitem [{\citenamefont {Epelbaum}\ \emph {et~al.}(2002)\citenamefont
  {Epelbaum}, \citenamefont {Nogga}, \citenamefont {Gl\"ockle}, \citenamefont
  {Kamada}, \citenamefont {Mei\ss{}ner},\ and\ \citenamefont
  {Wita\l{}a}}]{EpNo02}%
  \BibitemOpen
  \bibfield  {author} {\bibinfo {author} {\bibfnamefont {E.}~\bibnamefont
  {Epelbaum}}, \bibinfo {author} {\bibfnamefont {A.}~\bibnamefont {Nogga}},
  \bibinfo {author} {\bibfnamefont {W.}~\bibnamefont {Gl\"ockle}}, \bibinfo
  {author} {\bibfnamefont {H.}~\bibnamefont {Kamada}}, \bibinfo {author}
  {\bibfnamefont {U.-G.}\ \bibnamefont {Mei\ss{}ner}}, \ and\ \bibinfo {author}
  {\bibfnamefont {H.}~\bibnamefont {Wita\l{}a}},\ }\href@noop {} {\bibfield
  {journal} {\bibinfo  {journal} {Phys. Rev. C}\ }\textbf {\bibinfo {volume}
  {66}},\ \bibinfo {pages} {064001} (\bibinfo {year} {2002})}\BibitemShut
  {NoStop}%
\bibitem [{\citenamefont {Epelbaum}(2006)}]{Epelbaum06}%
  \BibitemOpen
  \bibfield  {author} {\bibinfo {author} {\bibfnamefont {E.}~\bibnamefont
  {Epelbaum}},\ }\href {\doibase
  http://dx.doi.org/10.1016/j.physletb.2006.06.046} {\bibfield  {journal}
  {\bibinfo  {journal} {Physics Letters B}\ }\textbf {\bibinfo {volume}
  {639}},\ \bibinfo {pages} {456 } (\bibinfo {year} {2006})}\BibitemShut
  {NoStop}%
\bibitem [{\citenamefont {Wegner}(1994)}]{Wegner1994}%
  \BibitemOpen
  \bibfield  {author} {\bibinfo {author} {\bibfnamefont {F.}~\bibnamefont
  {Wegner}},\ }\href {\doibase 10.1002/andp.19945060203} {\bibfield  {journal}
  {\bibinfo  {journal} {Ann. Phys.}\ }\textbf {\bibinfo {volume} {506}},\
  \bibinfo {pages} {77} (\bibinfo {year} {1994})}\BibitemShut {NoStop}%
\bibitem [{\citenamefont {Bogner}\ \emph {et~al.}(2007)\citenamefont {Bogner},
  \citenamefont {Furnstahl},\ and\ \citenamefont {Perry}}]{Bogner2007}%
  \BibitemOpen
  \bibfield  {author} {\bibinfo {author} {\bibfnamefont {S.~K.}\ \bibnamefont
  {Bogner}}, \bibinfo {author} {\bibfnamefont {R.~J.}\ \bibnamefont
  {Furnstahl}}, \ and\ \bibinfo {author} {\bibfnamefont {R.~J.}\ \bibnamefont
  {Perry}},\ }\href {\doibase 10.1103/PhysRevC.75.061001} {\bibfield  {journal}
  {\bibinfo  {journal} {Phys. Rev. C}\ }\textbf {\bibinfo {volume} {75}},\
  \bibinfo {pages} {061001} (\bibinfo {year} {2007})}\BibitemShut {NoStop}%
\bibitem [{\citenamefont {Roth}\ \emph {et~al.}(2008)\citenamefont {Roth},
  \citenamefont {Reinhardt},\ and\ \citenamefont
  {Hergert}}]{PhysRevC.77.064003}%
  \BibitemOpen
  \bibfield  {author} {\bibinfo {author} {\bibfnamefont {R.}~\bibnamefont
  {Roth}}, \bibinfo {author} {\bibfnamefont {S.}~\bibnamefont {Reinhardt}}, \
  and\ \bibinfo {author} {\bibfnamefont {H.}~\bibnamefont {Hergert}},\ }\href
  {\doibase 10.1103/PhysRevC.77.064003} {\bibfield  {journal} {\bibinfo
  {journal} {Phys. Rev. C}\ }\textbf {\bibinfo {volume} {77}},\ \bibinfo
  {pages} {064003} (\bibinfo {year} {2008})}\BibitemShut {NoStop}%
\bibitem [{\citenamefont {Bogner}\ \emph {et~al.}(2010)\citenamefont {Bogner},
  \citenamefont {Furnstahl},\ and\ \citenamefont {Schwenk}}]{Bogner201094}%
  \BibitemOpen
  \bibfield  {author} {\bibinfo {author} {\bibfnamefont {S.}~\bibnamefont
  {Bogner}}, \bibinfo {author} {\bibfnamefont {R.}~\bibnamefont {Furnstahl}}, \
  and\ \bibinfo {author} {\bibfnamefont {A.}~\bibnamefont {Schwenk}},\ }\href
  {\doibase http://dx.doi.org/10.1016/j.ppnp.2010.03.001} {\bibfield  {journal}
  {\bibinfo  {journal} {Progress in Particle and Nuclear Physics}\ }\textbf
  {\bibinfo {volume} {65}},\ \bibinfo {pages} {94 } (\bibinfo {year}
  {2010})}\BibitemShut {NoStop}%
\bibitem [{\citenamefont {Jurgenson}\ \emph {et~al.}(2009)\citenamefont
  {Jurgenson}, \citenamefont {Navr\'{a}til},\ and\ \citenamefont
  {Furnstahl}}]{Jurgenson2009}%
  \BibitemOpen
  \bibfield  {author} {\bibinfo {author} {\bibfnamefont {E.~D.}\ \bibnamefont
  {Jurgenson}}, \bibinfo {author} {\bibfnamefont {P.}~\bibnamefont
  {Navr\'{a}til}}, \ and\ \bibinfo {author} {\bibfnamefont {R.~J.}\
  \bibnamefont {Furnstahl}},\ }\href {\doibase 10.1103/PhysRevLett.103.082501}
  {\bibfield  {journal} {\bibinfo  {journal} {Phys. Rev. Lett.}\ }\textbf
  {\bibinfo {volume} {103}},\ \bibinfo {pages} {082501} (\bibinfo {year}
  {2009})}\BibitemShut {NoStop}%
\bibitem [{\citenamefont {Entem}\ and\ \citenamefont
  {Machleidt}(2003)}]{Entem2003}%
  \BibitemOpen
  \bibfield  {author} {\bibinfo {author} {\bibfnamefont {D.~R.}\ \bibnamefont
  {Entem}}\ and\ \bibinfo {author} {\bibfnamefont {R.}~\bibnamefont
  {Machleidt}},\ }\href {\doibase 10.1103/PhysRevC.68.041001} {\bibfield
  {journal} {\bibinfo  {journal} {Phys. Rev. C}\ }\textbf {\bibinfo {volume}
  {68}},\ \bibinfo {pages} {041001} (\bibinfo {year} {2003})}\BibitemShut
  {NoStop}%
\bibitem [{\citenamefont {Navr\'{a}til}(2007)}]{Navratil2007}%
  \BibitemOpen
  \bibfield  {author} {\bibinfo {author} {\bibfnamefont {P.}~\bibnamefont
  {Navr\'{a}til}},\ }\href {\doibase 10.1007/s00601-007-0193-3} {\bibfield
  {journal} {\bibinfo  {journal} {Few-Body Syst.}\ }\textbf {\bibinfo {volume}
  {41}},\ \bibinfo {pages} {117} (\bibinfo {year} {2007})}\BibitemShut
  {NoStop}%
\bibitem [{\citenamefont {Ekstr\"om}\ \emph {et~al.}(2015)\citenamefont
  {Ekstr\"om}, \citenamefont {Jansen}, \citenamefont {Wendt}, \citenamefont
  {Hagen}, \citenamefont {Papenbrock}, \citenamefont {Carlsson}, \citenamefont
  {Forss\'en}, \citenamefont {Hjorth-Jensen}, \citenamefont {Navr\'atil},\ and\
  \citenamefont {Nazarewicz}}]{EkJa15}%
  \BibitemOpen
  \bibfield  {author} {\bibinfo {author} {\bibfnamefont {A.}~\bibnamefont
  {Ekstr\"om}}, \bibinfo {author} {\bibfnamefont {G.~R.}\ \bibnamefont
  {Jansen}}, \bibinfo {author} {\bibfnamefont {K.~A.}\ \bibnamefont {Wendt}},
  \bibinfo {author} {\bibfnamefont {G.}~\bibnamefont {Hagen}}, \bibinfo
  {author} {\bibfnamefont {T.}~\bibnamefont {Papenbrock}}, \bibinfo {author}
  {\bibfnamefont {B.~D.}\ \bibnamefont {Carlsson}}, \bibinfo {author}
  {\bibfnamefont {C.}~\bibnamefont {Forss\'en}}, \bibinfo {author}
  {\bibfnamefont {M.}~\bibnamefont {Hjorth-Jensen}}, \bibinfo {author}
  {\bibfnamefont {P.}~\bibnamefont {Navr\'atil}}, \ and\ \bibinfo {author}
  {\bibfnamefont {W.}~\bibnamefont {Nazarewicz}},\ }\href {\doibase
  10.1103/PhysRevC.91.051301} {\bibfield  {journal} {\bibinfo  {journal} {Phys.
  Rev. C}\ }\textbf {\bibinfo {volume} {91}},\ \bibinfo {pages} {051301}
  (\bibinfo {year} {2015})}\BibitemShut {NoStop}%
\bibitem [{\citenamefont {Entem}\ \emph {et~al.}(2015)\citenamefont {Entem},
  \citenamefont {Kaiser}, \citenamefont {Machleidt},\ and\ \citenamefont
  {Nosyk}}]{PhysRevC.91.014002}%
  \BibitemOpen
  \bibfield  {author} {\bibinfo {author} {\bibfnamefont {D.~R.}\ \bibnamefont
  {Entem}}, \bibinfo {author} {\bibfnamefont {N.}~\bibnamefont {Kaiser}},
  \bibinfo {author} {\bibfnamefont {R.}~\bibnamefont {Machleidt}}, \ and\
  \bibinfo {author} {\bibfnamefont {Y.}~\bibnamefont {Nosyk}},\ }\href
  {\doibase 10.1103/PhysRevC.91.014002} {\bibfield  {journal} {\bibinfo
  {journal} {Phys. Rev. C}\ }\textbf {\bibinfo {volume} {91}},\ \bibinfo
  {pages} {014002} (\bibinfo {year} {2015})}\BibitemShut {NoStop}%
\bibitem [{\citenamefont {Entem}\ \emph {et~al.}(2017)\citenamefont {Entem},
  \citenamefont {Machleidt},\ and\ \citenamefont {Nosyk}}]{PhysRevC.96.024004}%
  \BibitemOpen
  \bibfield  {author} {\bibinfo {author} {\bibfnamefont {D.~R.}\ \bibnamefont
  {Entem}}, \bibinfo {author} {\bibfnamefont {R.}~\bibnamefont {Machleidt}}, \
  and\ \bibinfo {author} {\bibfnamefont {Y.}~\bibnamefont {Nosyk}},\ }\href
  {\doibase 10.1103/PhysRevC.96.024004} {\bibfield  {journal} {\bibinfo
  {journal} {Phys. Rev. C}\ }\textbf {\bibinfo {volume} {96}},\ \bibinfo
  {pages} {024004} (\bibinfo {year} {2017})}\BibitemShut {NoStop}%
\bibitem [{\citenamefont {Romero-Redondo}\ \emph {et~al.}(2016)\citenamefont
  {Romero-Redondo}, \citenamefont {Quaglioni}, \citenamefont {Navr\'atil},\
  and\ \citenamefont {Hupin}}]{PhysRevLett.117.222501}%
  \BibitemOpen
  \bibfield  {author} {\bibinfo {author} {\bibfnamefont {C.}~\bibnamefont
  {Romero-Redondo}}, \bibinfo {author} {\bibfnamefont {S.}~\bibnamefont
  {Quaglioni}}, \bibinfo {author} {\bibfnamefont {P.}~\bibnamefont
  {Navr\'atil}}, \ and\ \bibinfo {author} {\bibfnamefont {G.}~\bibnamefont
  {Hupin}},\ }\href {\doibase 10.1103/PhysRevLett.117.222501} {\bibfield
  {journal} {\bibinfo  {journal} {Phys. Rev. Lett.}\ }\textbf {\bibinfo
  {volume} {117}},\ \bibinfo {pages} {222501} (\bibinfo {year}
  {2016})}\BibitemShut {NoStop}%
\bibitem [{\citenamefont {Quaglioni}\ \emph {et~al.}(2017)\citenamefont
  {Quaglioni}, \citenamefont {Romero-Redondo}, \citenamefont {Navratil},\ and\
  \citenamefont {Hupin}}]{Quaglioni:2017vpa}%
  \BibitemOpen
  \bibfield  {author} {\bibinfo {author} {\bibfnamefont {S.}~\bibnamefont
  {Quaglioni}}, \bibinfo {author} {\bibfnamefont {C.}~\bibnamefont
  {Romero-Redondo}}, \bibinfo {author} {\bibfnamefont {P.}~\bibnamefont
  {Navratil}}, \ and\ \bibinfo {author} {\bibfnamefont {G.}~\bibnamefont
  {Hupin}},\ }\href@noop {} {\  (\bibinfo {year} {2017})},\ \Eprint
  {http://arxiv.org/abs/1710.07326} {arXiv:1710.07326 [nucl-th]} \BibitemShut
  {NoStop}%
%%CITATION = ARXIV:1710.07326;%%
\bibitem [{\citenamefont {Dohet-Eraly}\ \emph {et~al.}(2016)\citenamefont
  {Dohet-Eraly}, \citenamefont {Navr\'{a}til}, \citenamefont {Quaglioni},
  \citenamefont {Horiuchi}, \citenamefont {Hupin},\ and\ \citenamefont
  {Raimondi}}]{DOHETERALY2016430}%
  \BibitemOpen
  \bibfield  {author} {\bibinfo {author} {\bibfnamefont {J.}~\bibnamefont
  {Dohet-Eraly}}, \bibinfo {author} {\bibfnamefont {P.}~\bibnamefont
  {Navr\'{a}til}}, \bibinfo {author} {\bibfnamefont {S.}~\bibnamefont
  {Quaglioni}}, \bibinfo {author} {\bibfnamefont {W.}~\bibnamefont {Horiuchi}},
  \bibinfo {author} {\bibfnamefont {G.}~\bibnamefont {Hupin}}, \ and\ \bibinfo
  {author} {\bibfnamefont {F.}~\bibnamefont {Raimondi}},\ }\href {\doibase
  https://doi.org/10.1016/j.physletb.2016.04.021} {\bibfield  {journal}
  {\bibinfo  {journal} {Physics Letters B}\ }\textbf {\bibinfo {volume}
  {757}},\ \bibinfo {pages} {430 } (\bibinfo {year} {2016})}\BibitemShut
  {NoStop}%
\bibitem [{\citenamefont {Thompson}\ and\ \citenamefont
  {Nunes}(2009)}]{thompson2009nuclear}%
  \BibitemOpen
  \bibfield  {author} {\bibinfo {author} {\bibfnamefont {I.}~\bibnamefont
  {Thompson}}\ and\ \bibinfo {author} {\bibfnamefont {F.}~\bibnamefont
  {Nunes}},\ }\href {https://books.google.ca/books?id=sXH2BufDFyAC} {\emph
  {\bibinfo {title} {Nuclear Reactions for Astrophysics: Principles,
  Calculation and Applications of Low-Energy Reactions}}},\ Nuclear Reactions
  for Astrophysics: Principles, Calculation and Applications of Low-energy
  Reactions\ (\bibinfo  {publisher} {Cambridge University Press},\ \bibinfo
  {year} {2009})\BibitemShut {NoStop}%
\bibitem [{\citenamefont {Gebrerufael}\ \emph {et~al.}(2016)\citenamefont
  {Gebrerufael}, \citenamefont {Calci},\ and\ \citenamefont
  {Roth}}]{PhysRevC.93.031301}%
  \BibitemOpen
  \bibfield  {author} {\bibinfo {author} {\bibfnamefont {E.}~\bibnamefont
  {Gebrerufael}}, \bibinfo {author} {\bibfnamefont {A.}~\bibnamefont {Calci}},
  \ and\ \bibinfo {author} {\bibfnamefont {R.}~\bibnamefont {Roth}},\ }\href
  {\doibase 10.1103/PhysRevC.93.031301} {\bibfield  {journal} {\bibinfo
  {journal} {Phys. Rev. C}\ }\textbf {\bibinfo {volume} {93}},\ \bibinfo
  {pages} {031301} (\bibinfo {year} {2016})}\BibitemShut {NoStop}%
\end{thebibliography}


\end{document}